\documentclass{aastex631}
\usepackage{multirow}
\usepackage{subfigure}
\usepackage{CJK}
\usepackage{hyperref}
\hypersetup{hypertex=true,
	colorlinks=true,
	linkcolor=blue,
	anchorcolor=blue,
	citecolor=blue}

\shorttitle{Galactic CN/C$^{15}$N}
\shortauthors{Chen et al.}
\defcitealias{2021ApJS..257...39C}{Paper~I}
\graphicspath{{./}{figures/}}

\begin{document}
\begin{CJK*}{UTF8}{gbsn}
\title{Interstellar Nitrogen Isotope Ratios: Measurements on tracers of C$^{14}$N and C$^{15}$N}

\correspondingauthor{JiangShui Zhang}
\email{jszhang@gzhu.edu.cn}

\author[0000-0001-8980-9663]{J. L. Chen (陈家梁)}
\affil{Center for Astrophysics, Guangzhou University, Guangzhou, 510006, PR China} 

\author[0000-0002-5161-8180]{J. S. Zhang (张江水)}
\affil{Center for Astrophysics, Guangzhou University, Guangzhou, 510006, PR China}

\author[0000-0002-7495-4005]{C. Henkel}
\affil{Max-Planck-Institut f{\"u}r Radioastronomie, Auf dem H{\"u}gel 69, D-53121 Bonn, Germany}


\author[0000-0001-5574-0549]{Y. T. Yan (闫耀庭)}
\affil{Max-Planck-Institut f{\"u}r Radioastronomie, Auf dem H{\"u}gel 69, D-53121 Bonn, Germany}


\author[0000-0002-5634-131X]{H. Z. Yu (余鸿智)}
\affil{Ural Federal University, 19 Mira Street, 620002 Ekaterinburg, Russia} 

\author[0000-0001-9155-0777]{Y. X. Wang (汪友鑫)}
\affil{Center for Astrophysics, Guangzhou University, Guangzhou, 510006, PR China} 
\affil{Max-Planck-Institut f{\"u}r Radioastronomie, Auf dem H{\"u}gel 69, D-53121 Bonn, Germany}

\author[0000-0002-5230-8010]{Y. P. Zou (邹益鹏)}
\affil{Center for Astrophysics, Guangzhou University, Guangzhou, 510006, PR China} 

\author{J. Y. Zhao (赵洁渝)}
\affil{Center for Astrophysics, Guangzhou University, Guangzhou, 510006, PR China} 

\author[0009-0008-8866-0287]{X. Y. Wang (王雪莹)}
\affil{Center for Astrophysics, Guangzhou University, Guangzhou, 510006, PR China} 



\begin{abstract}

\end{abstract}



\begin{abstract}

	The nitrogen isotope ratio $^{14}$N/$^{15}$N is a powerful tool to trace Galactic stellar nucleosynthesis and constraining Galactic chemical evolution. Previous observations have found lower $^{14}$N/$^{15}$N ratios in the Galactic center and higher values in the Galactic disk. This is consistent with the inside-out formation scenario of our Milky Way.
	However, previous studies mostly utilized double isotope ratios also including $^{12}$C/$^{13}$C, which introduces additional uncertainties. Here we therefore present observations of C$^{14}$N and its rare isotopologue, C$^{15}$N, toward a sample of star forming regions, measured by the IRAM 30$\,$m and/or the ARO 12$\,$m telescope at $\lambda$ $\sim$ 3$\,$mm wavelength. For those 35 sources detected in both isotopologues, physical parameters are determined. Furthermore we have obtained nitrogen isotope ratios using the strongest hyperfine components of CN and C$^{15}$N. For those sources showing small deviations from Local Thermodynamical Equilibrium and/or self-absorption, the weakest hyperfine component, likely free of the latter effect, was used to obtain reliable $^{14}$N/$^{15}$N values. Our measured $^{14}$N/$^{15}$N isotope ratios from C$^{14}$N and C$^{15}$N measurements are compatible with those from our earlier measurements of NH$_{3}$ and $^{15}$NH$_{3}$ (\citetalias{2021ApJS..257...39C}), i.e., increasing ratios to a Galacticentric distance of $\sim$9 kpc. The unweighted second order polynomial fit yields $\frac{{\rm C^{14}N}}{{\rm C^{15}N}} =  (-4.85 \pm 1.89)\;{\rm kpc^{-2}} \times R_{\rm GC}^{2} + (82.11 \pm 31.93) \;{\rm kpc^{-1}} \times R_{\rm GC} - (28.12 \pm 126.62)$. Toward the outer galaxy, the isotope ratio tends to decrease, supporting an earlier finding by H$^{13}$CN/HC$^{15}$N. Galactic chemical evolution models are consistent with our measurements of the $^{14}$N/$^{15}$N isotope ratio, i.e. a rising trend from the Galactic center region to approximately 9 kpc, followed by a decreasing trend with increasing $R_{\rm GC}$ toward the outer Galaxy.

\end{abstract}

\section{Introduction} \label{sec:intro}

Metallicity enrichment in the Galactic disk occurs through stellar nucleosynthesis, which converts mainly hydrogen into heavier elements that are then ejected into the interstellar medium \citep[ISM; e.g.,][]{1994ARA&A..32..191W}. Radial metallicity gradients along the Galactic plane have been identified by observations of various kinds of objects such as planetary nebulae \cite[e.g.][]{2010ApJ...724..748H}, H{\sc ii} regions \cite[e.g.][]{2018MNRAS.478.2315E}, stars and circumstellar envelopes \cite[e.g.][]{2017MNRAS.464.3657X} and the ISM \cite[e.g.][]{2017MNRAS.471..987E}, revealing support for an inside-out formation scenario for our Galaxy \citep{1976MNRAS.176...31L}. 
Isotopic abundance ratios in the ISM provide important indicators for stellar nucleosynthesis, ejection patterns from different stars, and Galactic chemical evolution (GCE) \cite[e.g.][]{1994ARA&A..32..191W, 2005ApJ...634.1126M}. These ratios can be obtained accurately through observations of molecular clouds in the radio, mm, and submm bands, by analyzing molecular species with more than one stable isotopologue, including $^{12}$C/$^{13}$C \citep[e.g.,][]{2019ApJ...877..154Y,2023A&A...670A..98Y}, $^{14}$N/$^{15}$N \citep[e.g.,][hereafter \citetalias{2021ApJS..257...39C}]{2021ApJS..257...39C}, $^{18}$O/$^{17}$O \citep{1980Sci...208..663P,2008A&A...487..237W,2015ApJS..219...28Z, 2016RAA....16...47L, 2020IAUGA..30..278Z, 2020ApJS..249....6Z,2023ApJS..268...56Z}, $^{32}$S/$^{34}$S \citep{1996A&A...305..960C,2020ApJ...899..145Y,2020A&A...642A.222H,2023A&A...670A..98Y}, and isotope ratios with rare isotopologues, that is, $^{32}$S/$^{33}$S, $^{32}$S/$^{36}$S, $^{34}$S/$^{33}$S, $^{34}$S/$^{36}$S, and $^{33}$S/$^{36}$S \citep[e.g.,][]{1996A&A...313L...1M,2023A&A...670A..98Y}.

$^{14}$N/$^{15}$N is one of the most important tracers of stellar nucleosynthesis and metal enrichment in the ISM, since both stable isotopes have different synthesis pathways. In brief, both $^{14}$N and $^{15}$N can be produced in the CNO cycle during stellar hydrogen burning as a secondary product, while a primary component of $^{14}$N can also be formed in asymptotic giant branch stars \citep{2004MNRAS.350..407I} and low-metallicity, rotating massive stars \citep{2002A&A...390..561M,2018ApJS..237...13L}. Therefore, $^{14}$N could be considered a more primary product in comparison to $^{15}$N. This can result in an increasing $^{14}$N/$^{15}$N ratio with rising Galactocentric distance ($R_{\rm GC}$), as predicted by models of GCE (for a detailed description, see \citetalias{2021ApJS..257...39C}).

Previous studies of $^{14}$N/$^{15}$N in the ISM have utilized spectral radio lines of various molecular tracers revealing an increasing trend of $^{14}$N/$^{15}$N with $R_{\rm GC}$. However, those $^{14}$N/$^{15}$N measurements were mainly deduced from the method of double isotope ratios \cite[e.g., including HCN, or HNC;][]{1995A&A...295..194D,2012ApJ...744..194A,2018MNRAS.478.3693C,2020MNRAS.498.4663L,2022A&A...667A.151C} by additionally using $^{12}$C/$^{13}$C, which may lead to increased uncertainties. CN, NH$_{3}$ and N$_{2}$H$^{+}$ are the notable exceptions, which offer a direct assessment of line opacities in the crucial $^{14}$N-bearing main species. For this reason, we are starting a systematic study on Galactic interstellar $^{14}$N/$^{15}$N ratios in these three tracers. Observations of $^{14}$NH$_{3}$ and $^{15}$NH$_{3}$ toward a large sample of 210 sources with the Shanghai Tianma 65 m radio telescope (TMRT) and the Effelsberg 100 m telescope have been performed (\citetalias{2021ApJS..257...39C}). Our analysis of $^{14}$NH$_{3}$ and $^{15}$NH$_{3}$ shows a lower abundance ratio toward the Galactic center region with respect to molecular clouds in the Galactic disk, confirming the Galactic radial $^{14}$N/$^{15}$N gradient previously established \cite[e.g.,][]{2018MNRAS.478.3693C}. However, more data from the Galactic center and sources at large Galactocentric distances ($R_{\rm GC}$ \textgreater 8 kpc) are still urgently required to confirm and quantify this gradient more precisely.

In this paper we focus on the cyano radical (CN), which can be detected in a variety of sources with well determined distances (Section \ref{sec:distance}). Systematic discrepancies of nitrogen isotope ratios obtained from various molecular species can be related to the selected species to reveal and quantify for the first time chemical aspects caused by potential fractionation \citep{2015A&A...576A..99R,2019MNRAS.486.4805V} on a Galaxy wide scale. Observations of the $N$ = 1--0 lines of C$^{14}$N and C$^{15}$N are presented in Section \ref{sec:obser}. The main results of our measurements are illustrated in Section \ref{sec:result}. Section \ref{sec:discussion} discusses potential processes that could contaminate and affect the nitrogen ratios and provides a detailed comparison with results from earlier studies. Our main conclusions are summarized in Section \ref{sec:summary}.

\section{Sample and observations} \label{sec:obser}

\subsection{Sample and distances}\label{sec:distance}

Trigonometric parallax determination is a very direct and accurate method to measure the distances of sources from the Sun. Over the past decade, masers have been measured with high angular resolution in approximately 200 massive star-forming regions to determine trigonometric parallaxes and hence, distances as well as proper motions \citep[e.g.][]{2014ApJ...783..130R,2019ApJ...885..131R}. A total of 141 sources from previous studies \citep{1993A&AS...98..589W,2014ApJ...783..130R,2019ApJ...885..131R} are selected as targets for our measurement of nitrogen isotope ratios. Among our sample, the distances of 101 sources were obtained from trigonometric parallax measurements \citep{2014ApJ...783..130R,2019ApJ...885..131R}. For the other 40 sources, their distance values were derived from the Parallax-Based Distance Calculator V2 \footnote{\url{http://bessel.vlbi-astrometry.org/node/378}} \citep{2019ApJ...885..131R}. Sources are allocated to Galactic spiral arms using a Bayesian approach based on their ($l, b, v$) coordinates in relation to arm signatures seen in CO and H{\sc i} surveys \citep{1970IAUS...38..126W,1980ApJ...239L..53C}. The most reasonable distance (near or far) can be obtained by utilizing a full distance probability density function from the parallax-based distance calculator, which considers a source’s kinematic distance, displacement from the plane, and proximity to individual parallax sources. This is considered to be an important improvement to reveal radial variations of $^{14}$N/$^{15}$N in an unbiased way. The Galactocentric distances of targets were calculated via the heliocentric distance  \citep{2009ApJ...699.1153R},
\begin{equation}
	R_{\rm GC} = \sqrt{[R_{0}\cos(l)-d]^{2}+R_{0}^{2}\sin^{2}(l)}.
\end{equation}
Here $l$ is the Galactic longitude. $d$ and $R_0$ are the distance of the source from the Sun \citep{2014ApJ...783..130R,2019ApJ...885..131R} and the distance of the Sun from the Galactic center (8.122 $\pm$ 0.031 kpc) respectively. The latter is taken from the \cite{2018A&A...615L..15G}. The source list is presented in the Appendix.

\subsection{Observations}\label{sec:Obs}

\subsubsection{IRAM 30$\,$m observations}\label{sec:ObsIRAM}

Observations of the $N$ = 1--0 transitions of C$^{14}$N and C$^{15}$N lines were performed in 2016 June within project 013-16 (PI Zhiwei Liu), in 2020 August and September within project 004-20 (PI Yaoting Yan), and in 2021 April within project 125-20 (PI Yaoting Yan), with the Institut de Radio Astronomie Millimétrique (IRAM) 30 m single dish telescope\footnote{The IRAM 30 m is supported by the Institut National des Sciences de L’univers/Centre National de la Recherche Scientifique, (INSU/CNRS, France), Max-Planck-Gesellschaft (MPG, Germany), and Instituto Geográfico Nacional (IGN, Spain).}, at the Pico Veleta Observatory (Granada, Spain). 104 sources among our sample were observed. The center frequencies for C$^{14}$N and C$^{15}$N were set at 113490.98 and 110024.59 MHz, respectively, with a corresponding beam size of $\sim$23\arcsec \citep{2020ApJS..249....6Z}. An Eight Mixer Receiver (EMIR) with dual polarization and a Fourier Transform Spectrometer (FTS) backend was used, providing a spectral resolution of 192 kHz or $\sim$0.5 km s$^{-1}$ around 112 GHz. A standard position switching mode was employed with the off position at a (−30$^\prime$, 0$^\prime$) or (30$^\prime$, 0$^\prime$) offset in R.A. and decl., or in azimuth and elevation from the source. The on-source integration time depends on line strength, with an integration time ranging from 0.5-7 hr for each source. We checked the pointing every two hours toward nearby strong continuum sources (e.g., 3C 123, or NGC 7027). Focus calibrations were obtained at the beginning of the observations and during sunset and sunrise toward strong quasars \citep{2023A&A...670A..98Y}. The system temperatures were 200-300 K on an antenna temperature ($T_{\rm A}^{*}$) scale for the observations, with an root mean square (rms) noise of 10 -- 200\,mK in an unsmoothed channel. The main beam brightness temperature ($T_{\rm mb}$) was obtained from the antenna temperature $T_{\rm A}^{*}$ by multipling its value by the forward to the main beam efficiency (F$_{eff}$/B$_{eff}$ $\sim$ 0.94/0.78 = 1.21).

\subsubsection{{\it ARO} 12 m Observations}\label{sec:ObsARO}

Using the ARO 12 m telescope on Kitt Peak, Tuscon, AZ, USA\footnote{The ARO 12 m is supported by the Department of Astronomy and Steward Observatory of the University of Arizona (USA).}, with a corresponding beam size of $\sim$64\arcsec \citep{2015A&A...578A..70K}, we carried out  our observations of the $N$ = 1--0 transitions of C$^{14}$N and C$^{15}$N toward the remaining 37 sources in our sample.  For comparison, additional 10 sources targeted by the IRAM 30m were also observed by the ARO 12m. Observations were performed remotely from Koeln, Germany and Guangzhou University, China, in 2021 March and June within project Yan\_20A\_1 (PI: Yaoting Yan). The new dual-polarization receiver containing Atacama Large Millimeter/submillimeter Array (ALMA) Band 3 (83∼116 GHz) sideband-separating (SBS) mixers was employed. The new ARO Wideband Spectrometer (AROWS) backend was used, with a 120 MHz bandwidth (6400 channels), providing a spectral resolution of 18.75 kHz ($\sim$0.05 $\rm km \, \rm s^{-1}$). Observations were performed in position switching mode with an off-position 30$^\prime$ apart. System temperatures were 150 -- 250 K on a $T_{A}^{*}$ scale with an rms noise level of $\sim$30 mK. The $T_{\rm mb}$ scale can be obtained from the $T_{A}^{*}$ scale by $T_{\rm mb}$ = $T_{A}^{}$/$\eta_ {b}$, where $\eta_ {b}$ is the main beam efficiency correction factor \citep[e.g.,][]{2018ApJ...862...63C, 2020ApJS..249....6Z}, with a mean value of $\sim$0.83 during our observations.

In summary, 104 sources were observed by the IRAM and 47 sources by the ARO telescope, including 10 sources measured by both telescopes. Observational parameters of our sample are listed in the Appendix.


\section{Results}\label{sec:result}

\subsection{Spectral Fitting Results}\label{sec:detection}
\begin{figure*}[htbp]
	\centering
	{\includegraphics[width=4.4cm]{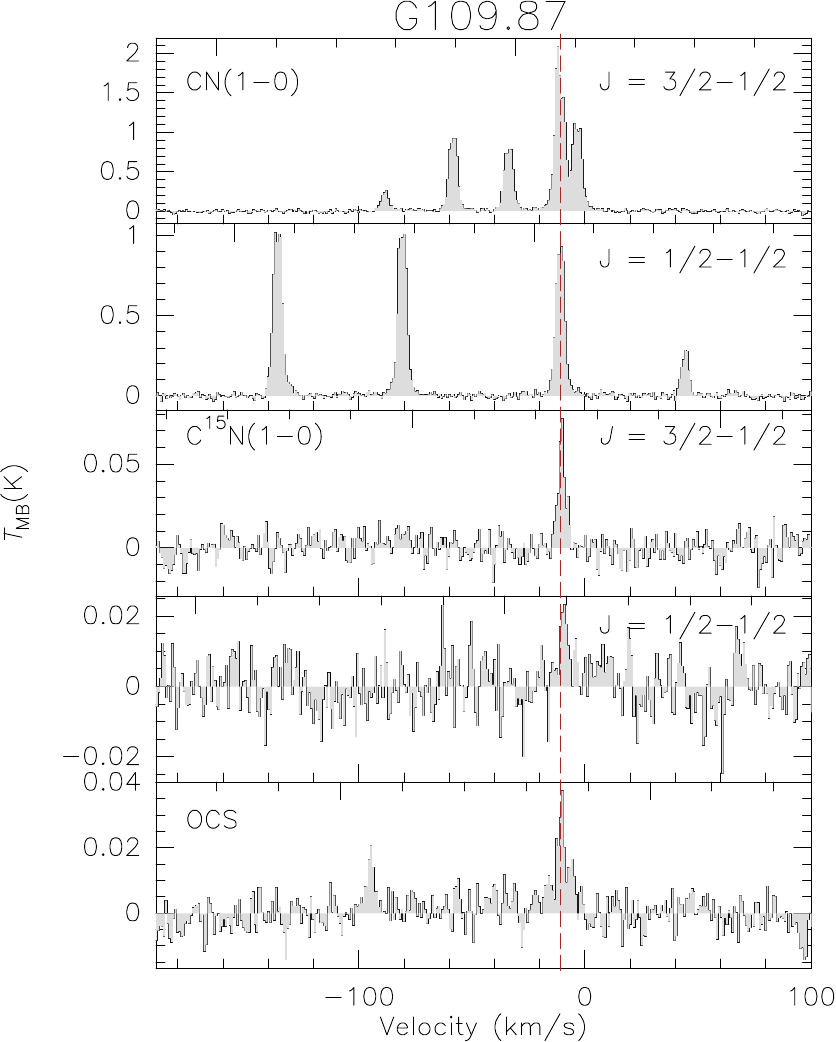}} 
	{\includegraphics[width=4.4cm]{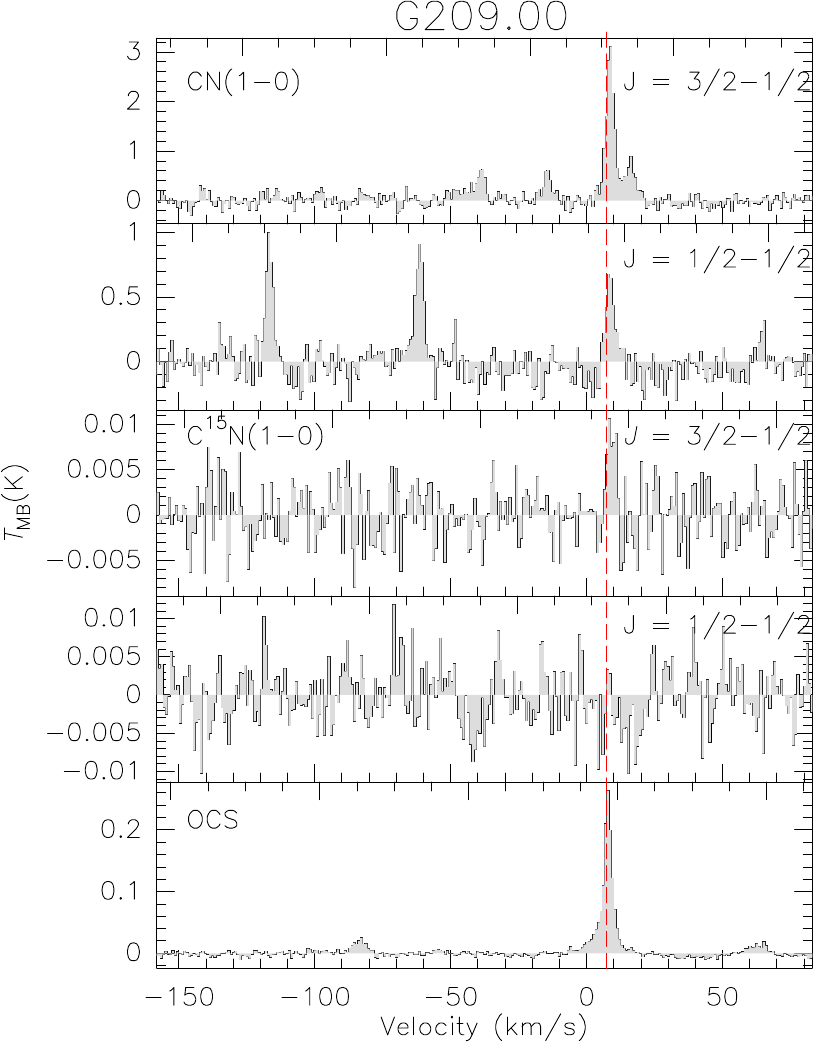}} 
	{\includegraphics[width=4.4cm]{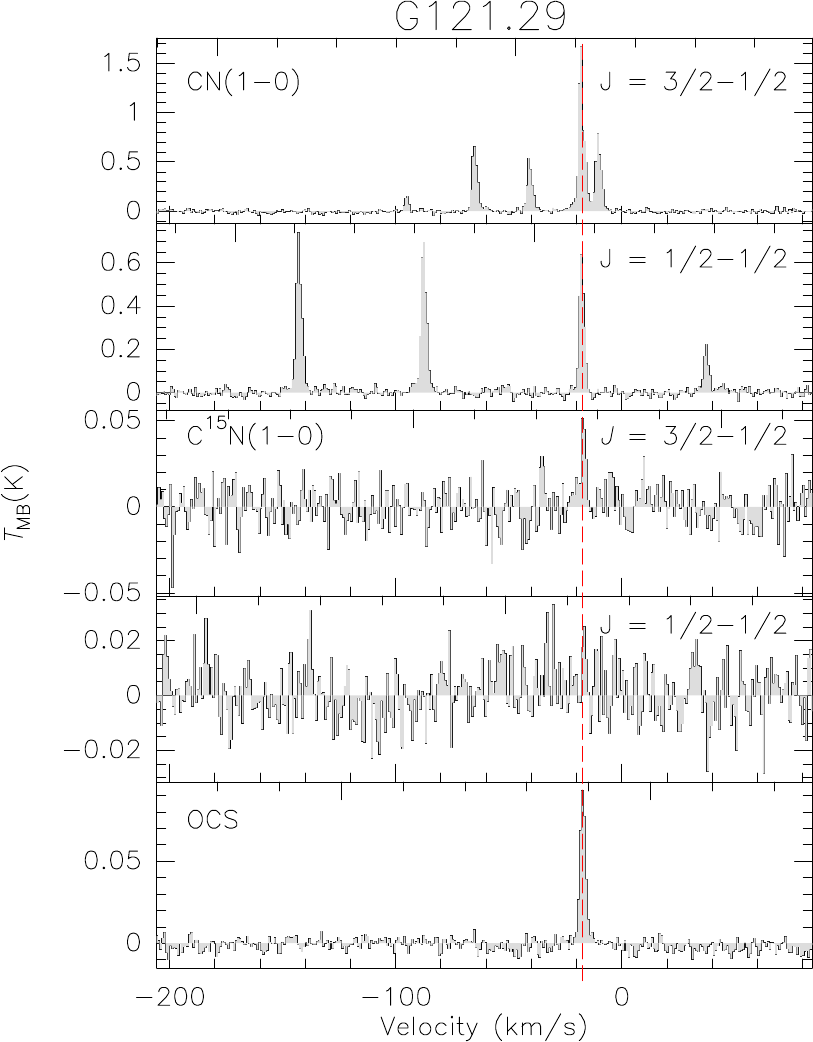}} 
	{\includegraphics[width=4.4cm]{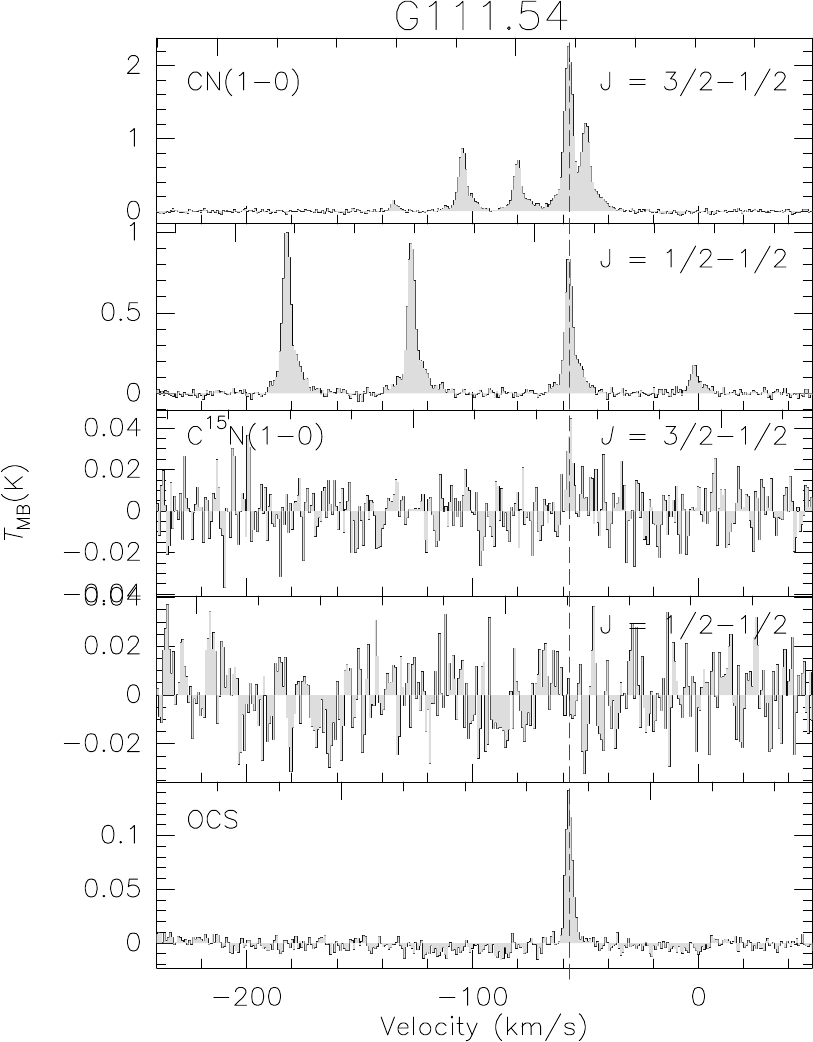}} 
	{\includegraphics[width=4.4cm]{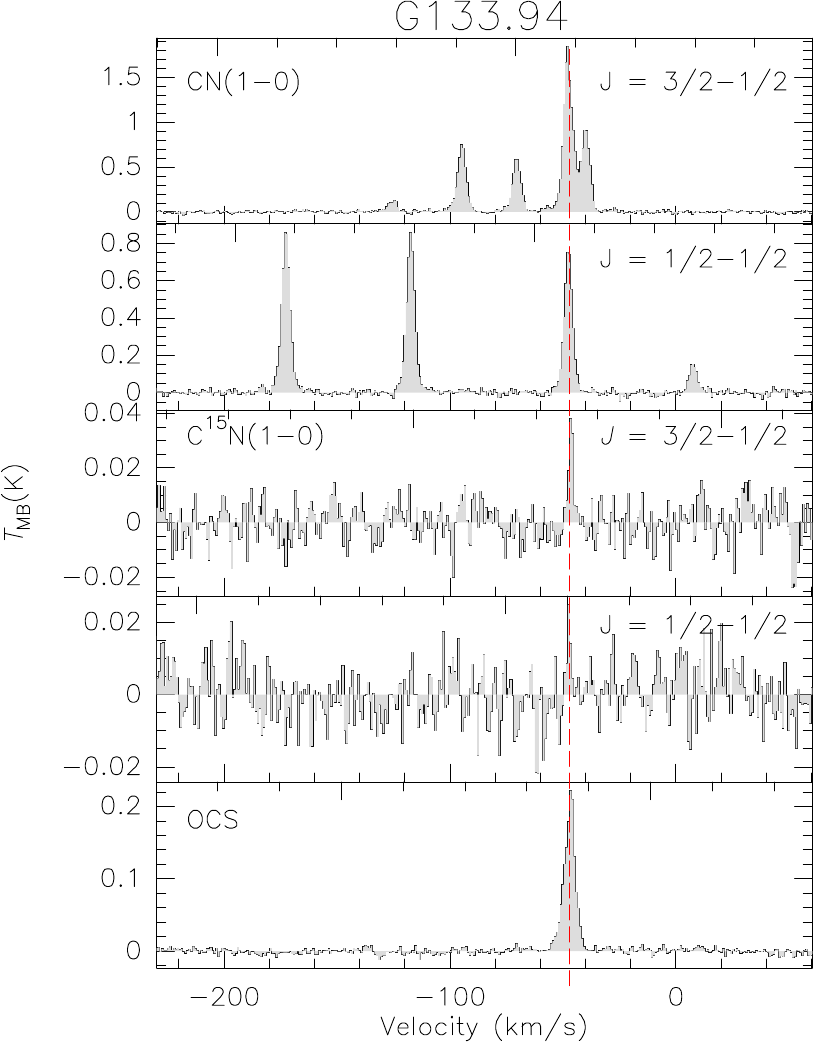}} 
	{\includegraphics[width=4.4cm]{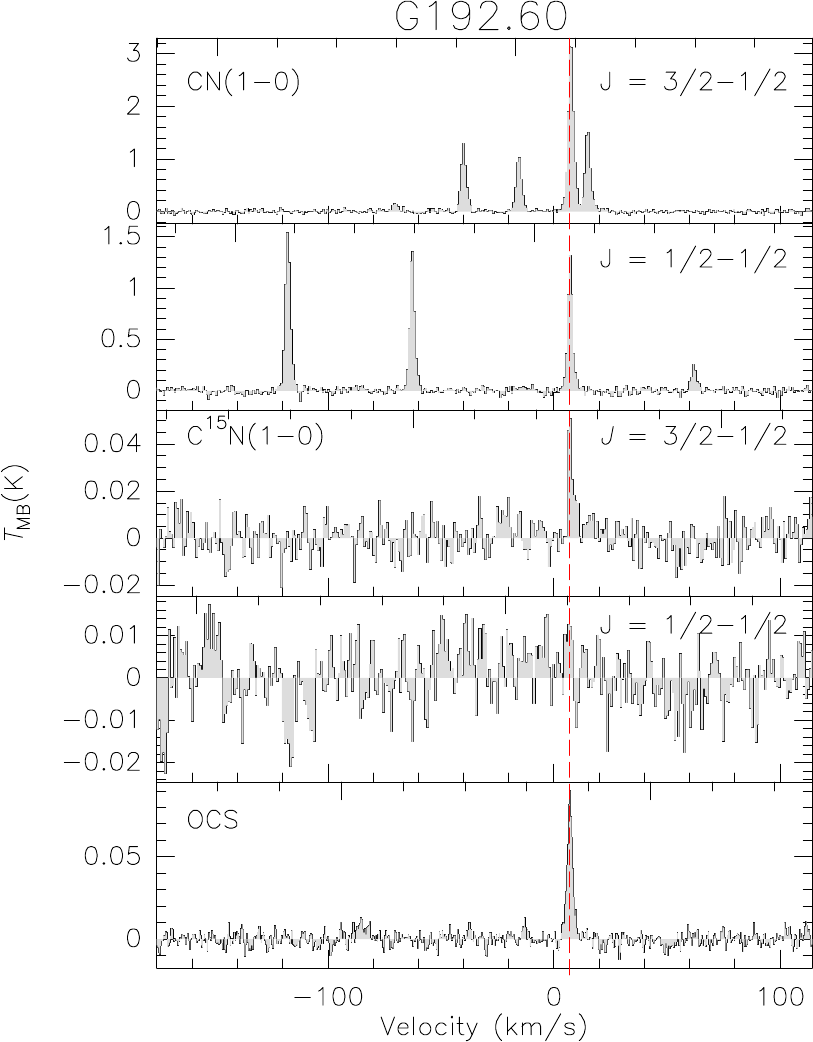}} 
	{\includegraphics[width=4.4cm]{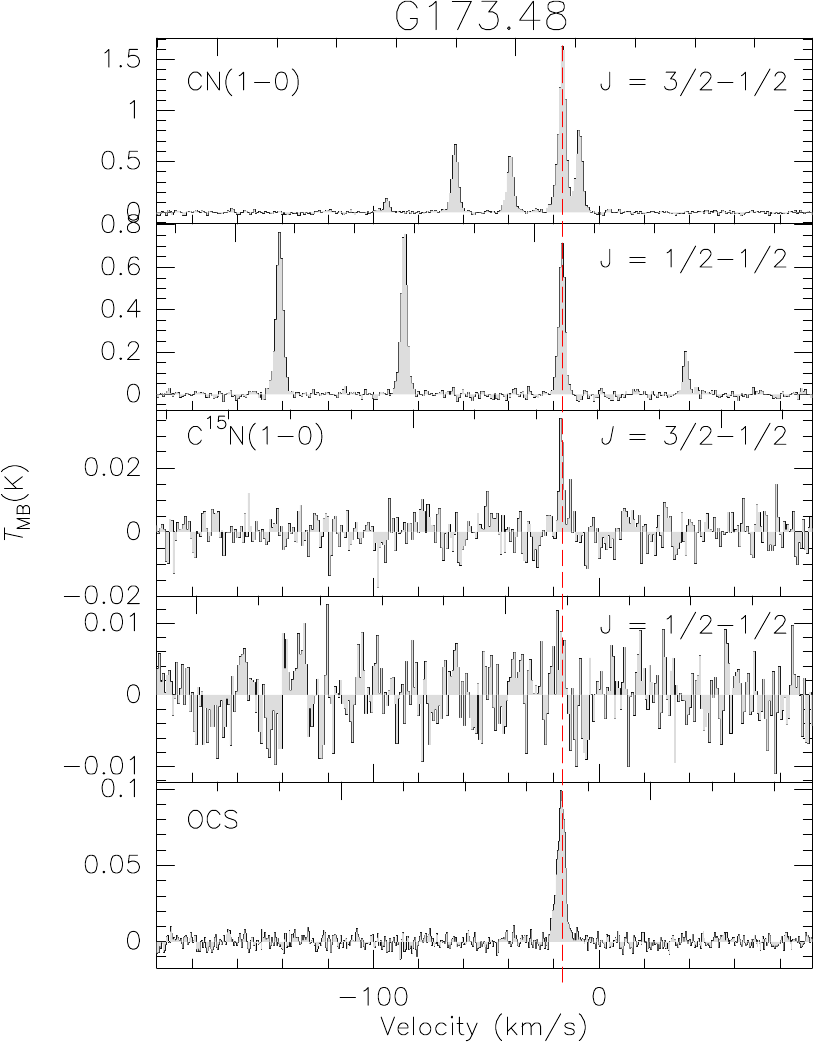}} 
	{\includegraphics[width=4.4cm]{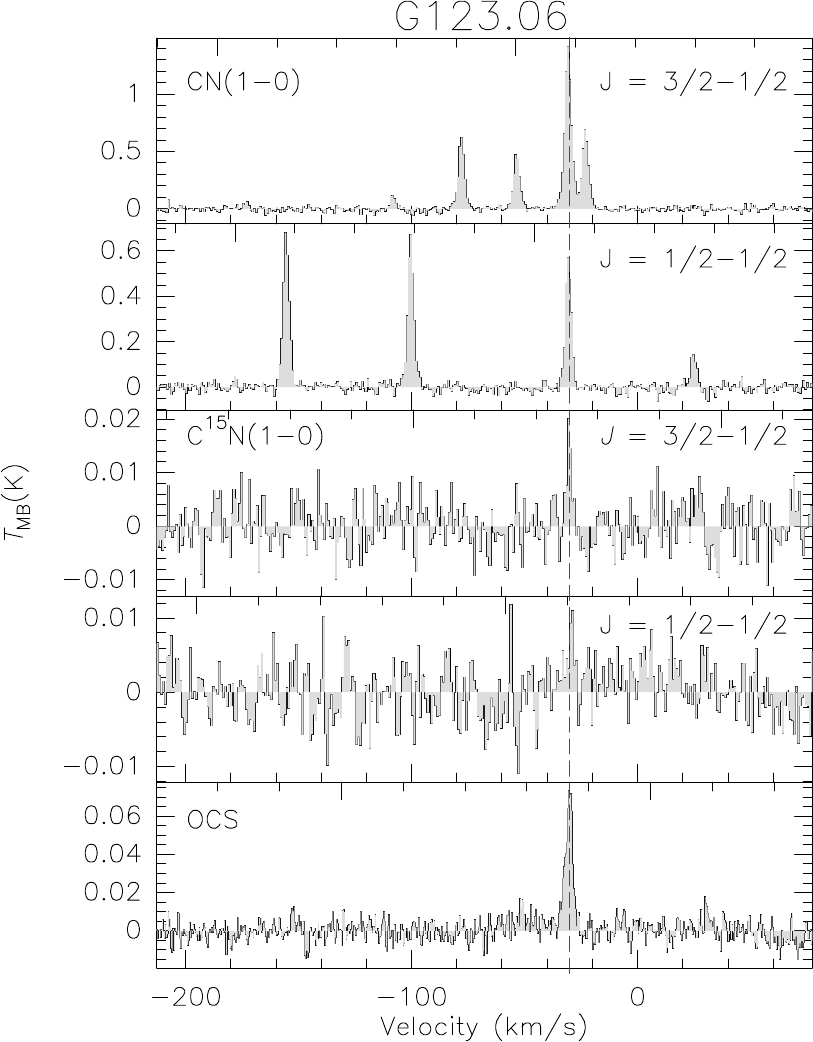}}  
	{\includegraphics[width=4.4cm]{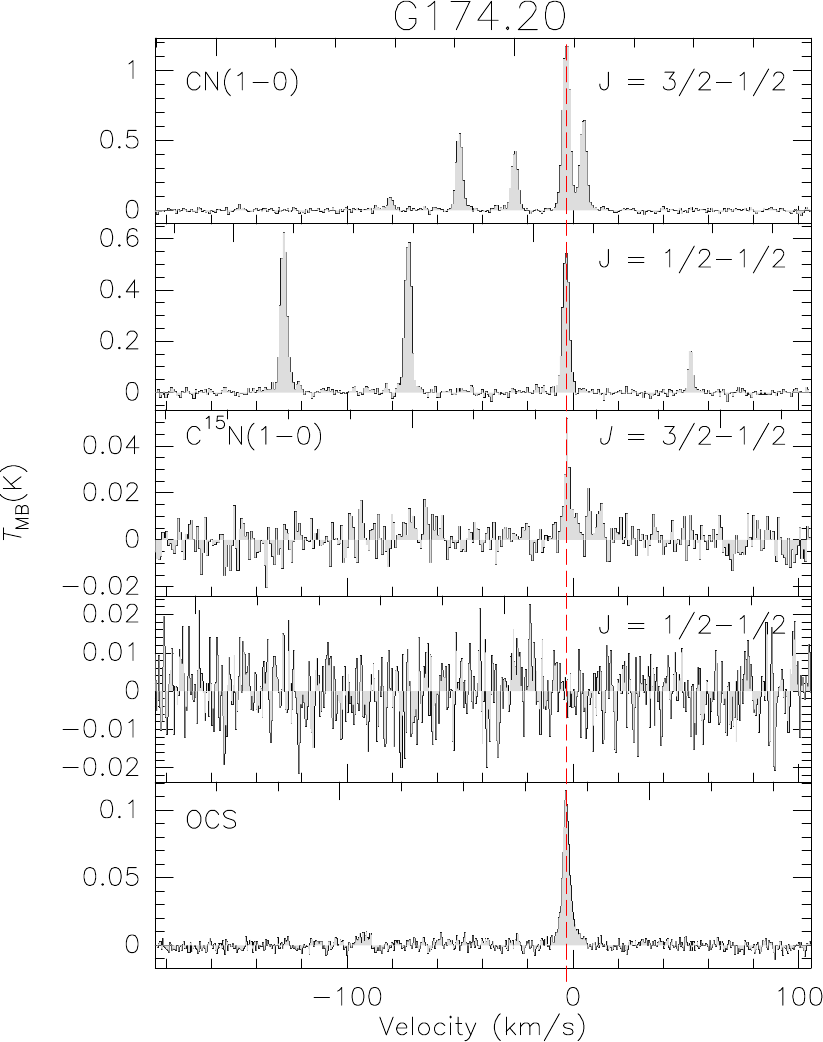}} 
	{\includegraphics[width=4.4cm]{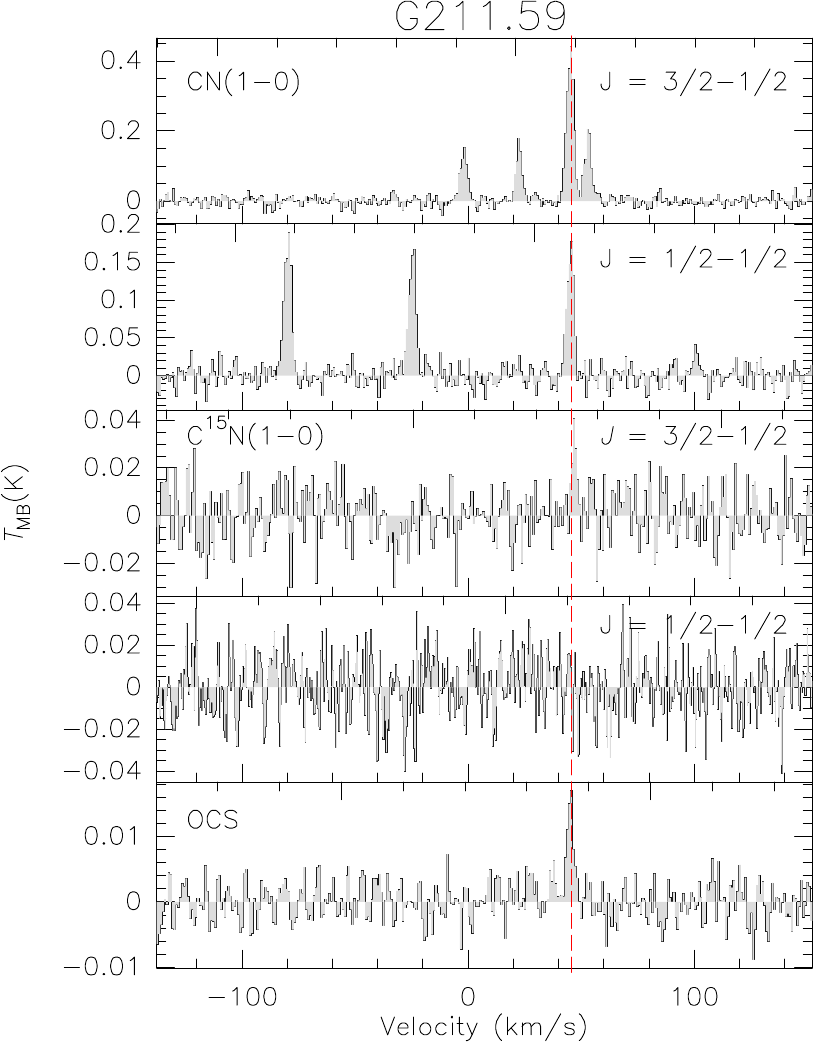}} 
	\caption{ARO spectra of those 10 sources with detected C$^{15}$N lines, after subtracting baselines and applying Hanning smoothing leading to  $\sim$0.8 km s$^{-1}$ wide channels. Vertical red dashed lines mark the line center velocity of the source measured from a Gaussian fit to the OCS J = 9 -- 8 line.}
	\label{ARO15CNfigself}
\end{figure*}

Among the 104 IRAM 30m targets, 28 sources were detected in both $N$ = 1--0 C$^{14}$N and C$^{15}$N,  while 10 out of 47 ARO targets were detected in both lines. Among the 10 sources observed by both telescopes, three sources (G109.87, G111.54, and G133.94) were detected in both lines. In total, we detected 35 sources in the C$^{14}$N and C$^{15}$N, $N$ = 1--0 lines within our sample of 141 Galactic molecular clouds. The main reason of low detection rate should be low abundance of C$^{15}$N, which is typically 100 times lower than that of C$^{14}$N. Selecting targets with strong CN emission from future CN survey would be good way to enhance the detection rate. The C$^{14}$N and C$^{15}$N spectra for our 35 sources from the IRAM 30 m and ARO 12 m telescopes are presented in Figures \ref{IRAM15CNfigself} and \ref{ARO15CNfigself}, respectively. 

The Continuum and Line Analysis Single-dish Software (CLASS) of the Grenoble Image and Line Data Analysis Software packages \footnote{\url{http://http://www.iram.fr/IRAMFR/GILDAS/}} \citep[GILDAS, e. g.,][]{2000ASPC..217..299G} was employed to reduce the spectral line data. After subtracting baselines and applying Hanning smoothing, the line parameters were obtained from Gaussian fits for detected lines (signal-to-noise (S/N) ratios \textgreater 3), with a spectral resolution of $\sim$1.1 km\,s$^{-1}$ for IRAM 30 m and 0.8 km\,s$^{-1}$ for ARO 12 m observations, respectively. Line parameters were obtained from Gaussian fits to the C$^{14}$N and C$^{15}$N lines.

11 sources show blended component features in their C$^{14}$N spectra , i.e., the strongest component ($J$ = 3/2 -- 1/2 $F$ = 5/2-3/2) is blended with the $J$ = 3/2 -- 1/2 $F$ = 3/2 -- 1/2 component (see Figures \ref{IRAM15CNfigself} and \ref{ARO15CNfigself}). 
For these sources, we summed line intensities over the entire velocity interval required to cover all component features (using the first moment by the "Print" command in CLASS), to determine the integrated intensities of the two spin-doublet transitions of CN ($J$ = 1/2 -- 1/2 and $J$ = 3/2 -- 1/2). Table \ref{tab:obser} summarizes the spectral line parameters of the 35 sources, including the integrated intensity and its uncertainty, the LSR velocity, and the main beam brightness peak temperatures of the C$^{14}$N and C$^{15}$N lines.

\startlongtable
\renewcommand\tabcolsep{7.0pt} 
		\begin{deluxetable*}{lccccllcclccl}
			\tablecaption{Observational parameters of the $N$ = 1 -- 0 rotationnal lines of C$^{14}$N and C$^{15}$N obtained from Gaussian fits \label{tab:obser}}
			\tablewidth{0pt}
			\tablehead{
				\colhead{Object}&
				\colhead{Telescope}&
				\colhead{Molecule}&
				\colhead{Transition}&
				\multicolumn{3}{c}{$ \int{T_{\rm mb}{\rm{d}} v}$}&
				\colhead{$V_{\rm LSR}$}&
				\colhead{$T_{\rm mb}$}\\
				\colhead{}&
				\colhead{}&
				\colhead{}&
				\colhead{}&
				\multicolumn{3}{c}{(${\rm K} \, {\rm km} \, {\rm s}^{-1}$)}&
				\colhead{($\rm km \, \rm s^{-1}$)}&
				\colhead{(K)}\\
				\colhead{(1)}&
				\colhead{(2)}&
				\colhead{(3)}&
				\colhead{(4)}&
				\multicolumn{3}{c}{(5)}&
				\colhead{(6)}&
				\colhead{(7)}
			}
			\startdata
			\multirow{2}*{G010.47}&\multirow{2}*{IRAM}&\multirow{2}*{C$^{14}$N, $N$=1-0 }&$I_{1}$&55.57 $\pm$ 3.70 &\multirow{2}*{$\bigl\}$}&\multirow{2}*{88.80 $\pm$ 3.70 }&\multirow{2}*{67.73 $\pm$ 0.10 }&\multirow{1}*{3.86 }\\
			&&&$I_{2}$&33.23 $\pm$ 0.30 && &&0.32 \\
			&&\multirow{2}*{C$^{15}$N, $N$=1-0 }&$I_{1}$&0.36 $\pm$ 0.20 &\multirow{2}*{$\bigl\}$}&\multirow{2}*{0.36 $\pm$ 0.20 }&\multirow{2}*{67.59 $\pm$ 0.23 }&\multirow{2}*{0.09 }\\
			&&&$I_{2}$&... &&&&\\
			\multirow{2}*{G009.62}&\multirow{2}*{IRAM}&\multirow{2}*{C$^{14}$N, $N$=1-0 }&$I_{1}$&55.76 $\pm$ 8.90 &\multirow{2}*{$\bigl\}$}&\multirow{2}*{98.03 $\pm$ 17.60 }&\multirow{2}*{4.79 $\pm$ 0.52 }&\multirow{1}*{5.16 }\\
			&&&$I_{2}$&42.90 $\pm$ 4.80 && &&0.57 \\
			&&\multirow{2}*{C$^{15}$N, $N$=1-0 }&$I_{1}$&0.31 $\pm$ 0.01 &\multirow{2}*{$\bigl\}$}&\multirow{2}*{0.31 $\pm$ 0.01 }&\multirow{2}*{4.78 $\pm$ 0.72 }&\multirow{2}*{0.07 }\\
			&&&$I_{2}$&... &&&&\\
			\multirow{2}*{G010.621}&\multirow{2}*{IRAM}&\multirow{2}*{C$^{14}$N, $N$=1-0 }&$I_{1}$&78.79 $\pm$ 1.40 &\multirow{2}*{$\bigl\}$}&\multirow{2}*{149.50 $\pm$ 1.60 }&\multirow{2}*{-3.06 $\pm$ 0.18 }&\multirow{1}*{8.92 }\\
			&&&$I_{2}$&70.71 $\pm$ 0.80 && &&0.82 \\
			&&\multirow{2}*{C$^{15}$N, $N$=1-0 }&$I_{1}$&0.64 $\pm$ 0.07 &\multirow{2}*{$\bigl\}$}&\multirow{2}*{0.64 $\pm$ 0.07 }&\multirow{2}*{-2.92 $\pm$ 0.31 }&\multirow{2}*{0.10 }\\
			&&&$I_{2}$&... &&&&\\
			\multirow{2}*{G023.43}&\multirow{2}*{IRAM}&\multirow{2}*{C$^{14}$N, $N$=1-0 }&$I_{1}$&18.31 $\pm$ 1.20 &\multirow{2}*{$\bigl\}$}&\multirow{2}*{42.93 $\pm$ 3.00 }&\multirow{2}*{101.63 $\pm$ 1.04 }&\multirow{1}*{1.42 }\\
			&&&$I_{2}$&21.13 $\pm$ 2.70 && &&0.45 \\
			&&\multirow{2}*{C$^{15}$N, $N$=1-0 }&$I_{1}$&0.22 $\pm$ 0.30 &\multirow{2}*{$\bigl\}$}&\multirow{2}*{0.22 $\pm$ 0.30 }&\multirow{2}*{101.41 $\pm$ 0.23 }&\multirow{2}*{0.06 }\\
			&&&$I_{2}$&... &&&&\\
			\multirow{2}*{G029.86}&\multirow{2}*{IRAM}&\multirow{2}*{C$^{14}$N, $N$=1-0 }&$I_{1}$&40.16 $\pm$ 6.80 &\multirow{2}*{$\bigl\}$}&\multirow{2}*{63.39 $\pm$ 7.70 }&\multirow{2}*{101.68 $\pm$ 1.03 }&\multirow{1}*{3.58 }\\
			&&&$I_{2}$&23.23 $\pm$ 3.50 && &&0.44 \\
			&&\multirow{2}*{C$^{15}$N, $N$=1-0 }&$I_{1}$&0.21 $\pm$ 0.20 &\multirow{2}*{$\bigl\}$}&\multirow{2}*{0.21 $\pm$ 0.20 }&\multirow{2}*{101.76 $\pm$ 0.47 }&\multirow{2}*{0.05 }\\
			&&&$I_{2}$&... &&&&\\
			\multirow{2}*{G013.87}&\multirow{2}*{IRAM}&\multirow{2}*{C$^{14}$N, $N$=1-0 }&$I_{1}$&34.60 $\pm$ 3.40 &\multirow{2}*{$\bigl\}$}&\multirow{2}*{55.79 $\pm$ 3.60 }&\multirow{2}*{50.04 $\pm$ 0.52 }&\multirow{1}*{2.27 }\\
			&&&$I_{2}$&21.19 $\pm$ 1.00 && &&0.42 \\
			&&\multirow{2}*{C$^{15}$N, $N$=1-0 }&$I_{1}$&0.16 $\pm$ 0.03 &\multirow{2}*{$\bigl\}$}&\multirow{2}*{0.16 $\pm$ 0.03 }&\multirow{2}*{50.95 $\pm$ 0.27 }&\multirow{2}*{0.03 }\\
			&&&$I_{2}$&... &&&&\\
			\multirow{2}*{G001.14}&\multirow{2}*{IRAM}&\multirow{2}*{C$^{14}$N, $N$=1-0 }&$I_{1}$&8.94 $\pm$ 1.30 &\multirow{2}*{$\bigl\}$}&\multirow{2}*{17.57 $\pm$ 1.00 }&\multirow{2}*{-17.70 $\pm$ 0.13 }&\multirow{1}*{1.28 }\\
			&&&$I_{2}$&8.63 $\pm$ 0.80 && &&0.44 \\
			&&\multirow{2}*{C$^{15}$N, $N$=1-0 }&$I_{1}$&0.37 $\pm$ 0.10 &\multirow{2}*{$\bigl\}$}&\multirow{2}*{0.37 $\pm$ 0.10 }&\multirow{2}*{-17.22 $\pm$ 0.20 }&\multirow{2}*{0.22 }\\
			&&&$I_{2}$&... &&&&\\
			\multirow{2}*{G029.95}&\multirow{2}*{IRAM}&\multirow{2}*{C$^{14}$N, $N$=1-0 }&$I_{1}$&40.90 $\pm$ 0.30 &\multirow{2}*{$\bigl\}$}&\multirow{2}*{66.69 $\pm$ 0.50 }&\multirow{2}*{96.83 $\pm$ 0.28 }&\multirow{1}*{5.49 }\\
			&&&$I_{2}$&25.78 $\pm$ 0.40 && &&0.50 \\
			&&\multirow{2}*{C$^{15}$N, $N$=1-0 }&$I_{1}$&0.22 $\pm$ 0.10 &\multirow{2}*{$\bigl\}$}&\multirow{2}*{0.22 $\pm$ 0.10 }&\multirow{2}*{98.65 $\pm$ 0.34 }&\multirow{2}*{0.05 }\\
			&&&$I_{2}$&... &&&&\\
			\multirow{2}*{G005.88}&\multirow{2}*{IRAM}&\multirow{2}*{C$^{14}$N, $N$=1-0 }&$I_{1}$&99.70 $\pm$ 9.60 &\multirow{2}*{$\bigl\}$}&\multirow{2}*{165.63 $\pm$ 10.80 }&\multirow{2}*{9.53 $\pm$ 0.02 }&\multirow{1}*{11.93 }\\
			&&&$I_{2}$&65.92 $\pm$ 5.10 && &&1.37 \\
			&&\multirow{2}*{C$^{15}$N, $N$=1-0 }&$I_{1}$&0.54 $\pm$ 0.10 &\multirow{2}*{$\bigl\}$}&\multirow{2}*{0.68 $\pm$ 0.10 }&\multirow{2}*{9.30 $\pm$ 0.12 }&\multirow{2}*{0.12 }\\
			&&&$I_{2}$&0.14 $\pm$ 0.03 &&&&\\
			\multirow{2}*{G012.81}&\multirow{2}*{IRAM}&\multirow{2}*{C$^{14}$N, $N$=1-0 }&$I_{1}$&79.39 $\pm$ 5.90 &\multirow{2}*{$\bigl\}$}&\multirow{2}*{133.96 $\pm$ 6.50 }&\multirow{2}*{35.35 $\pm$ 0.45 }&\multirow{1}*{6.13 }\\
			&&&$I_{2}$&54.56 $\pm$ 2.80 && &&1.17 \\
			&&\multirow{2}*{C$^{15}$N, $N$=1-0 }&$I_{1}$&0.74 $\pm$ 0.50 &\multirow{2}*{$\bigl\}$}&\multirow{2}*{0.74 $\pm$ 0.50 }&\multirow{2}*{35.50 $\pm$ 0.23 }&\multirow{2}*{0.13 }\\
			&&&$I_{2}$&... &&&&\\
			\multirow{2}*{G012.88}&\multirow{2}*{IRAM}&\multirow{2}*{C$^{14}$N, $N$=1-0 }&$I_{1}$&16.45 $\pm$ 1.60 &\multirow{2}*{$\bigl\}$}&\multirow{2}*{28.61 $\pm$ 1.90 }&\multirow{2}*{35.40 $\pm$ 0.02 }&\multirow{1}*{2.03 }\\
			&&&$I_{2}$&12.16 $\pm$ 0.90 && &&0.62 \\
			&&\multirow{2}*{C$^{15}$N, $N$=1-0 }&$I_{1}$&0.29 $\pm$ 0.10 &\multirow{2}*{$\bigl\}$}&\multirow{2}*{0.29 $\pm$ 0.10 }&\multirow{2}*{35.55 $\pm$ 0.21 }&\multirow{2}*{0.07 }\\
			&&&$I_{2}$&... &&&&\\
			\multirow{2}*{G049.49}&\multirow{2}*{IRAM}&\multirow{2}*{C$^{14}$N, $N$=1-0 }&$I_{1}$&107.62 $\pm$ 16.50 &\multirow{2}*{$\bigl\}$}&\multirow{2}*{176.60 $\pm$ 17.40 }&\multirow{2}*{61.13 $\pm$ 0.52 }&\multirow{1}*{9.12 }\\
			&&&$I_{2}$&68.98 $\pm$ 5.60 && &&0.48 \\
			&&\multirow{2}*{C$^{15}$N, $N$=1-0 }&$I_{1}$&0.92 $\pm$ 0.20 &\multirow{2}*{$\bigl\}$}&\multirow{2}*{0.92 $\pm$ 0.20 }&\multirow{2}*{61.90 $\pm$ 0.44 }&\multirow{2}*{0.07 }\\
			&&&$I_{2}$&... &&&&\\
			\multirow{2}*{G049.48}&\multirow{2}*{IRAM}&\multirow{2}*{C$^{14}$N, $N$=1-0 }&$I_{1}$&101.65 $\pm$ 8.70 &\multirow{2}*{$\bigl\}$}&\multirow{2}*{167.55 $\pm$ 9.30 }&\multirow{2}*{62.31 $\pm$ 0.10 }&\multirow{1}*{7.76 }\\
			&&&$I_{2}$&65.89 $\pm$ 3.30 && &&0.48 \\
			&&\multirow{2}*{C$^{15}$N, $N$=1-0 }&$I_{1}$&0.34 $\pm$ 0.10 &\multirow{2}*{$\bigl\}$}&\multirow{2}*{0.34 $\pm$ 0.10 }&\multirow{2}*{62.05 $\pm$ 0.37 }&\multirow{2}*{0.08 }\\
			&&&$I_{2}$&... &&&&\\
			\multirow{2}*{G035.02}&\multirow{2}*{IRAM}&\multirow{2}*{C$^{14}$N, $N$=1-0 }&$I_{1}$&19.68 $\pm$ 5.10 &\multirow{2}*{$\bigl\}$}&\multirow{2}*{34.46 $\pm$ 5.40 }&\multirow{2}*{52.98 $\pm$ 0.02 }&\multirow{1}*{2.69 }\\
			&&&$I_{2}$&14.78 $\pm$ 1.80 && &&0.34 \\
			&&\multirow{2}*{C$^{15}$N, $N$=1-0 }&$I_{1}$&0.17 $\pm$ 0.10 &\multirow{2}*{$\bigl\}$}&\multirow{2}*{0.17 $\pm$ 0.10 }&\multirow{2}*{52.75 $\pm$ 0.22 }&\multirow{2}*{0.12 }\\
			&&&$I_{2}$&... &&&&\\
			\multirow{2}*{G035.19}&\multirow{2}*{IRAM}&\multirow{2}*{C$^{14}$N, $N$=1-0 }&$I_{1}$&45.33 $\pm$ 2.71 &\multirow{2}*{$\bigl\}$}&\multirow{2}*{72.22 $\pm$ 2.92 }&\multirow{2}*{35.34 $\pm$ 1.27 }&\multirow{1}*{2.47 }\\
			&&&$I_{2}$&26.89 $\pm$ 1.09 && &&0.65 \\
			&&\multirow{2}*{C$^{15}$N, $N$=1-0 }&$I_{1}$&0.22 $\pm$ 0.20 &\multirow{2}*{$\bigl\}$}&\multirow{2}*{0.22 $\pm$ 0.20 }&\multirow{2}*{35.20 $\pm$ 0.38 }&\multirow{2}*{0.06 }\\
			&&&$I_{2}$&... &&&&\\
			\multirow{2}*{G035.14}&\multirow{2}*{IRAM}&\multirow{2}*{C$^{14}$N, $N$=1-0 }&$I_{1}$&31.73 $\pm$ 8.40 &\multirow{2}*{$\bigl\}$}&\multirow{2}*{52.27 $\pm$ 6.50 }&\multirow{2}*{31.41 $\pm$ 0.52 }&\multirow{1}*{2.78 }\\
			&&&$I_{2}$&20.54 $\pm$ 1.20 && &&0.75 \\
			&&\multirow{2}*{C$^{15}$N, $N$=1-0 }&$I_{1}$&0.22 $\pm$ 0.10 &\multirow{2}*{$\bigl\}$}&\multirow{2}*{0.22 $\pm$ 0.10 }&\multirow{2}*{31.89 $\pm$ 0.09 }&\multirow{2}*{0.07 }\\
			&&&$I_{2}$&... &&&&\\
			\multirow{2}*{G059.78}&\multirow{2}*{IRAM}&\multirow{2}*{C$^{14}$N, $N$=1-0 }&$I_{1}$&30.62 $\pm$ 0.10 &\multirow{2}*{$\bigl\}$}&\multirow{2}*{49.37 $\pm$ 0.40 }&\multirow{2}*{22.31 $\pm$ 0.01 }&\multirow{1}*{5.27 }\\
			&&&$I_{2}$&18.79 $\pm$ 0.40 && &&0.64 \\
			&&\multirow{2}*{C$^{15}$N, $N$=1-0 }&$I_{1}$&0.59 $\pm$ 0.10 &\multirow{2}*{$\bigl\}$}&\multirow{2}*{0.59 $\pm$ 0.10 }&\multirow{2}*{22.63 $\pm$ 0.21 }&\multirow{2}*{0.20 }\\
			&&&$I_{2}$&... &&&&\\
			\multirow{2}*{G069.54}&\multirow{2}*{IRAM}&\multirow{2}*{C$^{14}$N, $N$=1-0 }&$I_{1}$&38.11 $\pm$ 2.80 &\multirow{2}*{$\bigl\}$}&\multirow{2}*{64.37 $\pm$ 2.80 }&\multirow{2}*{12.59 $\pm$ 0.07 }&\multirow{1}*{4.23 }\\
			&&&$I_{2}$&26.26 $\pm$ 0.40 && &&0.55 \\
			&&\multirow{2}*{C$^{15}$N, $N$=1-0 }&$I_{1}$&0.35 $\pm$ 0.10 &\multirow{2}*{$\bigl\}$}&\multirow{2}*{0.35 $\pm$ 0.10 }&\multirow{2}*{12.02 $\pm$ 0.41 }&\multirow{2}*{0.07 }\\
			&&&$I_{2}$&... &&&&\\
			\multirow{2}*{G078.12}&\multirow{2}*{IRAM}&\multirow{2}*{C$^{14}$N, $N$=1-0 }&$I_{1}$&42.92 $\pm$ 3.30 &\multirow{2}*{$\bigl\}$}&\multirow{2}*{65.79 $\pm$ 3.30 }&\multirow{2}*{-3.64 $\pm$ 0.52 }&\multirow{1}*{7.41 }\\
			&&&$I_{2}$&22.86 $\pm$ 0.40 && &&0.81 \\
			&&\multirow{2}*{C$^{15}$N, $N$=1-0 }&$I_{1}$&0.14 $\pm$ 0.10 &\multirow{2}*{$\bigl\}$}&\multirow{2}*{0.14 $\pm$ 0.10 }&\multirow{2}*{-3.92 $\pm$ 0.17 }&\multirow{2}*{0.07 }\\
			&&&$I_{2}$&... &&&&\\
			\multirow{2}*{G081.75}&\multirow{2}*{IRAM}&\multirow{2}*{C$^{14}$N, $N$=1-0 }&$I_{1}$&33.16 $\pm$ 4.10 &\multirow{2}*{$\bigl\}$}&\multirow{2}*{54.80 $\pm$ 7.00 }&\multirow{2}*{-4.56 $\pm$ 0.07 }&\multirow{1}*{4.49 }\\
			&&&$I_{2}$&21.64 $\pm$ 5.70 && &&1.03 \\
			&&\multirow{2}*{C$^{15}$N, $N$=1-0 }&$I_{1}$&0.35 $\pm$ 0.10 &\multirow{2}*{$\bigl\}$}&\multirow{2}*{0.35 $\pm$ 0.10 }&\multirow{2}*{-4.08 $\pm$ 0.21 }&\multirow{2}*{0.10 }\\
			&&&$I_{2}$&... &&&&\\
			\multirow{2}*{G078.88}&\multirow{2}*{IRAM}&\multirow{2}*{C$^{14}$N, $N$=1-0 }&$I_{1}$&35.65 $\pm$ 4.28 &\multirow{2}*{$\bigl\}$}&\multirow{2}*{57.34 $\pm$ 5.19 }&\multirow{2}*{-4.99 $\pm$ 0.13 }&\multirow{1}*{3.99 }\\
			&&&$I_{2}$&21.69 $\pm$ 2.94 && &&0.23 \\
			&&\multirow{2}*{C$^{15}$N, $N$=1-0 }&$I_{1}$&0.29 $\pm$ 0.06 &\multirow{2}*{$\bigl\}$}&\multirow{2}*{0.29 $\pm$ 0.06 }&\multirow{2}*{-4.58 $\pm$ 0.60 }&\multirow{2}*{0.04 }\\
			&&&$I_{2}$&... &&&&\\
			\multirow{2}*{G092.67}&\multirow{2}*{IRAM}&\multirow{2}*{C$^{14}$N, $N$=1-0 }&$I_{1}$&28.61 $\pm$ 0.80 &\multirow{2}*{$\bigl\}$}&\multirow{2}*{46.86 $\pm$ 0.80 }&\multirow{2}*{-6.21 $\pm$ 0.06 }&\multirow{1}*{4.49 }\\
			&&&$I_{2}$&18.24 $\pm$ 0.40 && &&0.47 \\
			&&\multirow{2}*{C$^{15}$N, $N$=1-0 }&$I_{1}$&0.42 $\pm$ 0.10 &\multirow{2}*{$\bigl\}$}&\multirow{2}*{0.42 $\pm$ 0.10 }&\multirow{2}*{-5.95 $\pm$ 0.20 }&\multirow{2}*{0.18 }\\
			&&&$I_{2}$&... &&&&\\
			\multirow{2}*{G109.87}&\multirow{2}*{IRAM}&\multirow{2}*{C$^{14}$N, $N$=1-0 }&$I_{1}$&29.75 $\pm$ 6.20 &\multirow{2}*{$\bigl\}$}&\multirow{2}*{49.02 $\pm$ 6.30 }&\multirow{2}*{-10.68 $\pm$ 0.13 }&\multirow{1}*{2.57 }\\
			&&&$I_{2}$&19.27 $\pm$ 1.40 && &&0.37 \\
			&&\multirow{2}*{C$^{15}$N, $N$=1-0 }&$I_{1}$&0.38 $\pm$ 0.10 &\multirow{2}*{$\bigl\}$}&\multirow{2}*{0.38 $\pm$ 0.10 }&\multirow{2}*{-10.48 $\pm$ 0.21 }&\multirow{2}*{0.09 }\\
			&&&$I_{2}$&... &&&&\\
			&\multirow{2}*{ARO}&\multirow{2}*{C$^{14}$N, $N$=1-0 }&$I_{1}$&25.86 $\pm$ 1.30 &\multirow{2}*{$\bigl\}$}&\multirow{2}*{41.72 $\pm$ 3.80 }&\multirow{2}*{-11.32 $\pm$ 0.41 }&\multirow{1}*{2.21 }\\
			&&&$I_{2}$&15.86 $\pm$ 3.50 && &&0.32 \\
			&&\multirow{2}*{C$^{15}$N, $N$=1-0 }&$I_{1}$&0.30 $\pm$ 0.10 &\multirow{2}*{$\bigl\}$}&\multirow{2}*{0.38 $\pm$ 0.10 }&\multirow{2}*{-10.15 $\pm$ 0.12 }&\multirow{2}*{0.06 }\\
			&&&$I_{2}$&0.08 $\pm$ 0.01 &&&&\\
			\multirow{2}*{WB171}&\multirow{2}*{IRAM}&\multirow{2}*{C$^{14}$N, $N$=1-0 }&$I_{1}$&38.93 $\pm$ 3.40 &\multirow{2}*{$\bigl\}$}&\multirow{2}*{46.61 $\pm$ 5.10 }&\multirow{2}*{-6.60 $\pm$ 0.52 }&\multirow{1}*{6.49 }\\
			&&&$I_{2}$&7.68 $\pm$ 3.80 && &&0.32 \\
			&&\multirow{2}*{C$^{15}$N, $N$=1-0 }&$I_{1}$&0.25 $\pm$ 0.10 &\multirow{2}*{$\bigl\}$}&\multirow{2}*{0.25 $\pm$ 0.10 }&\multirow{2}*{-6.72 $\pm$ 0.23 }&\multirow{2}*{0.07 }\\
			&&&$I_{2}$&... &&&&\\
			\multirow{2}*{G209.00}&\multirow{2}*{ARO}&\multirow{2}*{C$^{14}$N, $N$=1-0 }&$I_{1}$&19.27 $\pm$ 0.80 &\multirow{2}*{$\bigl\}$}&\multirow{2}*{29.60 $\pm$ 1.00 }&\multirow{2}*{8.67 $\pm$ 0.05 }&\multirow{1}*{3.08 }\\
			&&&$I_{2}$&10.32 $\pm$ 0.60 && &&1.04 \\
			&&\multirow{2}*{C$^{15}$N, $N$=1-0 }&$I_{1}$&0.06 $\pm$ 0.10 &\multirow{2}*{$\bigl\}$}&\multirow{2}*{0.06 $\pm$ 0.10 }&\multirow{2}*{8.13 $\pm$ 0.29 }&\multirow{2}*{0.01 }\\
			&&&$I_{2}$&... &&&&\\
			\multirow{2}*{G121.29}&\multirow{2}*{ARO}&\multirow{2}*{C$^{14}$N, $N$=1-0 }&$I_{1}$&11.31 $\pm$ 0.90 &\multirow{2}*{$\bigl\}$}&\multirow{2}*{17.94 $\pm$ 0.90 }&\multirow{2}*{-18.02 $\pm$ 0.06 }&\multirow{1}*{1.82 }\\
			&&&$I_{2}$&6.63 $\pm$ 0.10 && &&0.25 \\
			&&\multirow{2}*{C$^{15}$N, $N$=1-0 }&$I_{1}$&0.10 $\pm$ 0.10 &\multirow{2}*{$\bigl\}$}&\multirow{2}*{0.10 $\pm$ 0.10 }&\multirow{2}*{-16.97 $\pm$ 0.18 }&\multirow{2}*{0.06 }\\
			&&&$I_{2}$&... &&&&\\
			\multirow{2}*{G111.54}&\multirow{2}*{IRAM}&\multirow{2}*{C$^{14}$N, $N$=1-0 }&$I_{1}$&38.79 $\pm$ 2.30 &\multirow{2}*{$\bigl\}$}&\multirow{2}*{62.77 $\pm$ 3.10 }&\multirow{2}*{-56.53 $\pm$ 1.03 }&\multirow{1}*{2.18 }\\
			&&&$I_{2}$&24.00 $\pm$ 2.10 && &&0.19 \\
			&&\multirow{2}*{C$^{15}$N, $N$=1-0 }&$I_{1}$&0.17 $\pm$ 0.10 &\multirow{2}*{$\bigl\}$}&\multirow{2}*{0.29 $\pm$ 0.10 }&\multirow{2}*{-56.80 $\pm$ 0.47 }&\multirow{2}*{0.04 }\\
			&&&$I_{2}$&0.12 $\pm$ 0.04 &&&&\\
			&\multirow{2}*{ARO}&\multirow{2}*{C$^{14}$N, $N$=1-0 }&$I_{1}$&28.68 $\pm$ 2.30 &\multirow{2}*{$\bigl\}$}&\multirow{2}*{47.65 $\pm$ 2.12 }&\multirow{2}*{-57.33 $\pm$ 1.03 }&\multirow{1}*{2.27 }\\
			&&&$I_{2}$&18.98 $\pm$ 2.21 && &&0.17 \\
			&&\multirow{2}*{C$^{15}$N, $N$=1-0 }&$I_{1}$&0.12 $\pm$ 0.03 &\multirow{2}*{$\bigl\}$}&\multirow{2}*{0.12 $\pm$ 0.03 }&\multirow{2}*{-56.88 $\pm$ 0.33 }&\multirow{2}*{0.04 }\\
			&&&$I_{2}$&... &&&&\\
			\multirow{2}*{G133.94}&\multirow{2}*{IRAM}&\multirow{2}*{C$^{14}$N, $N$=1-0 }&$I_{1}$&27.85 $\pm$ 2.10 &\multirow{2}*{$\bigl\}$}&\multirow{2}*{44.93 $\pm$ 2.10 }&\multirow{2}*{-46.59 $\pm$ 0.02 }&\multirow{1}*{2.84 }\\
			&&&$I_{2}$&17.08 $\pm$ 0.20 && &&0.25 \\
			&&\multirow{2}*{C$^{15}$N, $N$=1-0 }&$I_{1}$&0.20 $\pm$ 0.10 &\multirow{2}*{$\bigl\}$}&\multirow{2}*{0.10 $\pm$ 0.10 }&\multirow{2}*{-46.16 $\pm$ 0.18 }&\multirow{2}*{0.07 }\\
			&&&$I_{2}$&... &&&&\\
			&\multirow{2}*{ARO}&\multirow{2}*{C$^{14}$N, $N$=1-0 }&$I_{1}$&17.37 $\pm$ 0.30 &\multirow{2}*{$\bigl\}$}&\multirow{2}*{27.63 $\pm$ 0.30 }&\multirow{2}*{-47.81 $\pm$ 0.41 }&\multirow{1}*{1.92 }\\
			&&&$I_{2}$&10.26 $\pm$ 0.10 && &&0.17 \\
			&&\multirow{2}*{C$^{15}$N, $N$=1-0 }&$I_{1}$&0.09 $\pm$ 0.10 &\multirow{2}*{$\bigl\}$}&\multirow{2}*{0.15 $\pm$ 0.10 }&\multirow{2}*{-46.33 $\pm$ 0.13 }&\multirow{2}*{0.05 }\\
			&&&$I_{2}$&0.05 $\pm$ 0.01 &&&&\\
			\multirow{2}*{G192.60}&\multirow{2}*{ARO}&\multirow{2}*{C$^{14}$N, $N$=1-0 }&$I_{1}$&21.48 $\pm$ 0.20 &\multirow{2}*{$\bigl\}$}&\multirow{2}*{33.98 $\pm$ 0.30 }&\multirow{2}*{7.48 $\pm$ 0.04 }&\multirow{1}*{3.45 }\\
			&&&$I_{2}$&12.50 $\pm$ 0.20 && &&0.30 \\
			&&\multirow{2}*{C$^{15}$N, $N$=1-0 }&$I_{1}$&0.15 $\pm$ 0.10 &\multirow{2}*{$\bigl\}$}&\multirow{2}*{0.15 $\pm$ 0.10 }&\multirow{2}*{7.03 $\pm$ 0.10 }&\multirow{2}*{0.05 }\\
			&&&$I_{2}$&... &&&&\\
			\multirow{2}*{G173.48}&\multirow{2}*{ARO}&\multirow{2}*{C$^{14}$N, $N$=1-0 }&$I_{1}$&14.03 $\pm$ 0.20 &\multirow{2}*{$\bigl\}$}&\multirow{2}*{22.36 $\pm$ 0.30 }&\multirow{2}*{-16.21 $\pm$ 0.05 }&\multirow{1}*{1.71 }\\
			&&&$I_{2}$&8.33 $\pm$ 0.10 && &&0.22 \\
			&&\multirow{2}*{C$^{15}$N, $N$=1-0 }&$I_{1}$&0.13 $\pm$ 0.01 &\multirow{2}*{$\bigl\}$}&\multirow{2}*{0.13 $\pm$ 0.01 }&\multirow{2}*{-16.64 $\pm$ 0.12 }&\multirow{2}*{0.04 }\\
			&&&$I_{2}$&... &&&&\\
			\multirow{2}*{G123.06}&\multirow{2}*{ARO}&\multirow{2}*{C$^{14}$N, $N$=1-0 }&$I_{1}$&11.72 $\pm$ 0.10 &\multirow{2}*{$\bigl\}$}&\multirow{2}*{18.45 $\pm$ 0.10 }&\multirow{2}*{-30.72 $\pm$ 0.06 }&\multirow{1}*{1.46 }\\
			&&&$I_{2}$&6.73 $\pm$ 0.10 && &&0.17 \\
			&&\multirow{2}*{C$^{15}$N, $N$=1-0 }&$I_{1}$&0.03 $\pm$ 0.10 &\multirow{2}*{$\bigl\}$}&\multirow{2}*{0.05 $\pm$ 0.10 }&\multirow{2}*{-30.27 $\pm$ 0.17 }&\multirow{2}*{0.02 }\\
			&&&$I_{2}$&0.02 $\pm$ 0.01 &&&&\\
			\multirow{2}*{G174.20}&\multirow{2}*{ARO}&\multirow{2}*{C$^{14}$N, $N$=1-0 }&$I_{1}$&11.54 $\pm$ 0.20 &\multirow{2}*{$\bigl\}$}&\multirow{2}*{18.60 $\pm$ 0.20 }&\multirow{2}*{-2.88 $\pm$ 0.07 }&\multirow{1}*{1.21 }\\
			&&&$I_{2}$&7.05 $\pm$ 0.10 && &&0.18 \\
			&&\multirow{2}*{C$^{15}$N, $N$=1-0 }&$I_{1}$&0.10 $\pm$ 0.01 &\multirow{2}*{$\bigl\}$}&\multirow{2}*{0.10 $\pm$ 0.01 }&\multirow{2}*{-2.54 $\pm$ 0.10 }&\multirow{2}*{0.06 }\\
			&&&$I_{2}$&... &&&&\\
			\multirow{2}*{G211.59}&\multirow{2}*{ARO}&\multirow{2}*{C$^{14}$N, $N$=1-0 }&$I_{1}$&3.81 $\pm$ 0.10 &\multirow{2}*{$\bigl\}$}&\multirow{2}*{5.98 $\pm$ 0.10 }&\multirow{2}*{45.22 $\pm$ 0.07 }&\multirow{1}*{0.44 }\\
			&&&$I_{2}$&2.17 $\pm$ 0.10 && &&0.19 \\
			&&\multirow{2}*{C$^{15}$N, $N$=1-0 }&$I_{1}$&0.11 $\pm$ 0.01 &\multirow{2}*{$\bigl\}$}&\multirow{2}*{0.11 $\pm$ 0.01 }&\multirow{2}*{46.98 $\pm$ 0.29 }&\multirow{2}*{0.04 }\\
			&&&$I_{2}$&... &&&&\\
			\multirow{2}*{G135.27}&\multirow{2}*{IRAM}&\multirow{2}*{C$^{14}$N, $N$=1-0 }&$I_{1}$&4.05 $\pm$ 0.10 &\multirow{2}*{$\bigl\}$}&\multirow{2}*{6.13 $\pm$ 0.10 }&\multirow{2}*{-71.49 $\pm$ 0.02 }&\multirow{1}*{0.62 }\\
			&&&$I_{2}$&2.09 $\pm$ 0.10 && &&0.03 \\
			&&\multirow{2}*{C$^{15}$N, $N$=1-0 }&$I_{1}$&0.09 $\pm$ 0.02 &\multirow{2}*{$\bigl\}$}&\multirow{2}*{0.09 $\pm$ 0.02 }&\multirow{2}*{-71.66 $\pm$ 0.14 }&\multirow{2}*{0.03 }\\
			&&&$I_{2}$&... &&&&\\
			\multirow{2}*{WB380}&\multirow{2}*{IRAM}&\multirow{2}*{C$^{14}$N, $N$=1-0 }&$I_{1}$&4.50 $\pm$ 0.41 &\multirow{2}*{$\bigl\}$}&\multirow{2}*{7.14 $\pm$ 0.61 }&\multirow{2}*{-84.26 $\pm$ 0.11 }&\multirow{1}*{0.59 }\\
			&&&$I_{2}$&2.64 $\pm$ 0.42 && &&0.22 \\
			&&\multirow{2}*{C$^{15}$N, $N$=1-0 }&$I_{1}$&0.16 $\pm$ 0.04 &\multirow{2}*{$\bigl\}$}&\multirow{2}*{0.16 $\pm$ 0.04 }&\multirow{2}*{-84.51 $\pm$ 0.36 }&\multirow{2}*{0.07 }\\
			&&&$I_{2}$&... &&&&
			\enddata
			\tablecomments{
			Column(1): source name; column(2): used telescope; column(3): molecule; column(4): transition line, $I_{1}$ and $I_{2}$ are the transition line of $J$=3/2 -- 1/2 and $J$=1/2 -- 1/2, respectively; column(5): the integrated line intensities of the $I_{1}$ and $I_{2}$; column(6): LSR velocity; column(7): main beam brightness temperature for the strongest and weakest component of C$^{14}$N are presented in the upper two rows and the peak value of the $J$ = 3/2 -- 1/2 feature of C$^{15}$N, presented in the lower row.
			}
		\end{deluxetable*}

\begin{figure*}[htbp]
	\centering
	{\includegraphics[width=4.4cm]{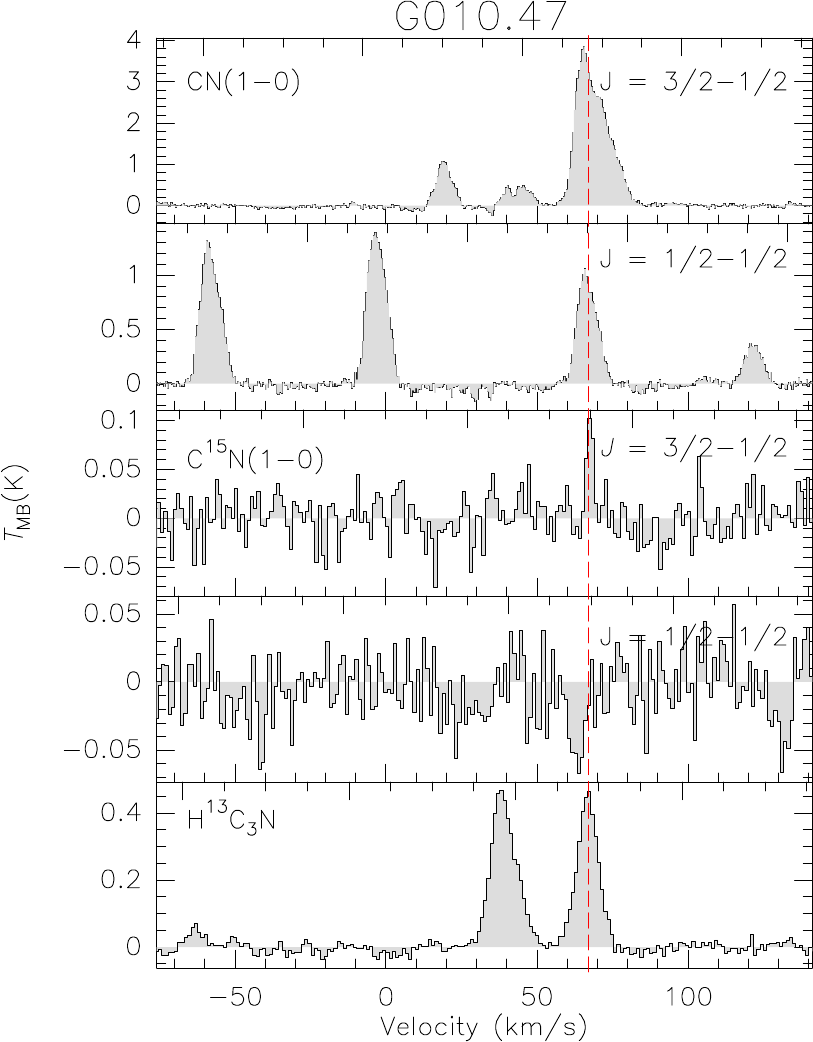}} 
	{\includegraphics[width=4.4cm]{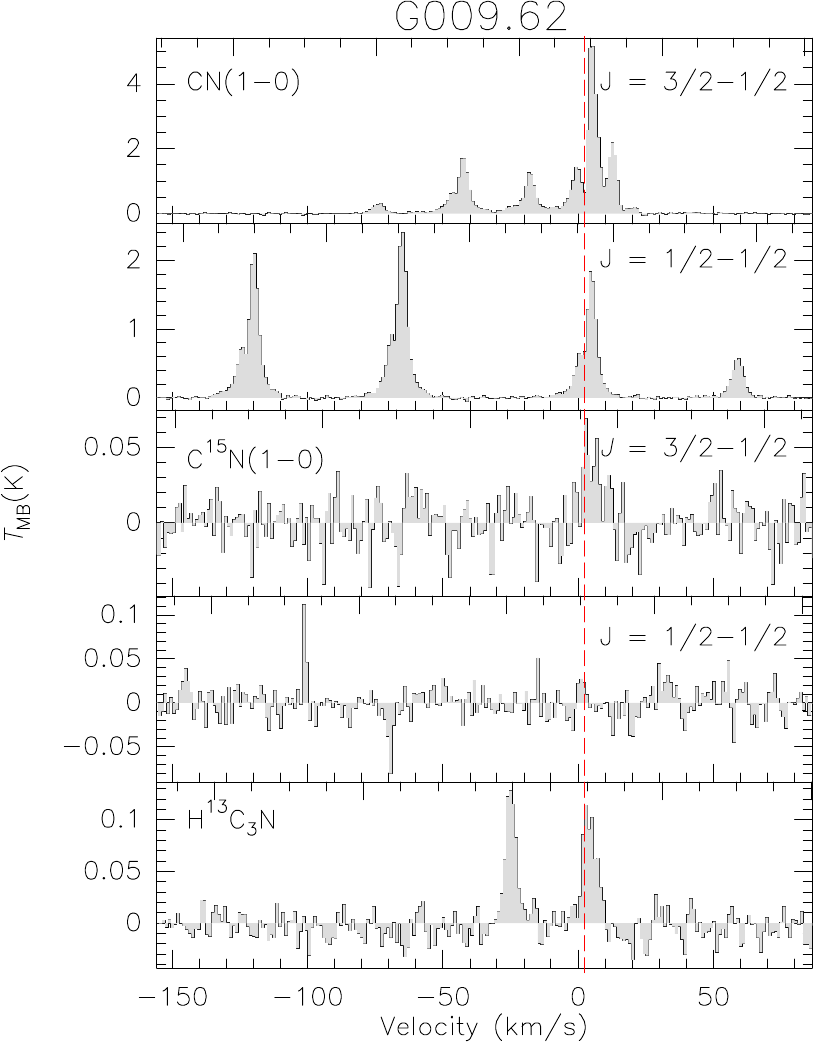}} 
	{\includegraphics[width=4.4cm]{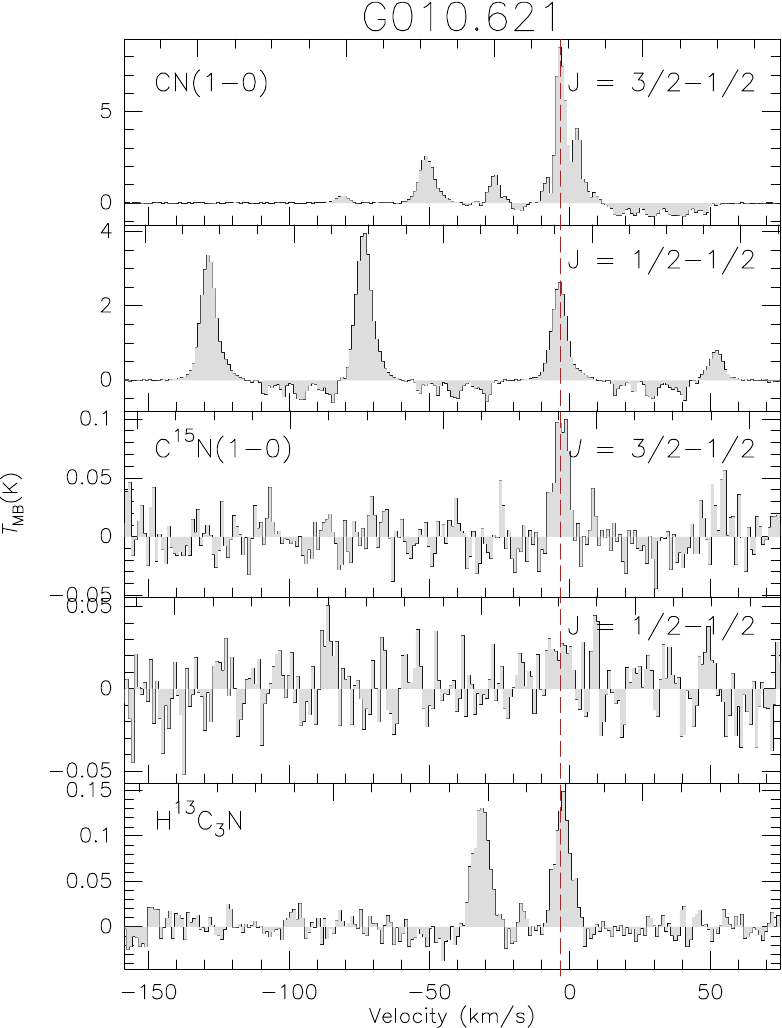}} 
	{\includegraphics[width=4.4cm]{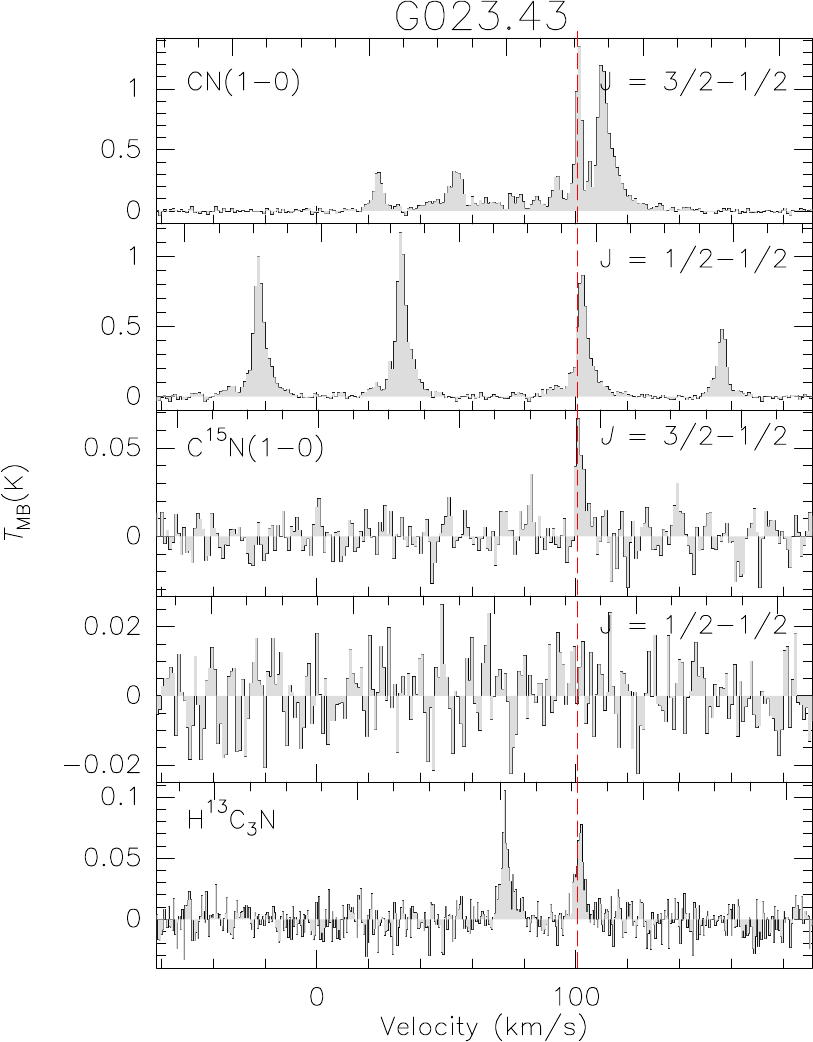}} 
	{\includegraphics[width=4.4cm]{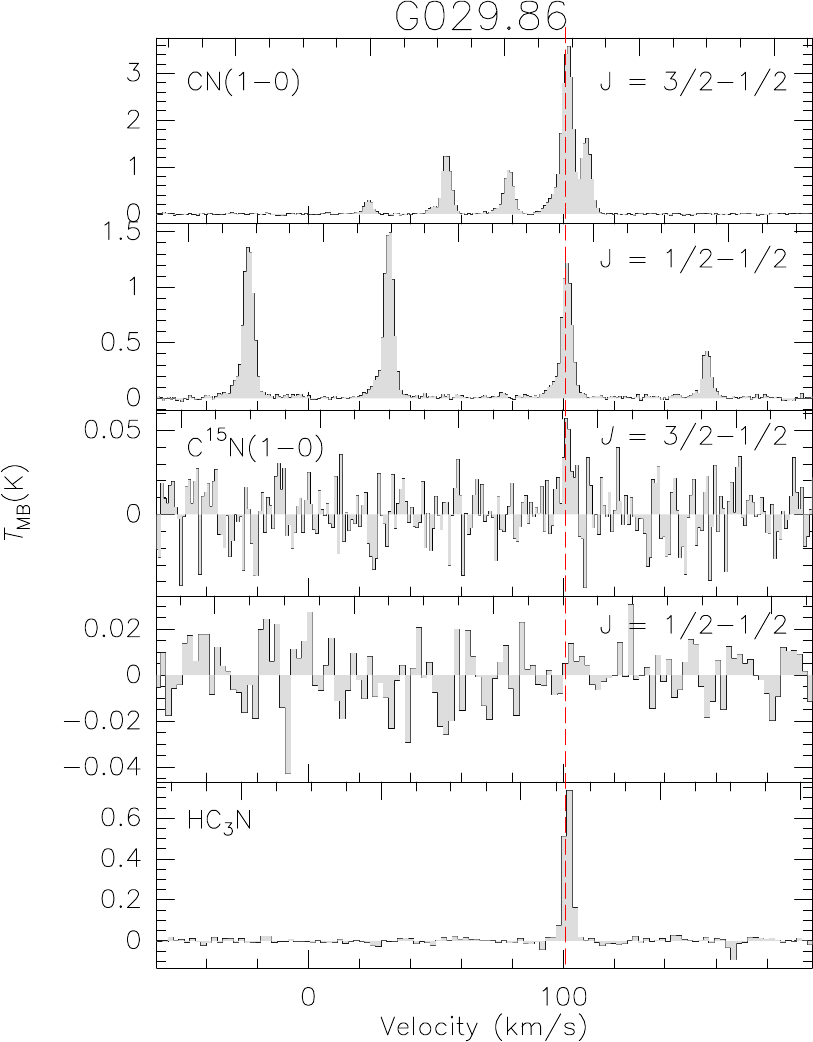}} 
	{\includegraphics[width=4.4cm]{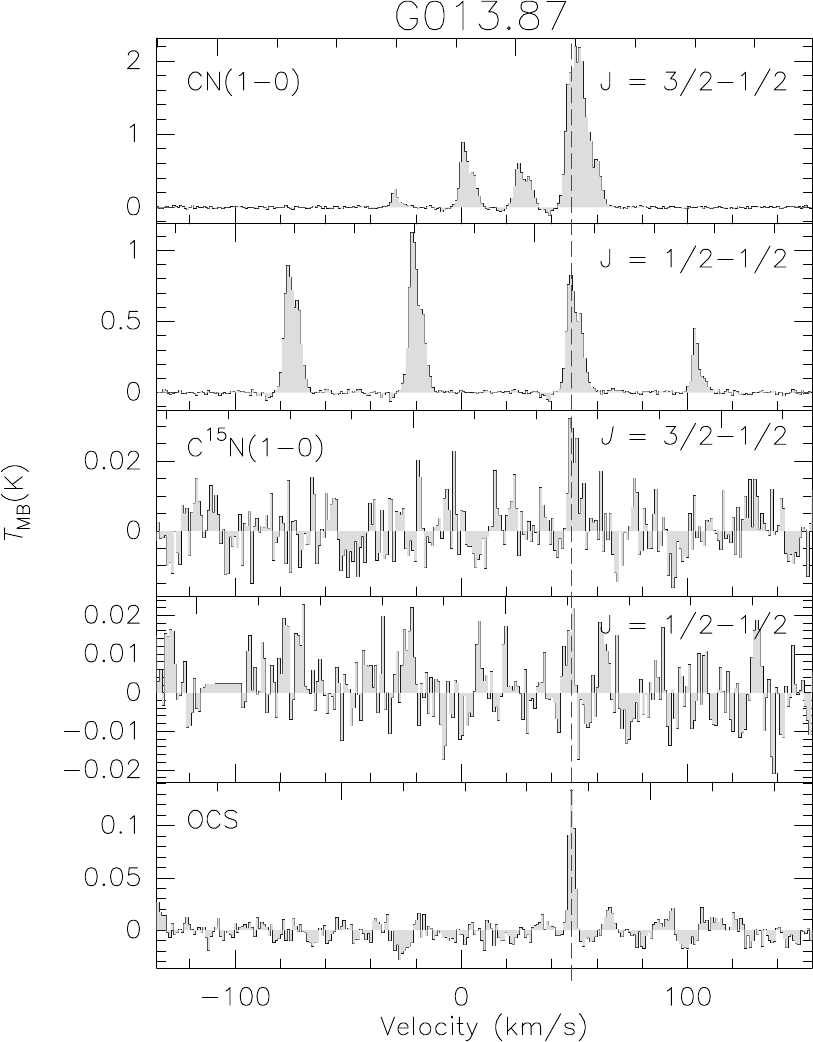}} 
	{\includegraphics[width=4.4cm]{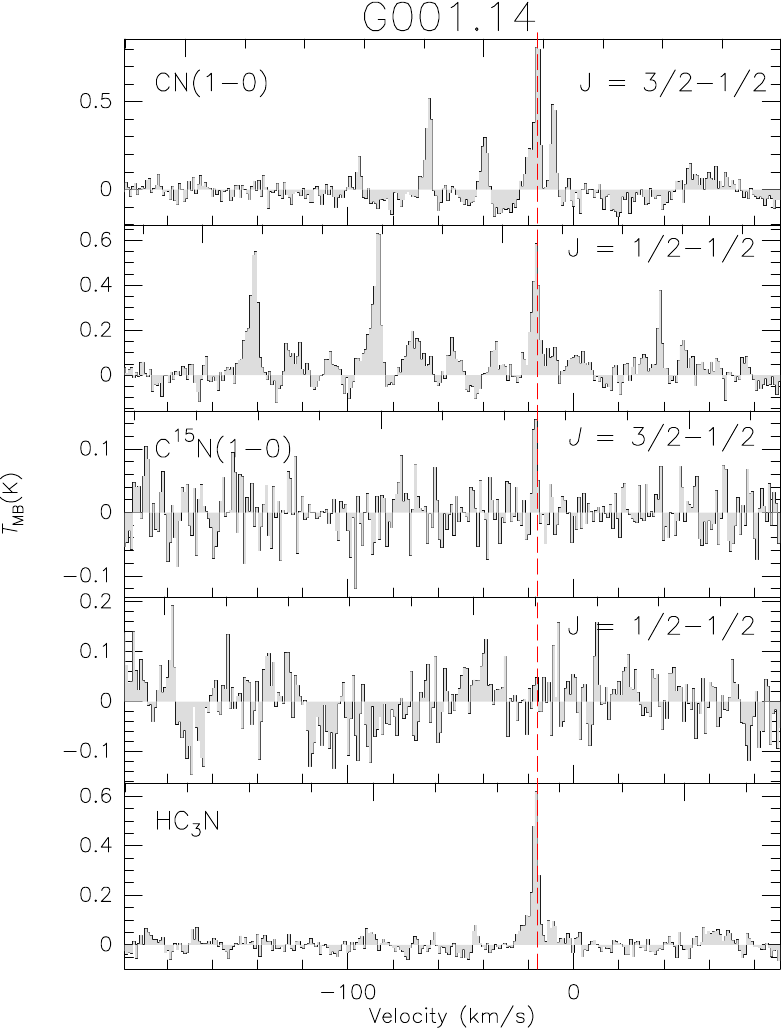}} 
	{\includegraphics[width=4.4cm]{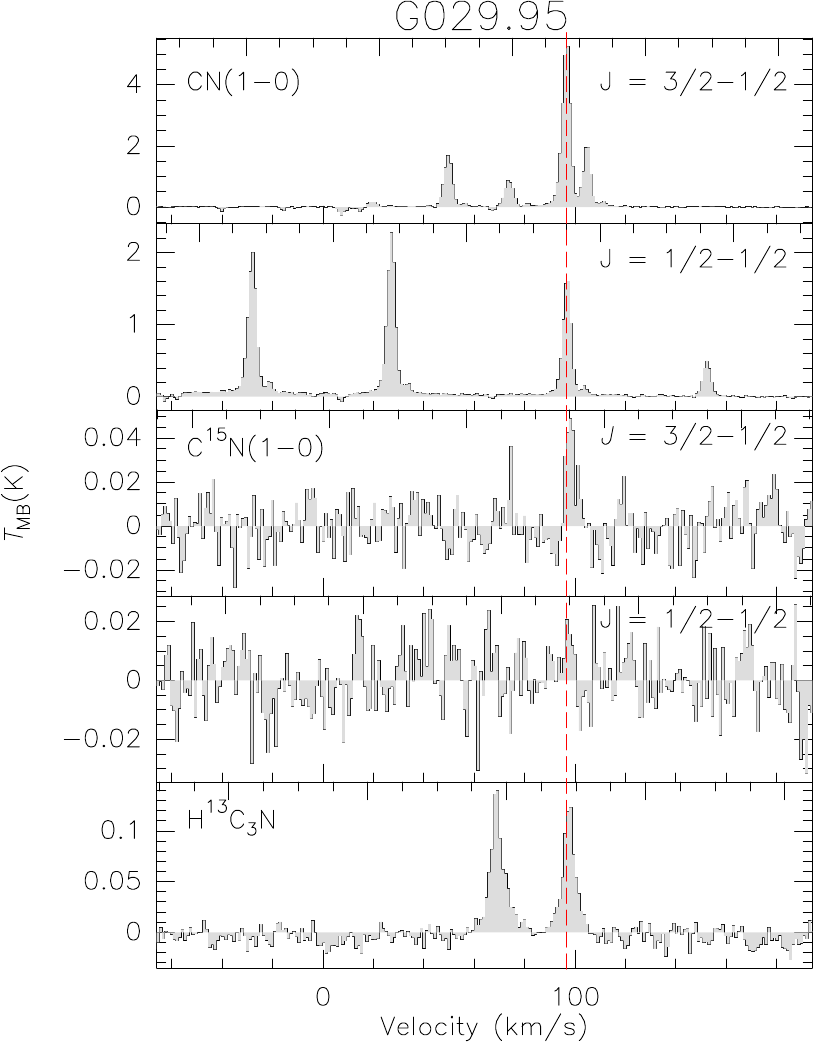}} 
	{\includegraphics[width=4.4cm]{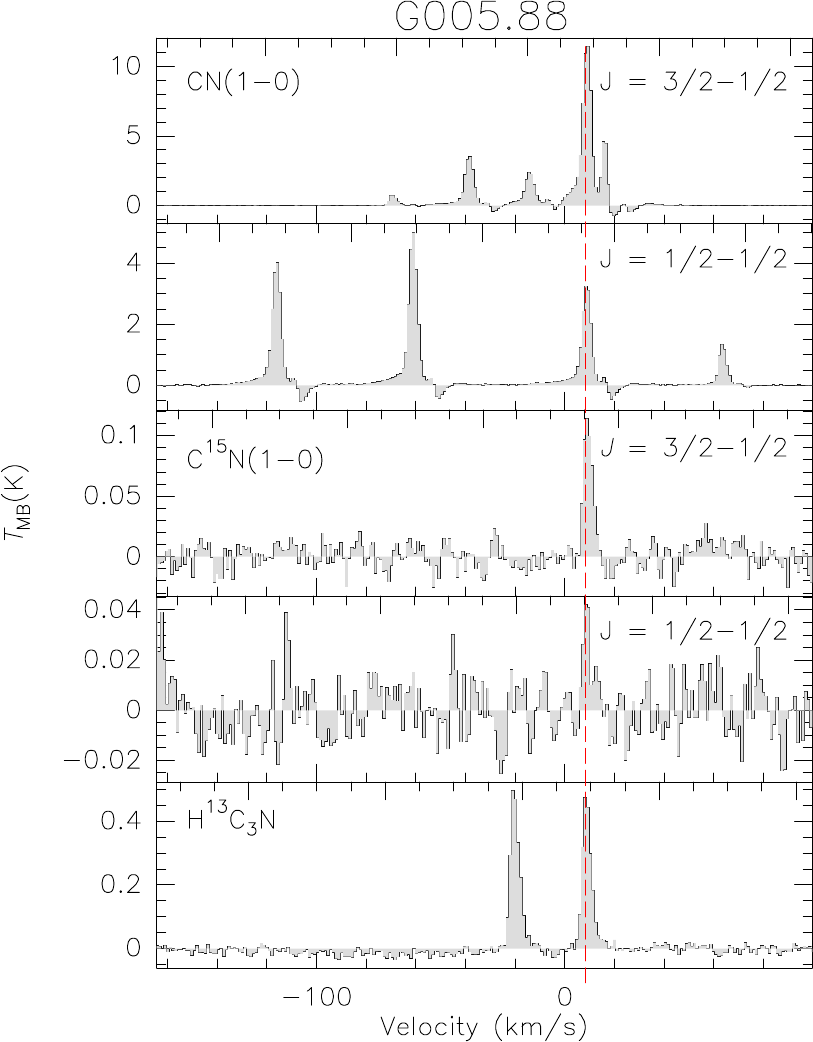}} 
	{\includegraphics[width=4.4cm]{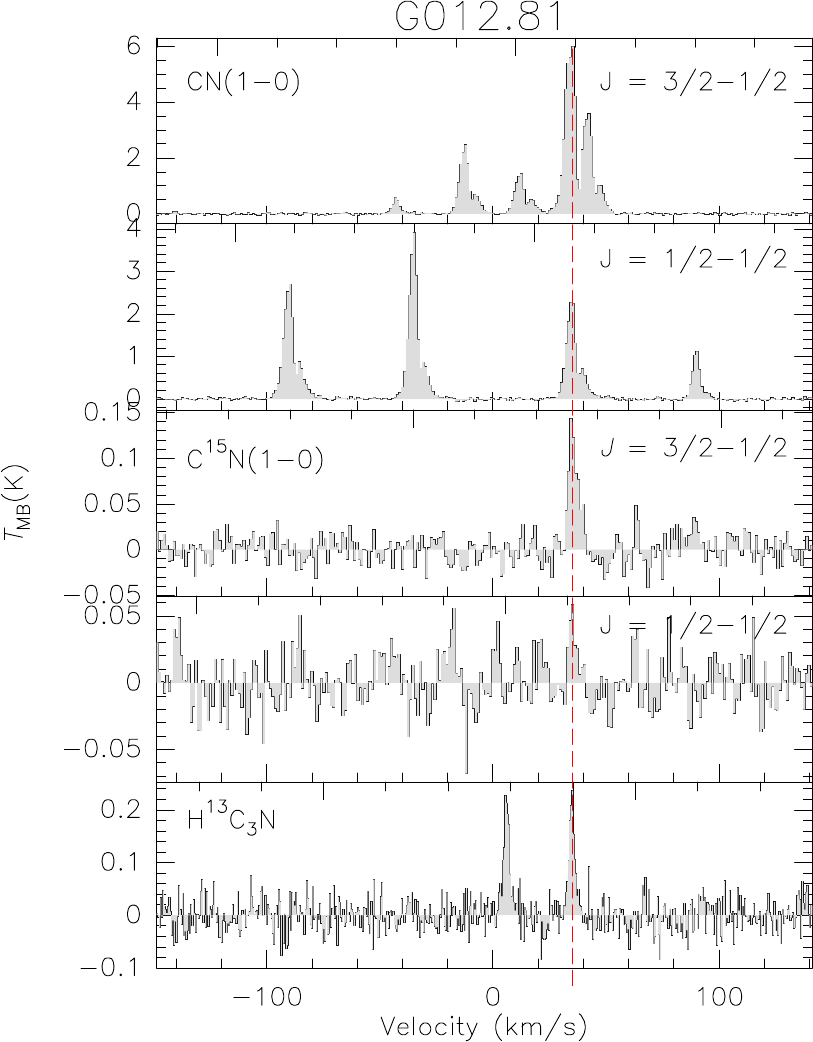}} 
	{\includegraphics[width=4.4cm]{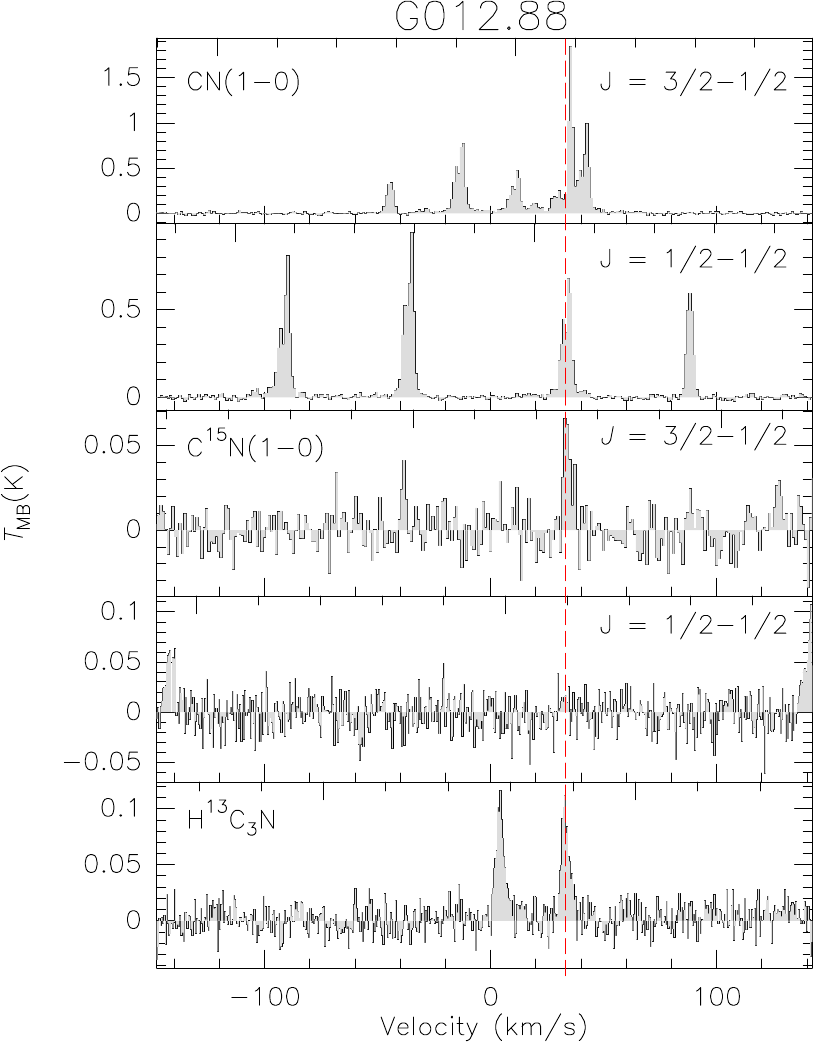}} 
	{\includegraphics[width=4.4cm]{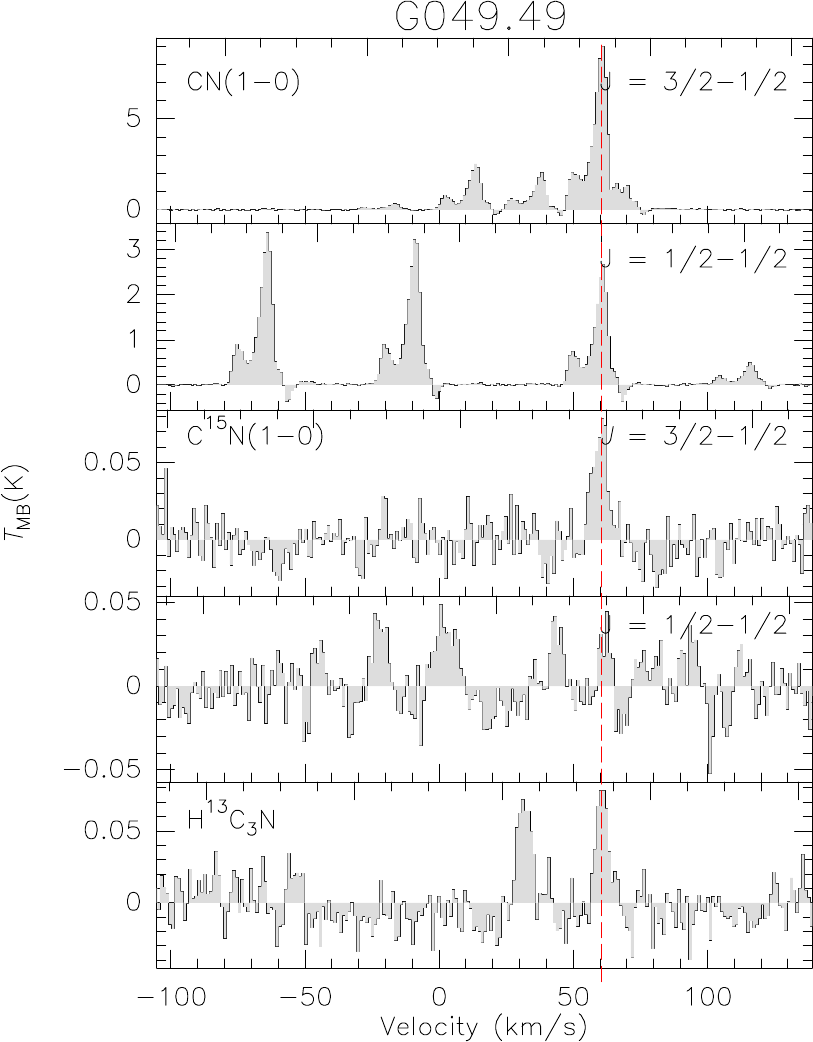}} 
	{\includegraphics[width=4.4cm]{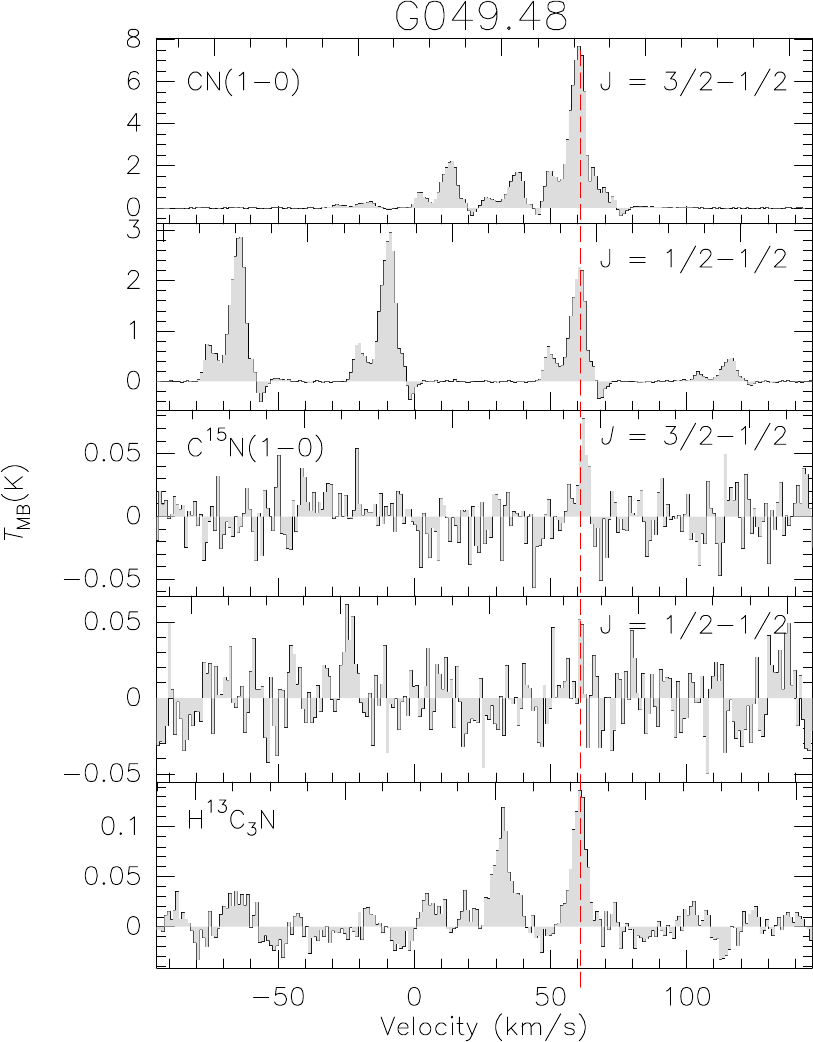}} 
	{\includegraphics[width=4.4cm]{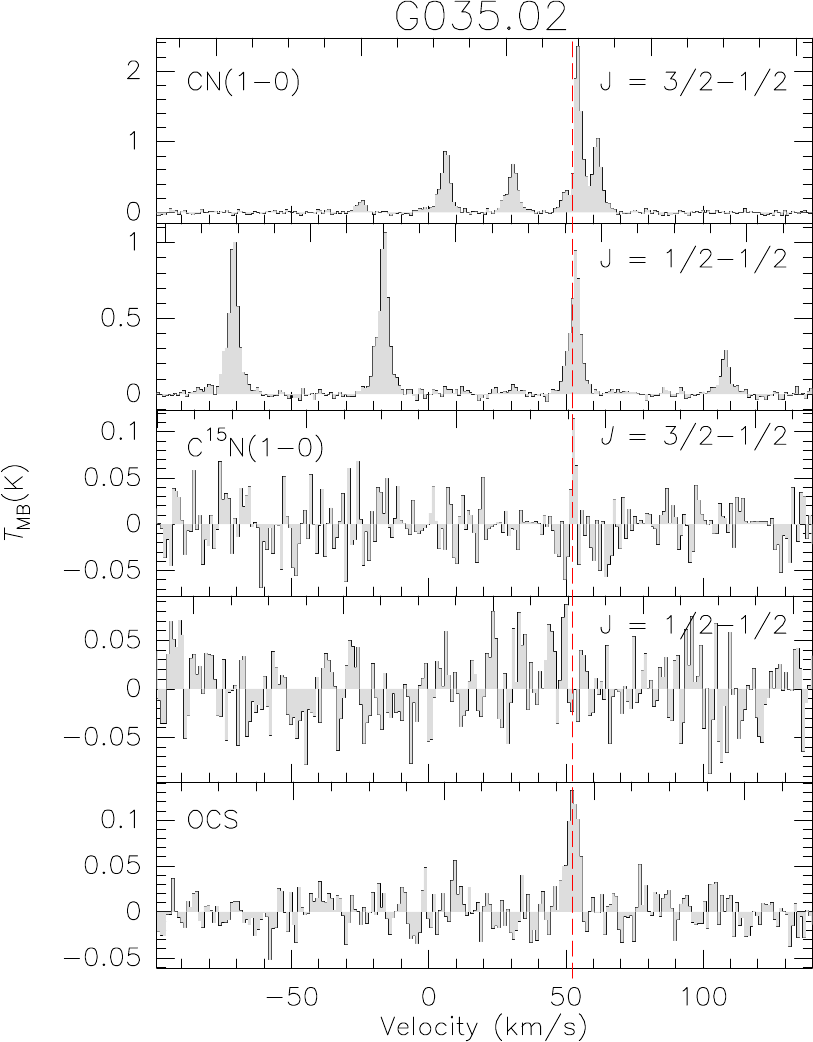}} 
	{\includegraphics[width=4.4cm]{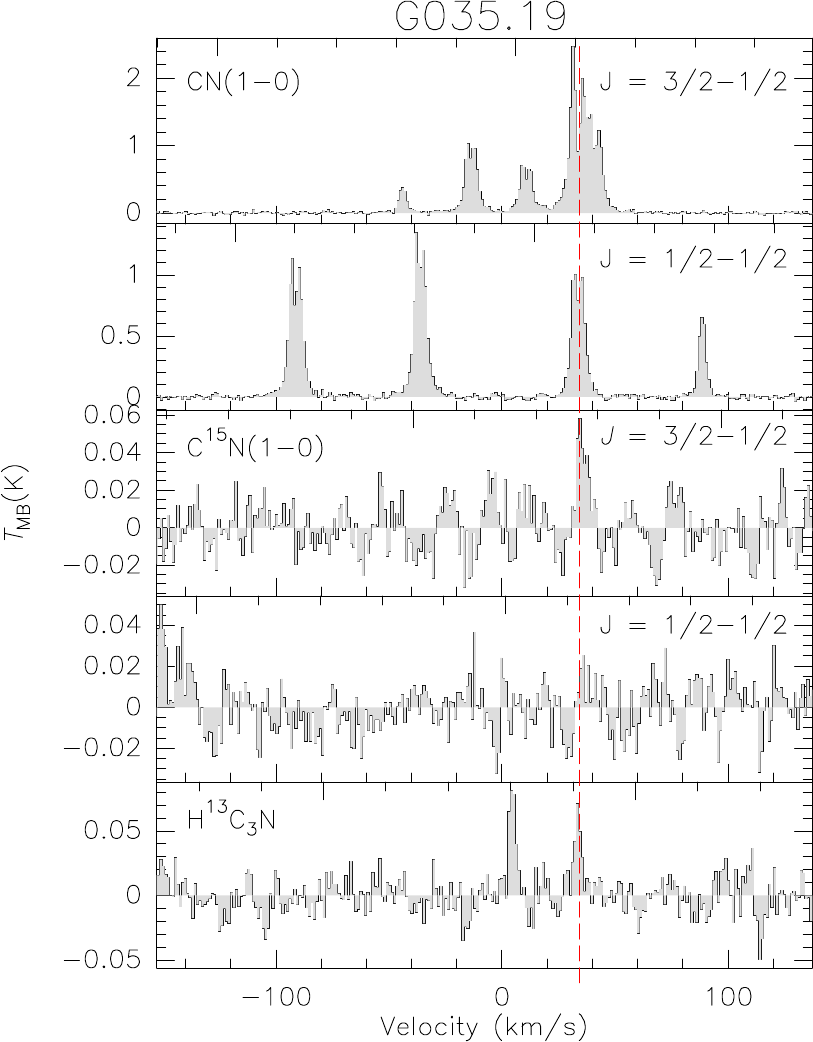}} 
	{\includegraphics[width=4.4cm]{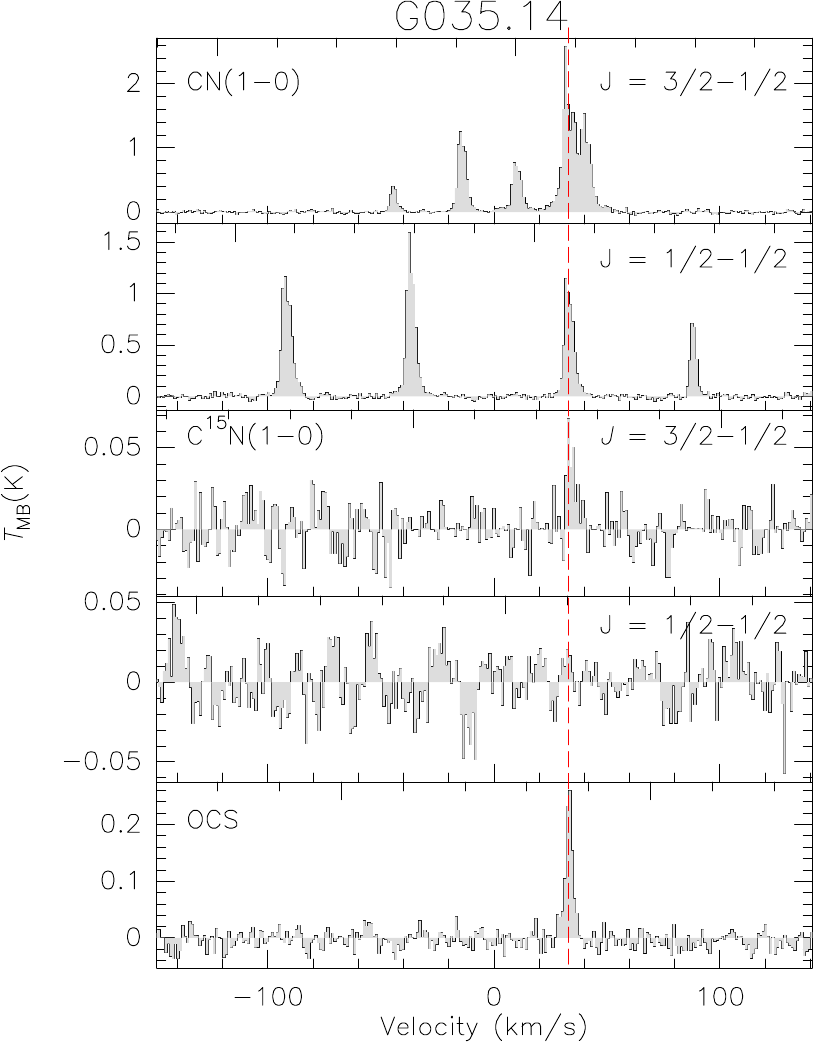}} 
	\caption{IRAM spectra of those 28 sources with detected C$^{15}$N lines, after subtracting baselines and applying Hanning smoothing leading to 1.1 km s$^{-1}$ wide channels. Vertical red dashed lines mark the line center velocity of the source, determined by a Gaussian fit to the HCC$^{13}$CN $J$ = 12--11, HC$_{3}$N $J$ = 12--11 or OCS 9--8 lines, which were observed simultaneously with CN.} 
	\label{IRAM15CNfigself}
\end{figure*}

\begin{figure*}[htbp]
	\addtocounter{figure}{-1} 
	\centering
	{\includegraphics[width=4.4cm]{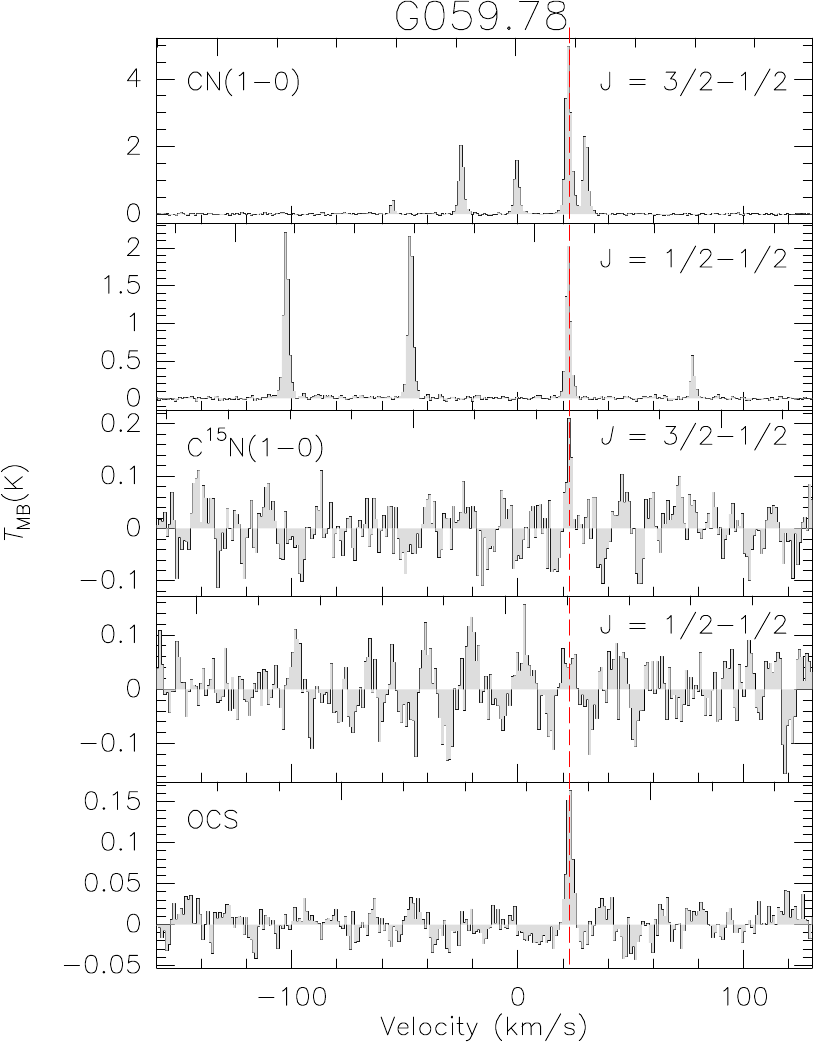}} 
	{\includegraphics[width=4.4cm]{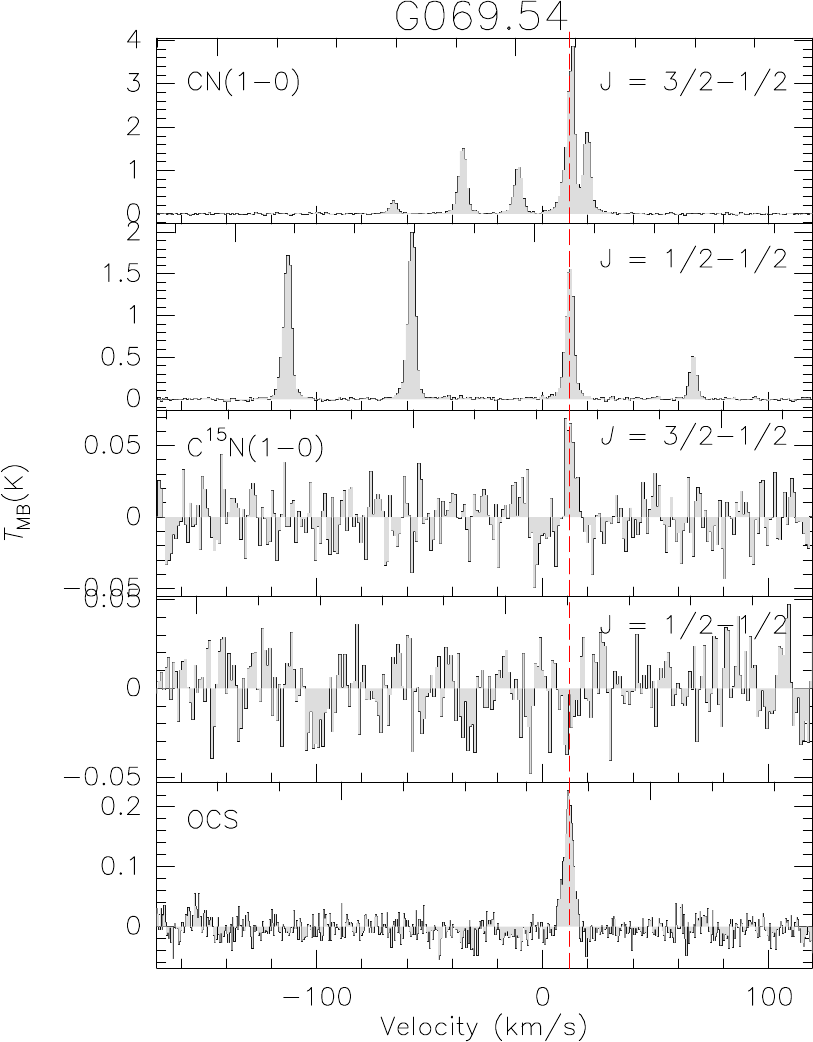}} 
	{\includegraphics[width=4.4cm]{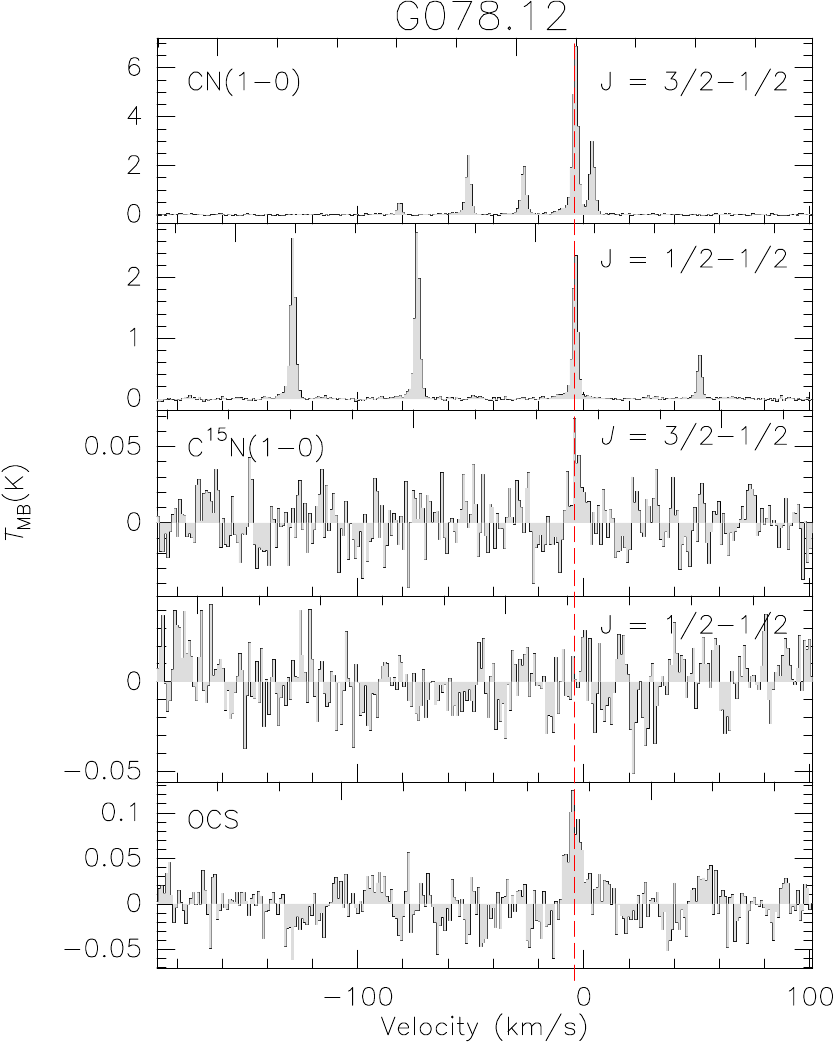}} 
	{\includegraphics[width=4.4cm]{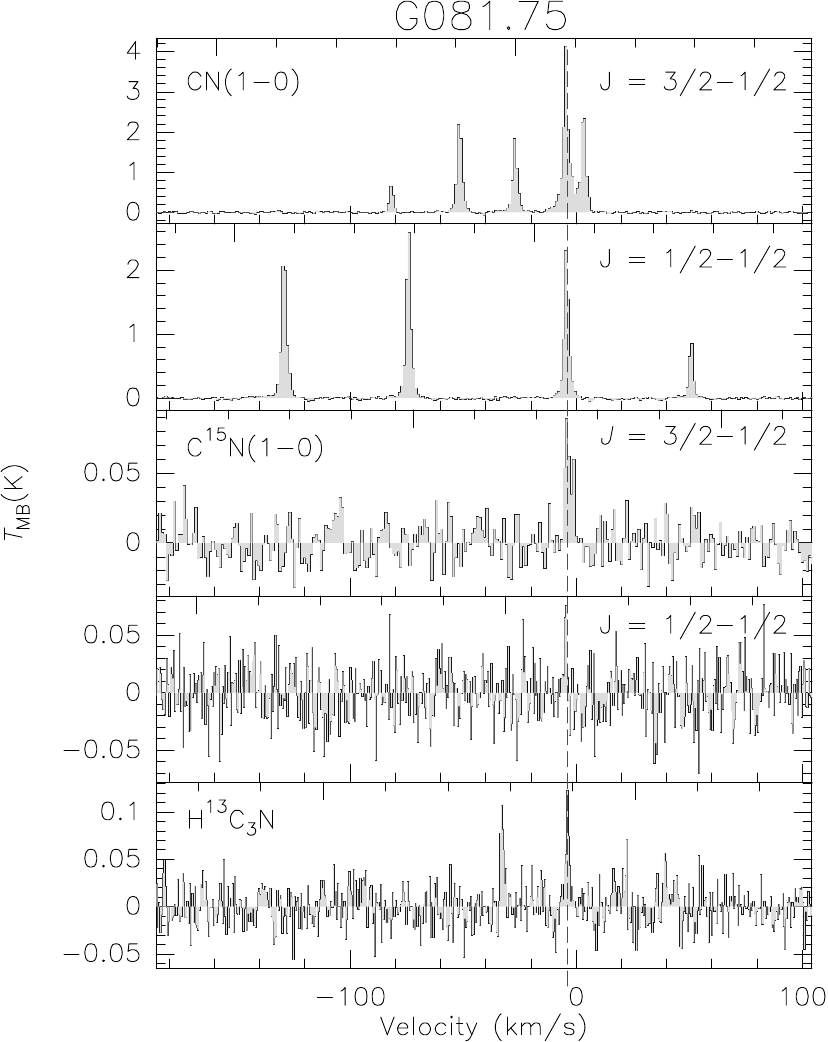}} 
	{\includegraphics[width=4.4cm]{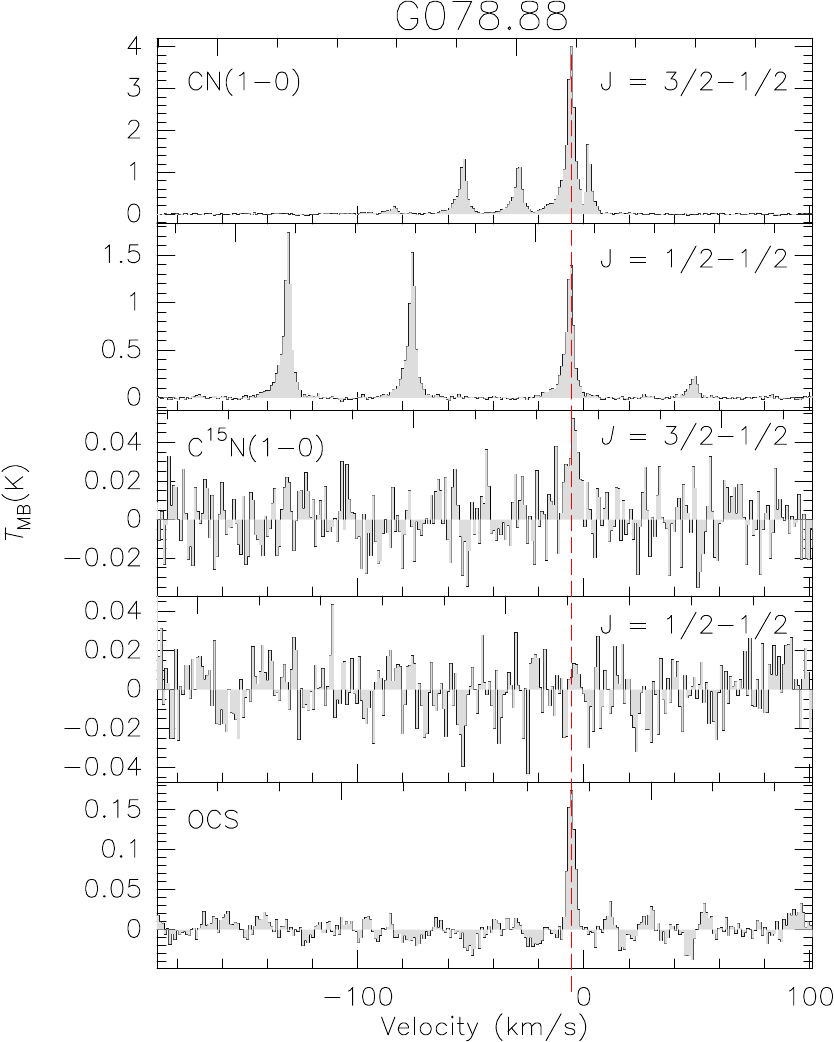}} 
	{\includegraphics[width=4.4cm]{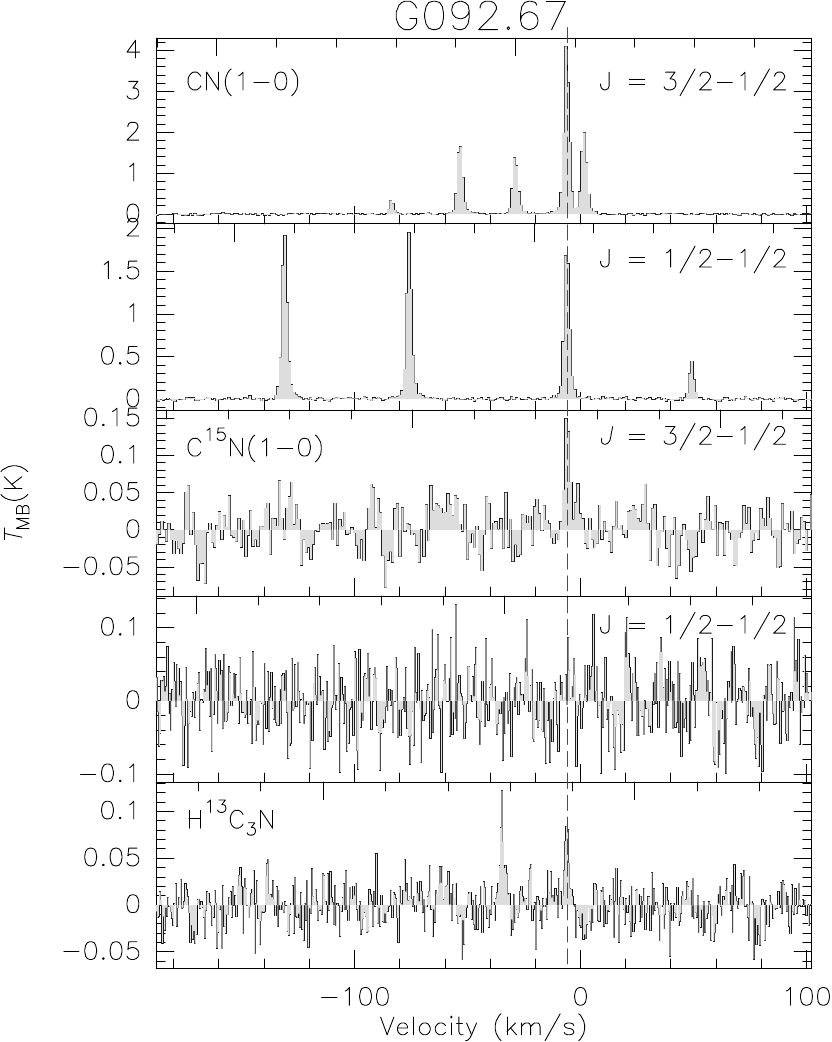}} 
	{\includegraphics[width=4.4cm]{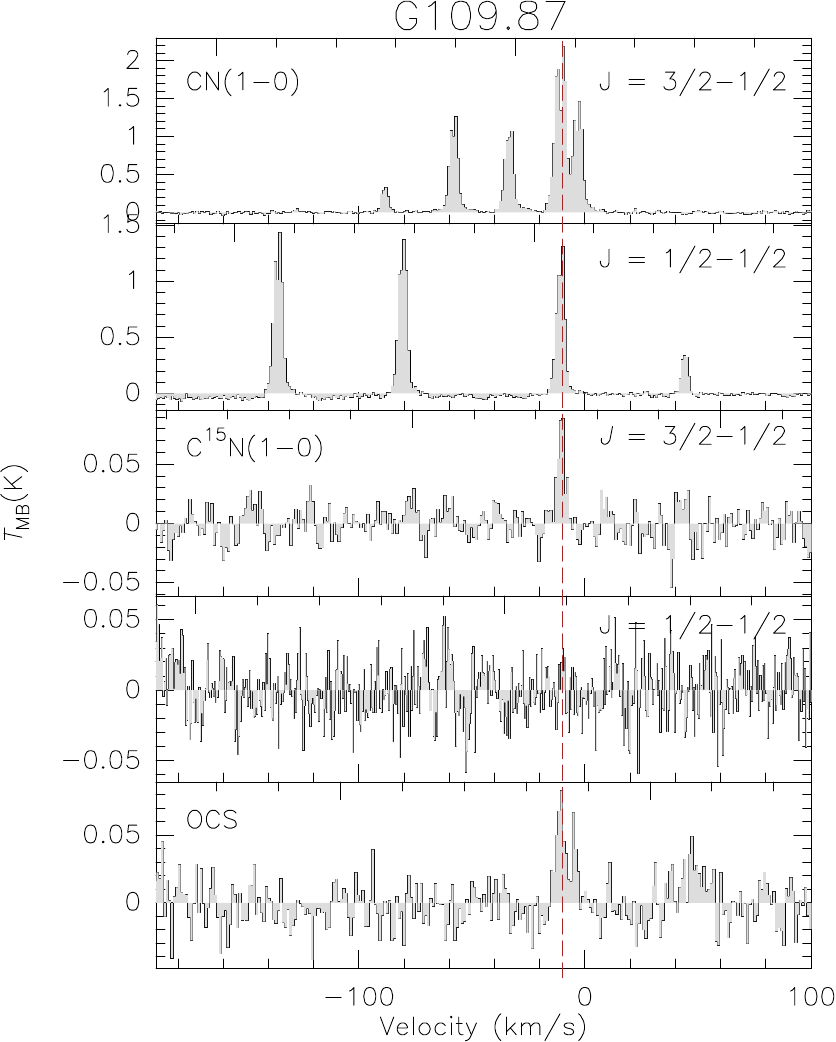}} 
	{\includegraphics[width=4.4cm]{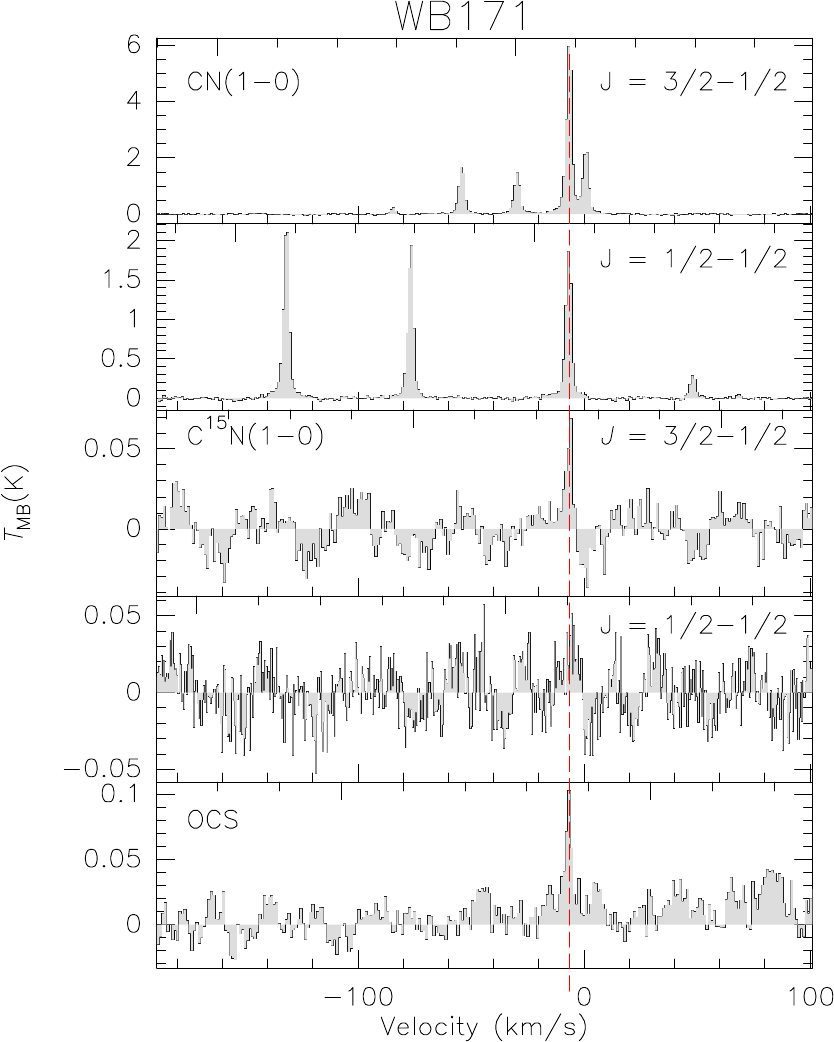}} 
	{\includegraphics[width=4.4cm]{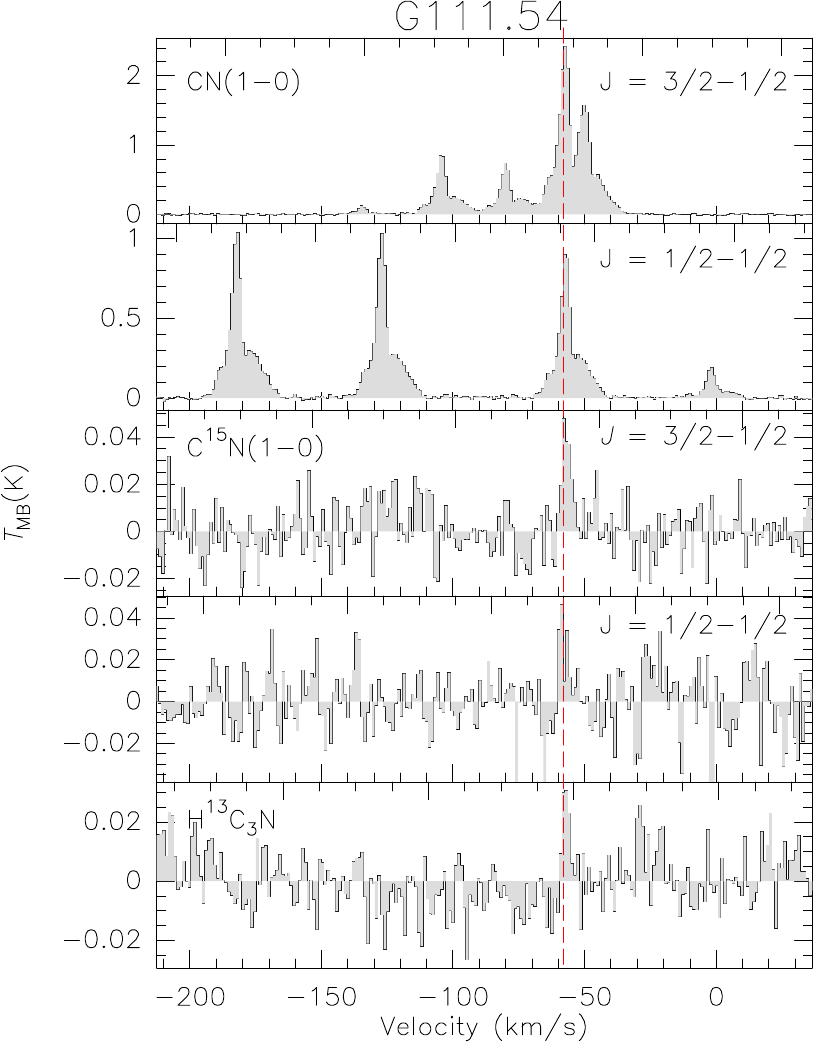}} 
	{\includegraphics[width=4.4cm]{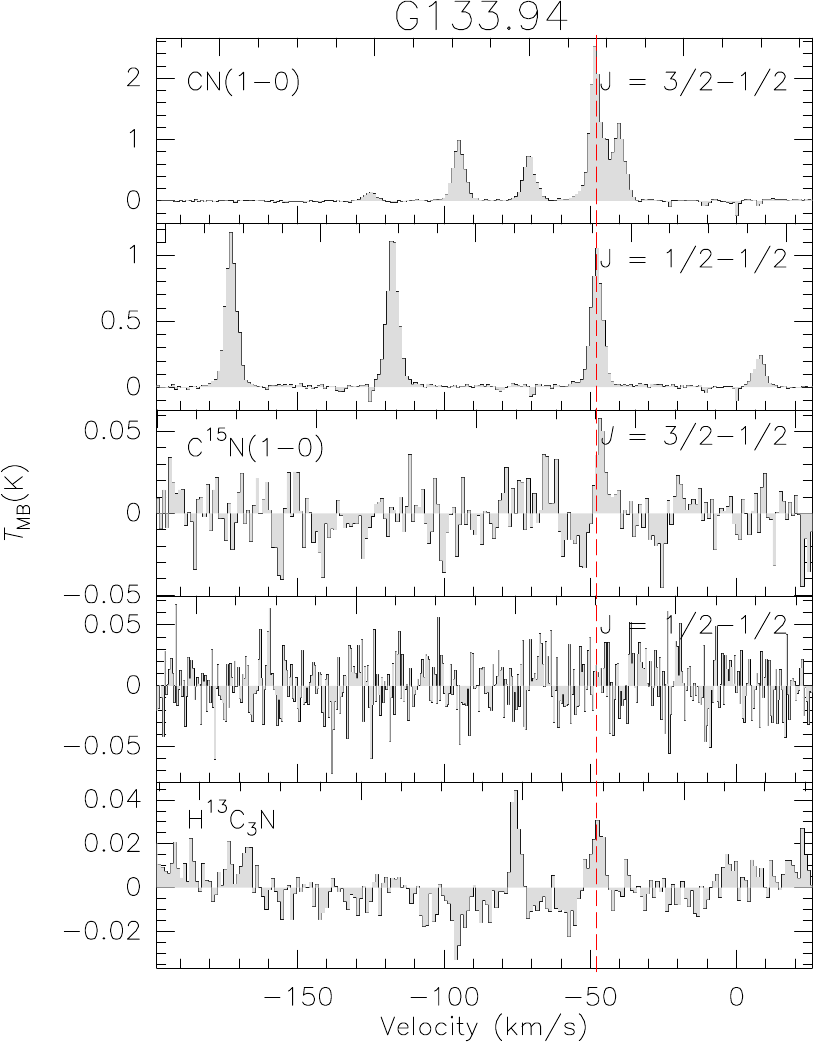}} 
	{\includegraphics[width=4.4cm]{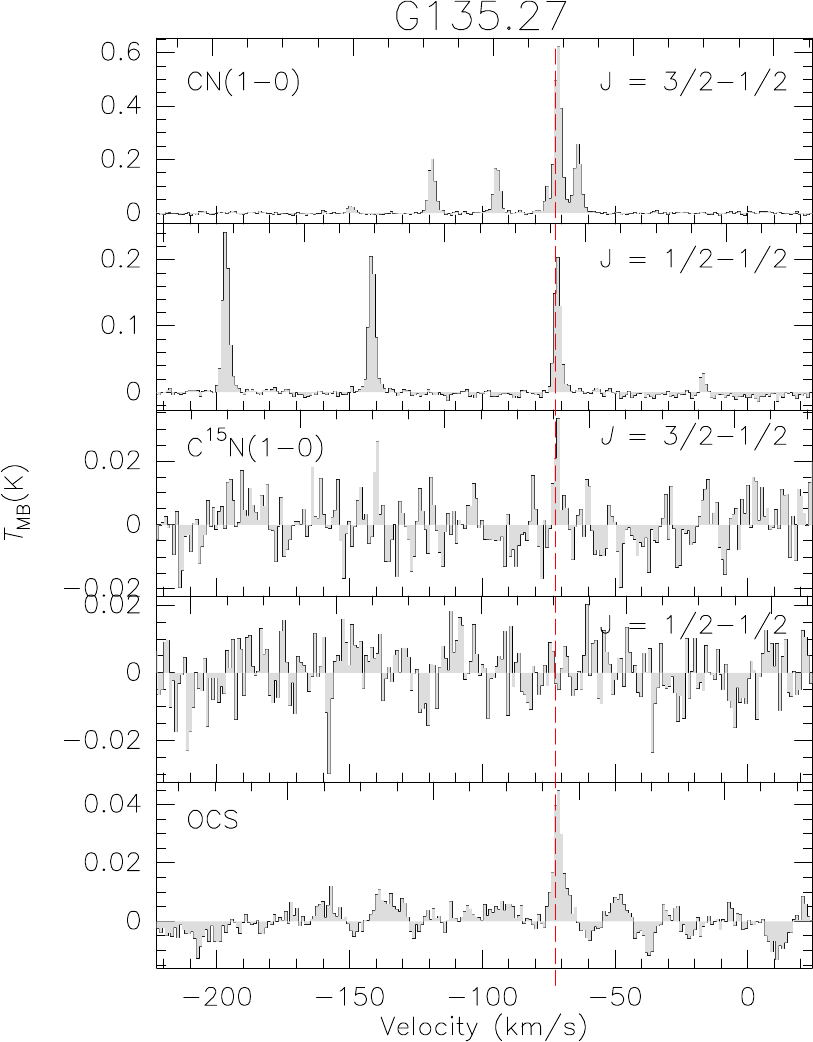}} 
	{\includegraphics[width=4.4cm]{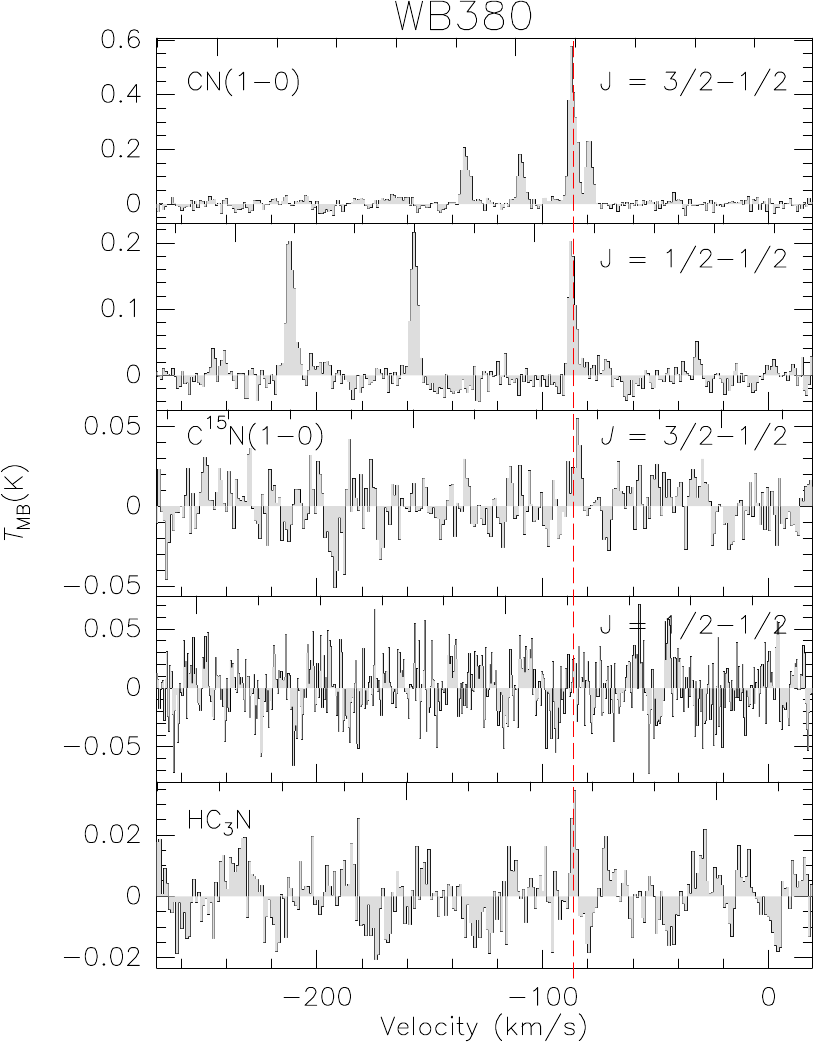}} 
	\caption{(Continued)}
	\label{}
\end{figure*}

\subsection{Line and physical parameters of our sample with detections of C$^{14}$N and C$^{15}$N}\label{sec:optical and Tex}

\begin{figure*}[htbp]
	\centering
	{\includegraphics[width=18cm]{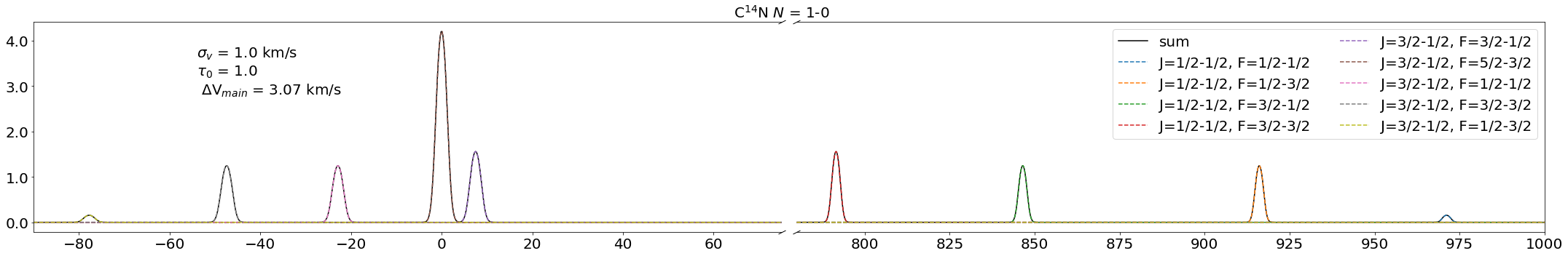}} 
	{\includegraphics[width=18cm]{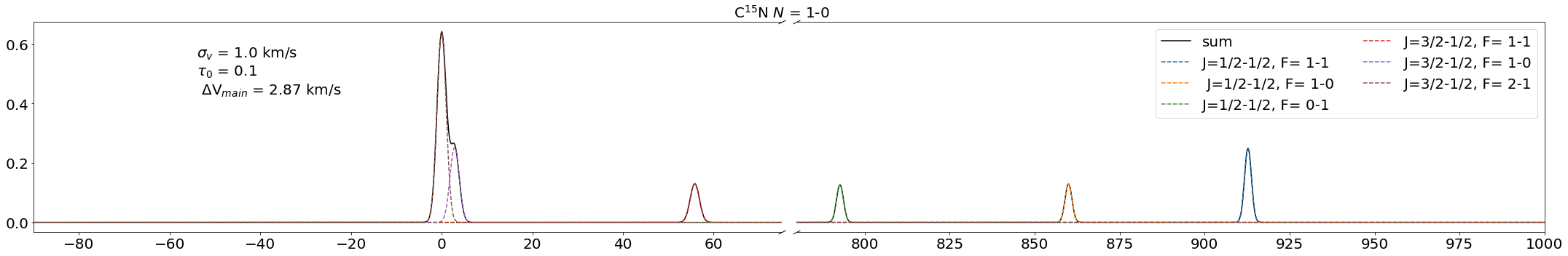}} 
	\caption{Synthetic C$^{14}$N, $N$ = 1--0 (upper panel) and C$^{15}$N, $N$ = 1--0 (lower panel) spectra for an intrinsic line width of 1.0 km s$^{-1}$ and a Gaussian line shape.}
	\label{fig:Synthetic_spectra}
\end{figure*}

Given the expected high nitrogen isotope ratios, the C$^{14}$N lines in clouds with measured C$^{15}$N emission may be optically thick, resulting in a non-linear link between the peak $T_{\rm mb}$ value and the column density of C$^{14}$N. Due to spin-rotation interaction, each C$^{14}$N rotational energy level with $N$ $>$ 0 is split into a doublet, the upper spin-doublet state, $J$ = 3/2 -- 1/2, and the lower spin-doublet state, $J$ = 1/2 -- 1/2. Each of these components is further split into a triplet of states by the spin of the nitrogen nucleus (I$_{1}$ = 1) \citep{1998A&A...329..443H,2012ApJ...744..194A}. The relative intensities of the resulting hyperfine (HF) components of C$^{14}$N are given in Table \ref{tab:linePara} of \cite{1998JQSRT..60..883P}.  Following the method in \cite{2021A&A...646A.170G}, and assuming LTE and that the intrinsic width of each HF line is 1 km s$^{-1}$, the synthetic line profiles of C$^{14}$N and C$^{15}$N, $N$ = 1--0 can be reproduced (see Figure  \ref{fig:Synthetic_spectra}). Among the 35 sources with C$^{15}$N detection, all HF components were detected in C$^{14}$N, which allows us to determine the optical depths of C$^{14}$N. Thus we determined the optical depth of the C$^{14}$N, $N$ = 1--0 lines, using HF fits (the “method” command in CLASS, e.g. \citetalias{2021ApJS..257...39C}). This method fits all hyperfine components of CN simultaneously. The optical depths of the strongest C$^{14}$N component are listed for those 35 sources with C$^{15}$N detections in column 3 of Table \ref{tab:result}.

\renewcommand\tabcolsep{9.0pt} 
\begin{deluxetable*}{lclccccc}
	\tablecaption{Line Parameters for C$^{14}$N and C$^{15}$N \label{tab:linePara}}
	\tablewidth{5pt}
	\tablehead{
		\colhead{Species}	&	\colhead{Transition}	&	\colhead{Hyperfine}	&	\colhead{Frequency}	&	\colhead{Hyperfine Relative}	&	\colhead{$A_{ul}$}	&	\colhead{$g_{u}$}	&	\colhead{$E_{U}$}	\\
		&		&	\colhead{Component}	&	\colhead{MHz}	&	\colhead{Intensity}	&	\colhead{(s$^{-1}$)}	&		&	\colhead{(K)}	
	}
	\decimalcolnumbers
	\startdata
	C$^{14}$N	&	$N$ = 1 -- 0,	&	$F$ = 1/2 -- 1/2	&	113123.34	&	0.0123	&	1.2866E-06	&	2	&	5.43004 	\\
	&	$J$ = 1/2 -- 1/2	&	$F$ = 1/2 -- 3/2	&	113144.19	&	0.0988	&	1.0529E-05	&	2	&	5.43003 	\\
	&		&	$F$ = 3/2 -- 1/2	&	113170.53	&	0.0988	&	5.1449E-06	&	4	&	5.43230 	\\
	&		&	$F$ = 3/2 -- 3/2	&	113191.32	&	0.1235	&	6.6828E-06	&	4	&	5.43229 	\\
	&	$N$ = 1 -- 0,	&	$F$ = 3/2 -- 1/2	&	113488.14	&	0.1235	&	6.7361E-06	&	4	&	5.44755 	\\
	&	$J$ = 3/2 -- 1/2	&	$F$ = 5/2 -- 3/2	&	113490.98	&	0.3333	&	1.1924E-05	&	6	&	5.44668 	\\
	&		&	$F$ = 1/2 -- 1/2	&	113499.64	&	0.0988	&	1.0629E-05	&	2	&	5.44810 	\\
	&		&	$F$ = 3/2 -- 3/2	&	113508.93	&	0.0988	&	5.1902E-06	&	4	&	5.44754 	\\
	&		&	$F$ = 1/2 -- 3/2	&	113520.41	&	0.0123	&	1.2995E-06	&	2	&	5.44809 	\\
	C$^{15}$N	&	$N$ = 1 -- 0,	&	$F$ = 1 -- 1	&	109689.61	&	0.1638	&	7.0958E-06	&	3	&	5.26510 	\\
	&	$J$ = 1/2 -- 1/2	&	$F$ = 1 -- 0	&	109708.99	&	0.0847	&	3.6717E-06	&	3	&	5.26517 	\\
	&		&	$F$ = 0 -- 1	&	109733.65	&	0.0829	&	1.0777E-05	&	1	&	5.26722 	\\
	&	$N$ = 1 -- 0,	&	$F$ = 1 -- 1	&	110004.09	&	0.0854	&	3.7021E-06	&	3	&	5.28020 	\\
	&	$J$ = 3/2 -- 1/2	&	$F$ = 1 -- 0	&	110023.54	&	0.1653	&	7.1603E-06	&	3	&	5.28027 	\\
	&		&	$F$ = 2 -- 1	&	110024.59	&	0.4179	&	1.0864E-05	&	5	&	5.28118 	\\
	\enddata
	\tablecomments{Column(1): species; column(2): transitions; column(3): HF components of C$^{14}$N and C$^{15}$N; column(4): frequency from the JPL Molecular Spectroscopy Catalog \citep{1998JQSRT..60..883P}; column(5): HF relative intensity of C$^{14}$N and C$^{15}$N; column(6): Einstein coefficient for spontaneous emission; column(7): upper state degeneracy; column(8): energy of the lower level above the ground state; column(9): energy of the upper level above the ground state.}
\end{deluxetable*}


Assuming conditions of Local Thermodynamic Equilibrium (LTE), the excitation temperature can be estimated using the expression with derived optical depth \citep{2015PASP..127..266M}:
\begin{equation}  \label{Temp:excitation}                                           
	T_{\rm mb}(main \; HF) =  f[J_{v}(T_{ex})\;-\;J_{v}(T_{bg})](1\;-\;e^{-\tau_{main}}),
\end{equation} 
where $T_{\rm mb}(main \; HF)$ is the $T_{\rm mb}$ of the strongest CN hyperfine component ($J$ = 3/2--1/2 $F$ = 5/2--3/2) and $f$ is the beam filling factor, assuming a value of unity. $T_{ex}$  is the excitation temperature and $T_{bg}$ is the cosmic background temperature. 
$J_{v}(T)$ is  the equivalent temperature of a black body at temperature T:
\begin{equation} \label{Temp:Jv}                                           
	J_{v}(T)=\frac{\frac{hv}{k}}{{\rm exp}({\frac{hv}{kT}})-1},
\end{equation} 
where $k$ is the Boltzmann constant and $h$ is the Planck constant. 
The resulting excitation temperatures are listed in column 4 of Table \ref{tab:result}.

\subsection{Measured abundance ratios}\label{sec:ratio}

In Section \ref{sec:optical and Tex}, we have calculated the optical depths and excitation temperatures of C$^{14}$N for the 35 sources with detections of both C$^{14}$N and C$^{15}$N. Those results can be used to derive the $^{14}$N/$^{15}$N ratio, combining the brightness temperature of  spectral lines of C$^{14}$N and C$^{15}$N. Under the conditions of LTE and optically thin line emission, the strongest HF components in CN and C$^{15}$N were utilized to determine the $^{14}$N/$^{15}$N ratios directly, with relative HF intensities of C$^{14}$N and C$^{15}$N ($R_{HF}$ = 1.25) as weights \citep{2002ApJ...578..211S,2012ApJ...744..194A}.
\begin{equation} \label{Ratio:Tpeak}                                     
	\frac{^{14}N}{^{15}N} = \frac{T_{mb}({\rm CN})}{T_{mb}({\rm C^{15}N})}\times R_{HF}.
\end{equation} 
In the optically thick case, $T_{\rm mb}(CN)$ in Equation (\ref{Ratio:Tpeak}) is replaced by $\tau_{main} \times T_{ex}$. 
Since the main HF line $F$ = 2-1 of C$^{15}$N is partly blended with the weaker $F$ = 1-0 component (with a relative intensity ratio of 0.418 : 0.165 under (realistic) optically thin conditions (for details, see Table \ref{tab:linePara}), \cite{2012ApJ...744..194A} modeled the observed line to account for the contribution of the weaker component, giving typical corrections on the order of 10\% to 20\%. Here we adopted an uncertainty of 15\% when calculating $^{14}$N/$^{15}$N ratios with Equation (\ref{Ratio:Tpeak}). The corrected $^{14}$N/$^{15}$N results for our 35 targets are listed in column 10 of Table \ref{tab:result}.  For those three sources (G109.87, G111.54 and G133.94) measured by both the IRAM 30 m and ARO 12 m telescope, the unweighted mean value of their C$^{14}$N/C$^{15}$N ratios from the two telescopes was taken for our later analysis, including their errors in the overall error budgets. There is less than a 20\% margin of error associated with the C$^{14}$N/C$^{15}$N ratio measured by both telescopes (see details in Section \ref{sec:obseffect}). Possible physical, chemical and observational contaminating effects influencing the abundance ratios are discussed in Secttions \ref{sec:obseffect} to \ref{sec:fract}. 

\renewcommand\tabcolsep{2.8pt} 
\begin{deluxetable*}{lccccccccccc} 
	\tablecaption{Line Parameters for C$^{14}$N and C$^{15}$N \label{tab:result}}
	\tablewidth{0pt}
	\tablehead{
		\colhead{Object}&
		\colhead{Telescope}&
		\colhead{$\tau_{main}$}&
		\colhead{$T_{ex}$}&
		\colhead{$T_{kin}$}&
		\colhead{$D_{sun}$}&
		\colhead{$R_{\rm GC}$}&
		\colhead{$R_{12}(\mathrm{CN})$}&
		\colhead{$R_{12}(\mathrm{C^{15}N})$}&
		\colhead{$R_{mean}$}&
		\colhead{$\rm \frac{C^{14}N}{C^{15}N}$}&
		\colhead{$\rm \frac{C^{14}N}{C^{15}N}^{N}$}\\
		&
		&
		&
		\colhead{(K)}&
		\colhead{(K)}&
		\colhead{(kpc)}&
		\colhead{(kpc)}&
		&
		&
		&
		&
	}
	\decimalcolnumbers
	\startdata
	G010.47&IRAM&0.1(0.1)&6.9(0.6)&32.4&8.2(0.1)&1.6(0.1)&0.60(0.04)&&1.51(0.11)&45.8(15.9)&104.4(36.2)\\
	G009.62&IRAM&1.0(0.1)&11.5(0.1)&22.6&3.9(0.3)&3.2(0.3)&0.90(0.02)&&1.43(0.12)&179.9(67.7)&241.4(90.8)\\
	G010.621&IRAM&0.8(0.1)&20.1(0.1)&13.1&4.8(0.2)&3.4(0.2)&1.15(0.17)&&2.15(0.28)&92.1(32.0)&230.0(80.0)\\
	G023.43&IRAM&13.0(0.2)&4.5(0.1)&...&5.5(0.8)&3.6(0.4)&0.58(0.13)&&0.96(0.05)&974.6(546.8)&204.7(168.4)\\
	G029.86&IRAM&1.0(0.1)&8.9(0.1)&23.1&6.9(0.7)&4.1(0.1)&0.63(0.01)&&1.24(0.09)&198.2(107.6)&\\
	G013.87&IRAM&0.8(0.1)&7.4(0.1)&86.0&3.9(0.2)&4.4(0.2)&0.77(0.15)&&1.27(0.03)&76.8(25.6)&383.8(127.8)\\
	G001.14&IRAM&10.8(1.3)&4.3(0.2)&47.0&3.7(0.3)&4.4(0.3)&0.97(0.17)&&1.30(0.16)&228.4(113.7)&\\
	G029.95&IRAM&0.1(0.1)&8.6(0.1)&33.0&6.9(0.7)&4.4(0.1)&0.61(0.07)&&1.56(0.05)&108.2(55.6)&\\
	G005.88&IRAM&0.1(0.1)&15.1(0.4)&39.7&2.9(0.2)&5.2(0.2)&0.66(0.08)&0.26(0.08)&1.32(0.09)&104.6(24.8)&324.4(76.8)\\
	G012.81&IRAM&3.5(0.1)&9.7(0.1)&15.6&2.9(0.3)&5.3(0.3)&0.69(0.06)&&1.53(0.04)&282.2(110.2)&267.7(104.5)\\
	G012.88&IRAM&4.5(0.1)&5.2(0.1)&23.0&2.9(0.3)&5.7(0.3)&0.74(0.09)&&1.46(0.05)&368.2(214.4)&266.8(155.4)\\
	G049.49&IRAM&0.1(0.1)&12.3(0.3)&24.4&5.5(0.3)&6.2(0.1)&0.75(0.22)&&1.29(0.11)&131.3(55.9)&188.5(80.2)\\
	G049.48&IRAM&0.1(0.1)&10.9(0.3)&25.6&5.5(0.3)&6.2(0.1)&0.65(0.06)&&1.35(0.07)&100.9(40.3)&170.1(68.0)\\
	G035.02&IRAM&2.5(0.1)&6.1(0.1)&23.7&2.9(0.7)&6.4(0.4)&0.64(0.11)&&1.12(0.05)&133.2(69.1)&\\
	G035.19&IRAM&7.4(0.1)&5.6(0.1)&17.1&2.2(0.2)&6.5(0.2)&0.59(0.04)&&1.31(0.08)&765.4(394.8)&322.9(166.6)\\
	G035.14&IRAM&8.5(0.1)&6.0(0.1)&17.1&2.2(0.2)&6.5(0.1)&0.65(0.18)&&1.40(0.11)&819.7(437.8)&330.7(176.6)\\
	G059.78&IRAM&2.8(0.1)&8.9(0.1)&25.7&2.2(0.1)&7.3(0.1)&0.61(0.01)&&1.23(0.09)&128.7(49.5)&\\
	G069.54&IRAM&4.4(0.1)&7.5(0.1)&14.6&4.1(0.6)&7.6(0.1)&0.69(0.05)&&1.29(0.10)&490.0(276.5)&\\
	G078.12&IRAM&1.1(0.1)&14.5(0.1)&21.9&1.1(6.2)&7.9(0.4)&0.53(0.04)&&1.16(0.12)&252.1(169.4)&\\
	G081.75&IRAM&7.9(0.1)&7.8(0.1)&19.4&2.3(0.4)&8.0(0.1)&0.61(0.11)&&1.25(0.11)&641.2(272.7)&\\
	G078.88&IRAM&0.5(0.1)&13.6(0.1)&45.2&2.9(1.2)&8.2(0.2)&0.65(0.19)&&1.06(0.06)&161.4(83.3)&\\
	G092.67&IRAM&3.8(0.1)&7.9(0.1)&28.9&1.6(0.1)&8.4(0.1)&0.64(0.02)&&1.16(0.10)&176.5(69.3)&\\
	G109.87&IRAM&10.8(0.1)&5.7(0.1)&19.5&0.8(0.1)&8.4(0.1)&0.65(0.14)&&1.09(0.03)&742.2(301.4)&\\
	&ARO&7.0(0.1)&5.3(0.1)&19.5&0.8(0.1)&8.4(0.1)&0.61(0.14)&0.26(0.10)&1.07(0.05)&616.5(208.3)&\\
	WB171&IRAM&0.1(0.1)&9.6(0.3)&...&0.8(0.1)&8.4(0.1)&0.53(0.10)&&1.09(0.08)&95.0(39.4)&\\
	G209.00&ARO&0.1(0.1)&6.1(1.4)&112.1&0.4(0.1)&8.5(0.1)&0.54(0.04)&&1.35(0.24)&273.9(224.8)&\\
	G121.29&ARO&3.3(0.1)&5.0(0.1)&18.0&2.4(0.9)&8.6(0.6)&0.62(0.07)&&1.25(0.08)&314.2(147.8)&\\
	G111.54&IRAM&2.1(0.1)&5.6(0.1)&22.3&2.7(0.1)&9.4(0.1)&0.66(0.09)&&1.08(0.07)&326.4(192.8)&\\
	&ARO&2.4(0.1)&5.7(0.1)&22.3&2.7(0.1)&9.4(0.1)&0.59(0.05)&&1.18(0.04)&336.1(169.2)&\\
	G133.94&IRAM&3.1(0.1)&6.2(0.1)&21.2&2.0(0.1)&9.6(0.1)&0.61(0.05)&&1.22(0.05)&291.5(109.3)&\\
	&ARO&3.4(0.1)&5.1(0.1)&21.2&2.0(0.1)&9.6(0.1)&0.59(0.01)&&1.15(0.07)&357.3(131.0)&\\
	G192.60&ARO&2.0(0.1)&7.2(0.1)&16.1&1.7(0.1)&9.7(0.1)&0.58(0.01)&&1.14(0.02)&279.6(98.4)&\\
	G173.48&ARO&2.7(0.1)&4.9(0.1)&25.7&1.7(0.1)&9.8(0.1)&0.58(0.01)&&1.12(0.03)&345.4(120.1)&\\
	G123.06&ARO&3.1(0.1)&4.6(0.1)&21.8&2.5(0.3)&9.9(0.2)&0.59(0.01)&&1.24(0.04)&671.4(287.3)&\\
	G174.20&ARO&3.9(0.1)&4.2(0.1)&22.6&2.1(0.1)&10.2(0.1)&0.61(0.01)&&1.18(0.02)&306.9(103.9)&\\
	G211.59&ARO&2.3(0.3)&3.4(0.1)&...&4.3(0.2)&12.1(0.2)&0.57(0.03)&&1.01(0.06)&205.2(117.5)&\\
	G135.27&IRAM&0.3(0.1)&5.6(0.1)&...&5.6(0.4)&13.1(0.4)&0.52(0.03)&&1.06(0.08)&20.3(16.0)&\\
	WB380&IRAM&1.9(0.3)&3.6(0.1)&...&9.8(0.7)&15.9(0.6)&0.59(0.11)&&1.20(0.08)&98.7(47.8)&\\
	\enddata
	\tablecomments{Column(1): The 35 sources; column(2): used telescope; column(3): the peak optical depth of the main group of HF components of C$^{14}$N, from the intensity ratio method \citep{2021ApJS..257...39C}; column(4): $T_{ex}$ from the radiative transfer function; column(5): the kinetic temperatures for our sources derived by \cite{2010MNRAS.402.2682H}, \cite{2011ApJ...741..110D}, \cite{2016ApJ...822...59S} and \citetalias{2021ApJS..257...39C}, which were estimated from the para-NH$_{3}$ (1, 1) and (2, 2) transitions; column(6): heliocentric distance with error. For two sources (WB171 and WB380) without parallax data, kinematic distances were estimated (see details in Section \ref{sec:distance}); column(7): Galactocentric distance with error from the heliocentric distance; column(8)-(9): $R_{12}(\mathrm{CN})$ and $R_{12}(\mathrm{C^{15}N})$ are the integrated intensity ratios of the $J$=1/2 – 1/2 to $J$=3/2 – 1/2 spin-doublet lines, with values of 0.5 and 0.28, respectively, in the optically thin case under LTE conditions (see Section \ref{sec:LTE deviations in CN}); column(10): the opacity corrected C$^{14}$N/C$^{15}$N ratios of our sources, making use of the strongest HF component of CN. column(11): Modified C$^{14}$N/C$^{15}$N results for those 12 sources, which suffer from LTE deviations and/or self-absorption in C$^{14}$N. In these cases the weakest HF component of C$^{14}$N was used to estimate the $^{14}$N/$^{15}$N ratio, which will be used in later analysis (see details in Section \ref{sec:obseffect} and \ref{sec:self-absorptions}).}
\end{deluxetable*}

\section{Discussion}\label{sec:discussion}

\subsection{Possible effects on abundance ratios}\label{sec:Possible effects on abundance ratios}


\subsubsection{Observational effects}\label{sec:obseffect}

\begin{figure*} [htbp]
	\centering
	{\includegraphics[width=17cm]{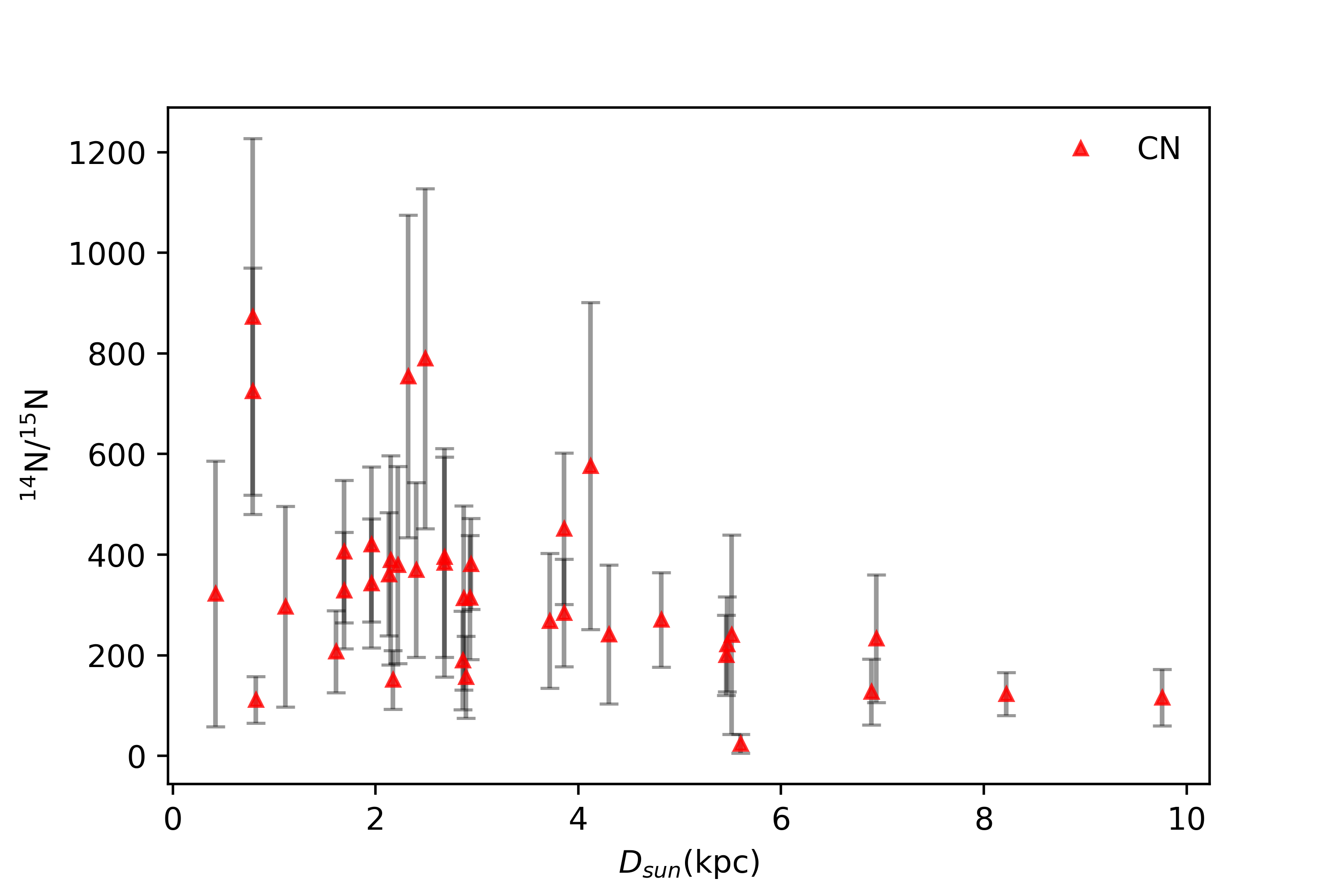}}
	\caption{Our $^{14}$N/$^{15}$N isotope ratios from C$^{14}$N and C$^{15}$N, plotted against the heliocentric distance.} 
	\label{CN_Dsun}
\end{figure*}

The linear beam sizes of sources vary as a function of distance and the presence of more dispersed low-density gas within larger beam sizes of sources at greater distance may impact the resulting isotope ratios. Additionally, there could be a potential bias towards brighter and more massive sources that could consistently have higher opacities \citep{2008A&A...487..237W,2020ApJS..249....6Z,2021ApJS..257...39C,2023ApJS..268...56Z}, which could lead to an underestimation of the C$^{14}$N/C$^{15}$N ratios. To evaluate potential impacts of beam dilution on our ratios, we present the isotope ratio against the heliocentric distance in Figure \ref{CN_Dsun}. While there is a large scatter in the ratios at small distances, only low values are found in the few sources at large distances, which will be explained in Section \ref{sec:gradient}. There is at least no strong dependence between the ratio and the distance.

\begin{figure*}[htbp]
	\centering
	{\includegraphics[width=6cm]{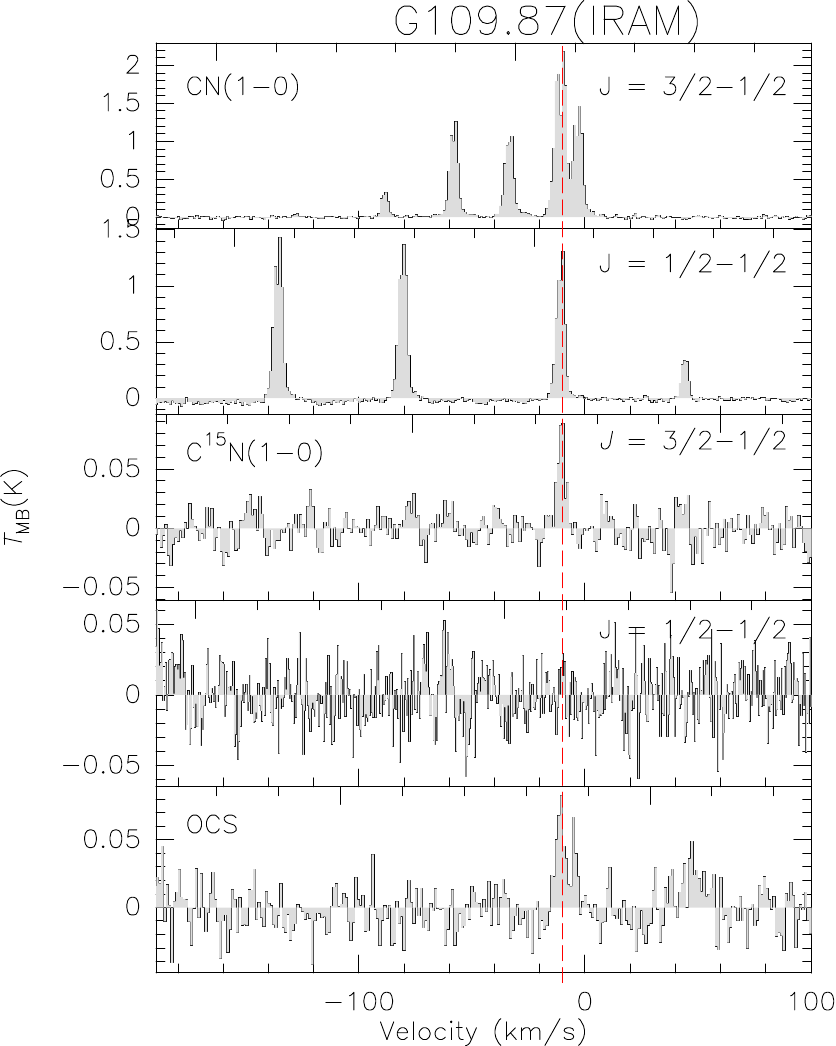}} 
	{\includegraphics[width=6cm]{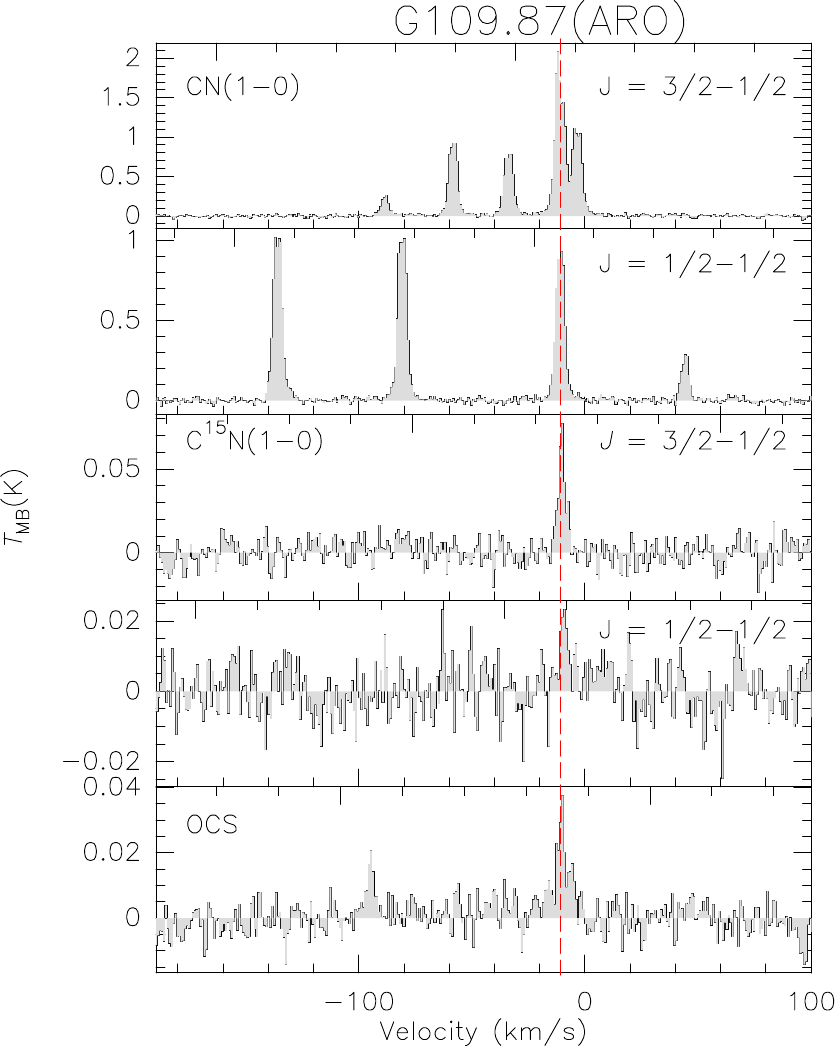}} 
	{\includegraphics[width=6cm]{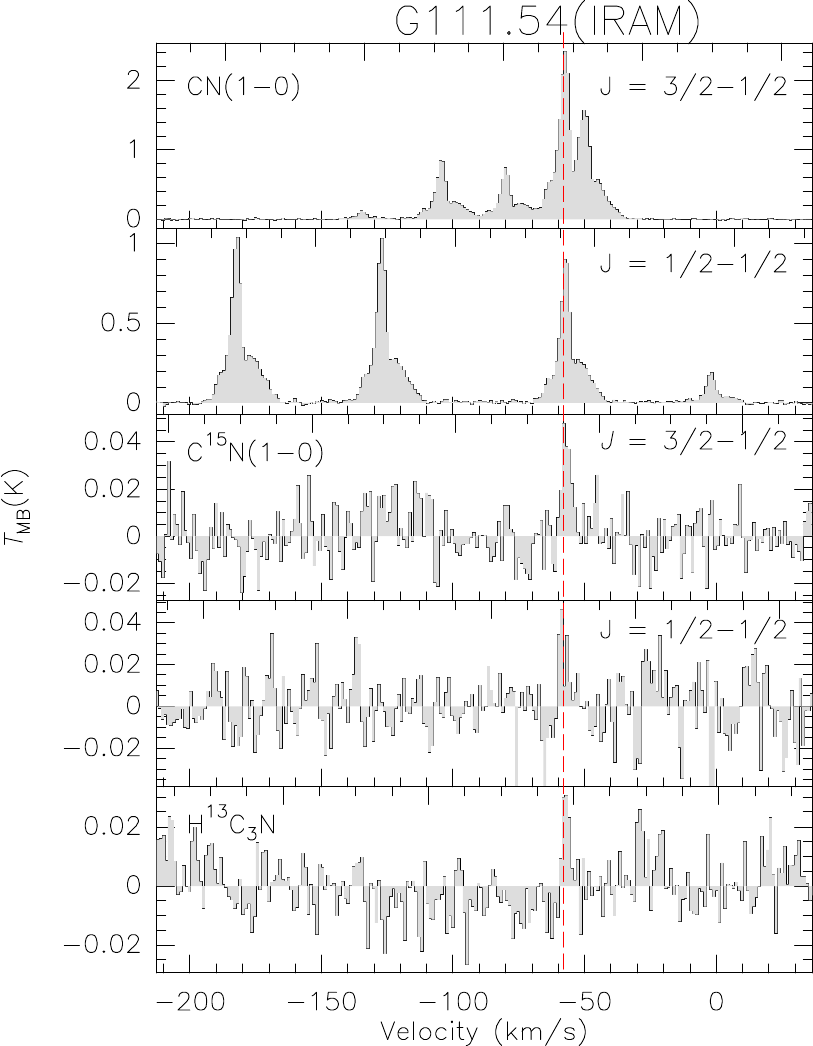}} 
	{\includegraphics[width=6cm]{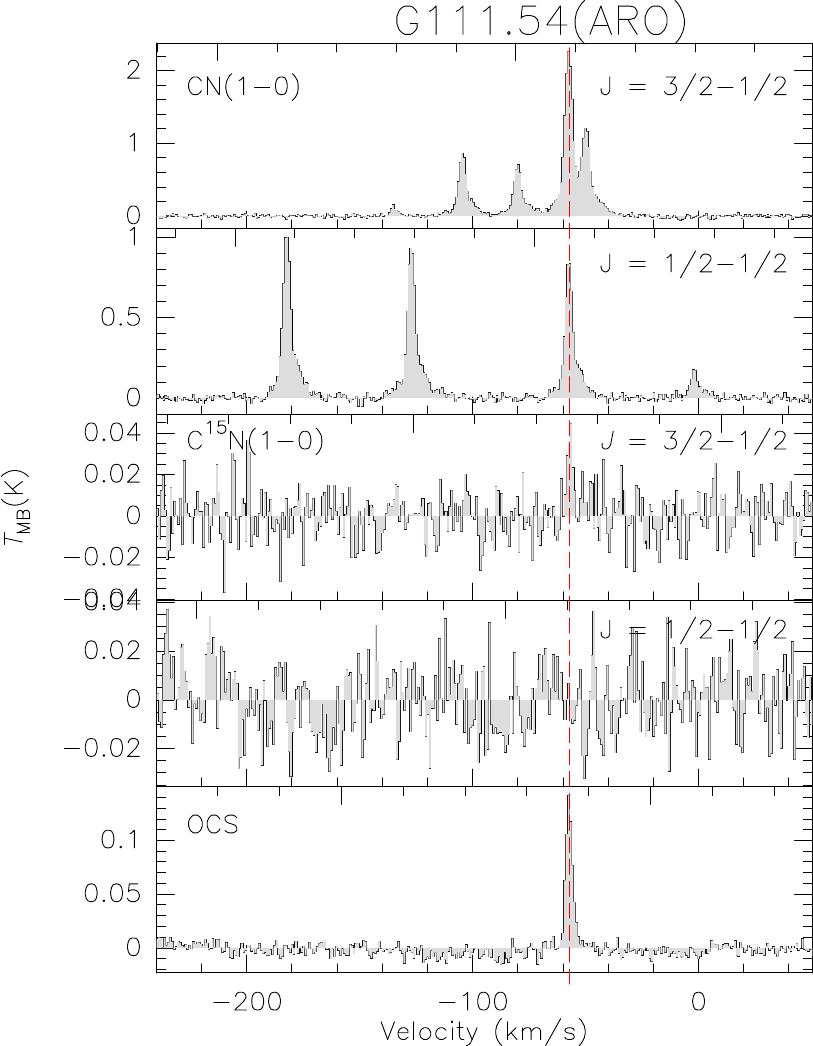}} 
	{\includegraphics[width=6cm]{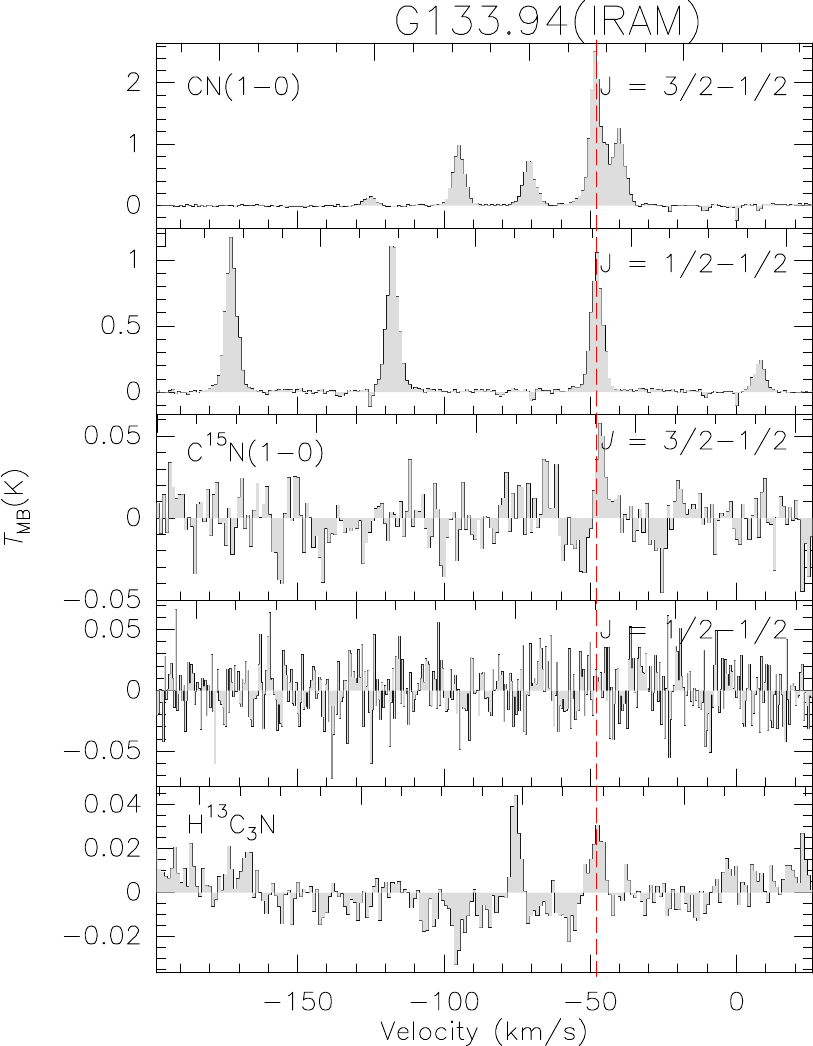}} 
	{\includegraphics[width=6cm]{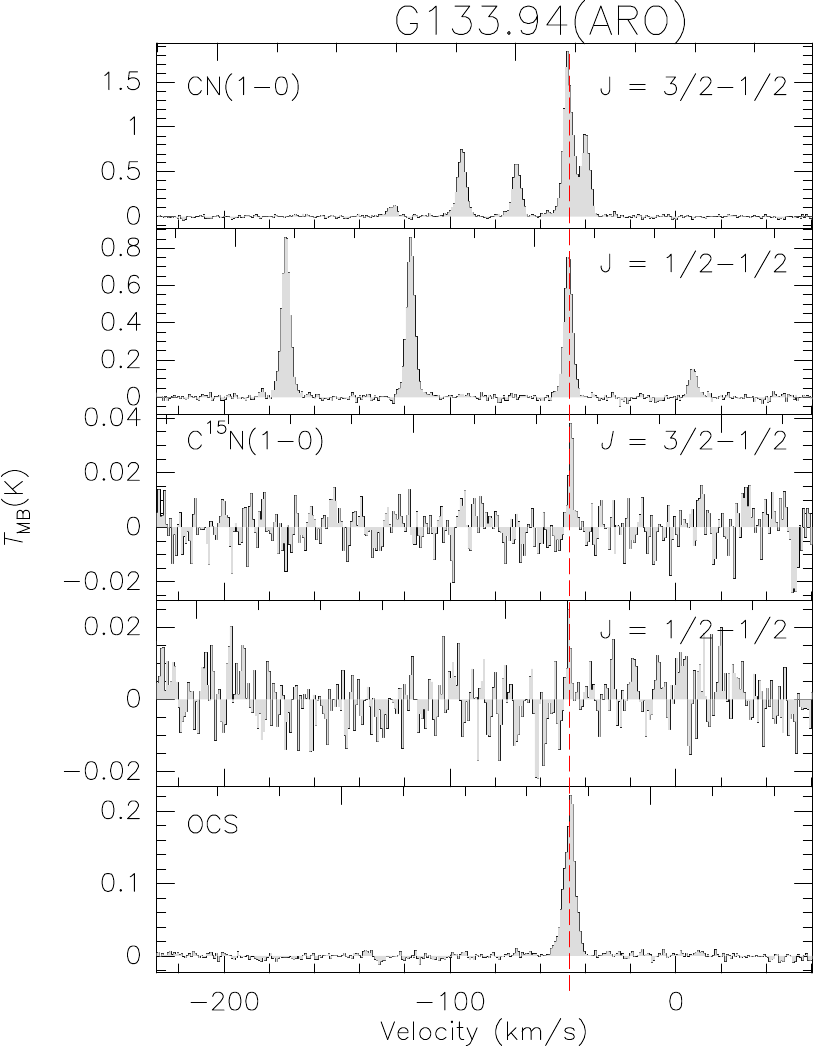}} 
	\caption{The spectra of C$^{14}$N (upper panels) and C$^{15}$N (lower panels) of those three sources detected by both the IRAM 30$\,$m (right column) and the ARO 12$\,$m telescopes (left column). Vertical red dashed lines mark the line center velocity of the source measured from Gaussian fits to the OCS 9--8 or HCC$^{13}$CN $J$ = 12--11 line.}
	\label{fig:iram/aro}
\end{figure*}

As mentioned in Section \ref{sec:detection}, three sources  out of our sample (G109.87, G111.54, and G133.94) were detected by the IRAM 30$\,$m and ARO 12 m telescopes in both isotopologues, which have a beam area difference of approximately eight. This provides an opportunity to investigate the effects of beam dilution between data from the two telescopes. Comparisons show that the spectra of G109.87 and G111.54 are similar, i.e., all individual spectral lines from the same source show similar line profiles and peak temperatures on both telescopes. It suggests that these two sources are extended, i.e., their sizes are larger than both beam sizes. For G133.94, all individual spectral lines of C$^{14}$N and C$^{15}$N show a stronger signal in the IRAM 30 m observations, with respect to the ARO 12 m measurement (see Figure  \ref{fig:iram/aro}). This means that the source size should be smaller than the ARO beam size, and the dilution effect on individual spectral lines is not negligible.

The abundance ratio results of the three sources (G109.87, G111.54, and G133.94) detected by both telescopes were compared. Within the error limits identical results can be found toward these three sources (see Table \ref{tab:result}). This suggests that the beam dilution effect has no significant impact on our ratio results.

\subsubsection{LTE deviations in C$^{14}$N and C$^{15}$N}\label{sec:LTE deviations in CN}

LTE deviations in CN may affect the reliability of using the strongest HF components to determine the $^{14}$N/$^{15}$N ratio. Therefore, evaluating the LTE deviations for our 35 targets is necessary. The integrated intensity ratio of the $J$ = 1/2 – 1/2 to $J$ = 3/2 – 1/2 spin-doublet lines, $R_{12}(\mathrm{CN})$ can be used to estimate the LTE deviation effect \citep{1998A&A...329..443H}:
\begin{equation} \label{equ:LTE}                                     
	R_{12}(\mathrm{CN}) = \frac{I(J = 1/2 - 1/2)}{I(J = 3/2 - 1/2)},
\end{equation} 
where $I$ represents the intensity of a given line and $R_{12}(\mathrm{CN})$ has a theoretical value of 0.5 in the case of optical thin lines under LTE conditions, reaching the limit of unity in case of extremely high opacities. Using our observational data, we calculated the values of $R_{12}(\mathrm{CN})$ for all 35 targets (Table \ref{tab:result}). Except one source (G010.621), all $R_{12}(\mathrm{CN})$ values of our sample are smaller than unity but slightly larger than 0.5, which may provide evidence for C$^{14}$N line saturation \citep{1998A&A...329..443H}.




Four slightly stronger HF components of C$^{14}$N, namely the $J$ = 1/2 -- 1/2 $F$ = 1/2 -- 3/2 and $F$ = 3/2 -- 1/2, $J$ = 3/2 – 1/2 $F$ = 1/2 -- 1/2 and $F$ = 3/2 -- 3/2 features (Figure \ref{fig:Synthetic_spectra}), are likely all optically thin and should have the same intensities under LTE conditions (without including the component $J$ = 1/2 -- 1/2 $F$ = 3/2 -- 3/2, which is stronger than these  four components). We find that those four HF components show slightly different intensities from spectra, which are likely caused by non-significant LTE deviations. Thus, to quantify "non-significant" LTE deviations for our sources, we calculated two integrated intensity ratios of $J$ = 1/2 – 1/2 $F$ = 1/2 -- 3/2 to $F$ = 3/2 -- 1/2 and $J$ = 3/2 – 1/2 $F$ = 1/2 -- 1/2 to $F$ = 3/2 -- 3/2, and then derived the mean value ($R_{mean}$) of these two ratios (Table \ref{tab:result}). The $R_{mean}$ values of those 34 sources range from 0.95 -- 1.56, which is consistent with the theoretical value of unity in the case of optical thin lines under LTE conditions. This supports non-significant LTE deviations in them. 

\begin{figure*}[htbp]
	\centering
	{\includegraphics[width=9cm]{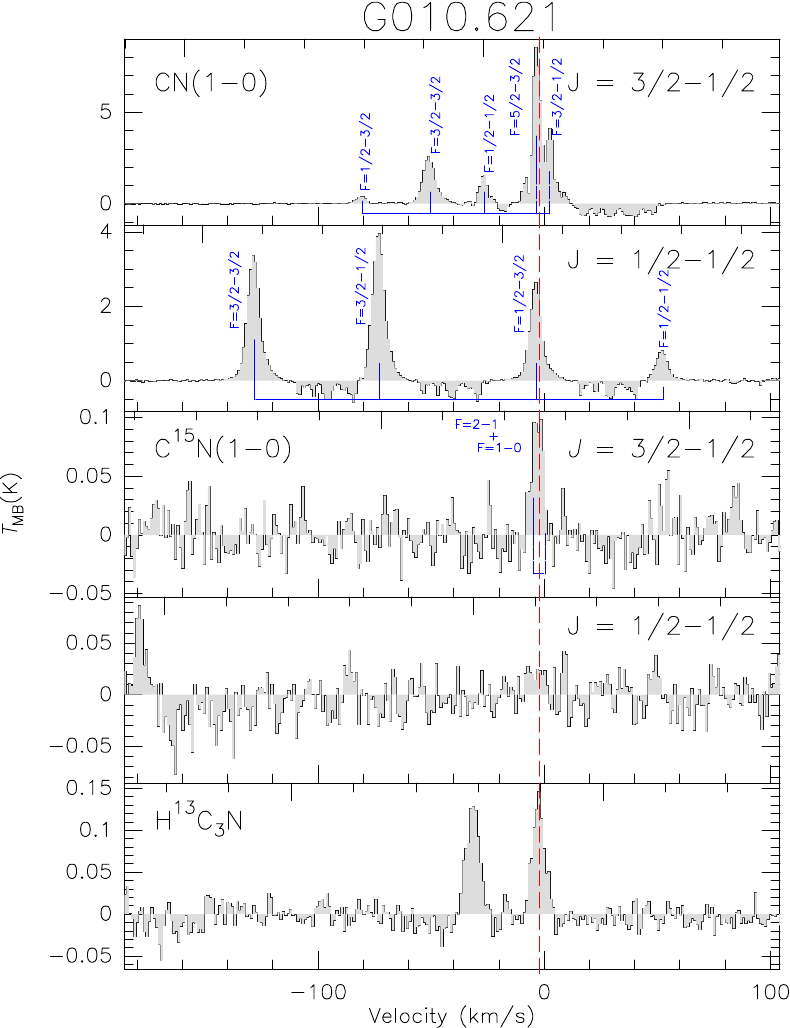}} 
	\caption{IRAM spectra of G010.621 with detected C$^{15}$N lines, demonstrating LTE deviations in CN. Vertical red dashed lines mark the line center velocity of the source for the strongest 
		hyperfine component derived from a Gaussian fit to HCC$^{13}$CN $J$ = 12--11 line.}
	\label{fig:LTE deviations}
\end{figure*}

Toward G010.621, the remaining target out of our sample of 35 detected sources, $R_{12}(\mathrm{CN})$ is 1.15 with low optical depth ($\tau_{main}$ is approximately 0.8, see Table \ref{tab:result}), suggesting severe deviations from LTE \citep{1998A&A...329..443H}. And this is supported by its $R_{mean}$ of $\sim$2.15, which is much larger than the LTE theoretical value of unity. The spectral features also support significant LTE deviations in this source (Figure \ref{fig:LTE deviations}). Under LTE conditions, the $J$ = 1/2 -- 1/2 $F$ = 3/2 -- 3/2 component  is the strongest one and the $J$ =1/2 -- 1/2 $F$ = 3/2--1/2 line has the same intensity as the $J$ = 1/2 -- 1/2 $F$ = 1/2 -- 3/2 component (see details in Figure \ref{fig:Synthetic_spectra}).  However, the CN spectra in this source show the $J$ =1/2 -- 1/2 $F$ = 3/2 -- 1/2 component is stronger than the $J$ = 1/2 -- 1/2 $F$ = 3/2 -- 3/2 and the $J$ = 1/2 -- 1/2 $F$ = 1/2 -- 3/2 component. 

For C$^{15}$N, two out of 35 sources (G005.88, G109.87) were detected in both the $J$ = 1/2 -- 1/2 and $J$ = 3/2 -- 1/2 spin-doublet lines. Thus, we also evaluated the LTE deviations in C$^{15}$N of these two sources. The estimated $R_{12}(\mathrm{C^{15}N})$ values of these sources are 0.26 $\pm$ 0.08 and 0.26 $\pm$ 0.10 (see Table \ref{tab:result}), which are consistent with the theoretical value of 0.28. This indicates that C$^{15}$N LTE-deviations are not significant.


\subsubsection{Self-absorption}\label{sec:self-absorptions}

Self-absorption of CN lines would be significant if clouds contain outer envelopes with properties similar to those of diffuse clouds, where the optical extinction $A_{v}$ approximately equals unity \citep{1984ApJ...283..668C}. This can influence the optical depth when using HF line ratios to derive it (for details, see Section \ref{sec:optical and Tex}) and challenge the validity of opacity-corrected C$^{14}$N/C$^{15}$N ratios. Observationally, effects of self-absorption include apparent shifts in line velocity and non-LTE HF line ratios \citep{2006MNRAS.367..553P,2010ApJ...710..150C,2013ApJS..206...22C}. To access the self-absorption of CN lines in our sample, we compared our line center velocities of CN with those of other molecular lines, e.g. those of HC$_3$N, HCC$^{13}$CN $J$ = 12--11 and OCS 9--8, which were detected simultaneously with the CN lines. These molecular transitions should be considered as optically thin, due to the low intensity with a main beam brightness peak temperature below 0.6$\,$K (see Figure \ref{IRAM15CNfigself} and \ref{ARO15CNfigself}). Comparisons show that 12 (including G010.62, the source with most notable LTE deviations) of our 35 sources display a shift in line center velocity between CN and other molecular lines, ranging from from 1.02 to 2.99$\,$$\rm km \, \rm s^{-1}$ (see Table \ref{tab:self-absorption}). Toward these 12 sources, self-absorption, i.e. an apparent self-absorption dip, can also be found (Figures \ref{IRAM15CNfigself} and \ref{ARO15CNfigself}). The CN line profile for seven sources among them is asymmetric and blue-shifted with respect to optically thin lines, and the other 5 sources have asymmetic line profiles with redshifted line velocity. Therefore we suggest that these 12 of our 35 sources should have suffered from self-absorption through multiple emitting regions along the line of sight, giving rise to complex line profiles \citep{2006MNRAS.367..553P,2013ApJS..206...22C}. To account for this effect, we have chosen a special cautionary procedure for the evaluation of $^{14}$N/$^{15}$N abundance ratios that is described in Section \ref{sec:fract}.

\renewcommand\tabcolsep{35.5pt} 
\begin{deluxetable}{lccc}
	
	\tablecaption{Comparsions of shifts in line velocity \label{tab:self-absorption}}
	\tablehead{	
		\colhead{Object}&\colhead{$V_{\rm LSR}^{\rm CN}$}&\colhead{$V_{\rm LSR}$}&\colhead{shifts in line velocity}\\
		\colhead{}&\colhead{($\rm km \, \rm s^{-1}$)}&\colhead{($\rm km \, \rm s^{-1}$)}&\colhead{($\rm km \, \rm s^{-1}$)}
	}
	\colnumbers
	\startdata
	G010.47$^{a}$&64.88&67.36&2.48\\
	G009.62$^{b}$&4.93&3.29&1.65\\
	G010.621$^{a}$&-3.66&-2.00&1.65\\
	G023.43$^{b}$&104.33&101.34&2.99\\
	G013.87$^{b}$&51.06&48.58&2.48\\
	G005.88$^{b}$&10.11&8.44&1.67\\
	G012.81$^{a}$&33.85&35.84&1.99\\
	G012.88$^{b}$&35.85&32.86&2.99\\
	G049.49$^{a}$&59.03&61.06&2.03\\
	G049.48$^{a}$&59.33&61.32&1.99\\
	G035.02$^{b}$&32.26&33.28&1.02\\
	G035.19$^{a}$&31.35&34.34&2.99\\
	\enddata
	\tablecomments{Column(1): source name; column(2): LSR velocity of the strongest HF component of C$^{14}$N; column(3): LSR velocity of optically thin lines, e.g. HCC$^{13}$CN $J$ = 12--11, HC$_{3}$N $J$ = 12--11 and OCS 9--8; column(4): absolute shifts in line velocity of C$^{14}$N compared to optically thin lines. $^{a}$ The line profile is asymmetric and blue-shifted. $^{b}$ The line profile is asymmetric and red-shifted.}
\end{deluxetable}

\subsubsection{Isotope selective photodissociation}\label{sec:isotope selective photodissociation}

Additionally, isotope-selective photodissociation has been suggested as the predominant mechanism of N-fractionation \citep{2014A&A...562A..61H,2018A&A...615A..75V,2018ApJ...857..105F}. This was supported by recent observation results toward the prototypical pre-stellar core L1544  \citep{2022A&A...664L...2S}. 
However, it is not supported by measurements toward the Orion molecular clouds. The measured $^{14}$N/$^{15}$N ratio in the Orion Bar (photodissociation region, 361 $\pm$ 141) is somehat higher but within the large error bars consistent with that in the star-forming region Orion-KL (234 $\pm$ 47, \cite{2012ApJ...744..194A}), and the Orion Nebula Cluster (G209.00, 322 $\pm$ 264, our results).

\subsubsection{Nitrogen fractionation}\label{sec:fract}

The potential of chemical nitrogen fractionation should be briefly discussed in order to determine correct isotopic abundance ratios from observed C$^{14}$N/C$^{15}$N line intensities. With respect to astrochemical models, nitrogen fractionation remains inconclusive \citep{2008MNRAS.385L..48R, 2010A&A...521L..26L, 2015A&A...576A..99R, 2018A&A...609A.129C, 2018MNRAS.474.3720W, 2019MNRAS.486.4805V, 2019MNRAS.484.2747L}. 

Isotope-exchange reactions were considered as the main mechanism to cause nitrogen fractionation, which leads to a $^{15}$N enhancement in CO-depleted dense gas at low temperatures of \textless10 K \citep{1981ApJ...247L.123A, 2000MNRAS.317..563T,2002ApJ...569L.133C,2008MNRAS.385L..48R,2015ApJ...808L..46F,2018MNRAS.478.3693C,2019MNRAS.484.2747L}. However, such reactions  at low temperatures may be hindered by an entrance barrier, as this barrier impedes the interaction of atoms, ions, and molecules, leading to an unchanged $^{14}$N/$^{15}$N ratio \citep{2015A&A...576A..99R,2018MNRAS.474.3720W}. This has been demonstrated observationally by \cite{2015ApJ...808L..46F} and \cite{2018A&A...609A.129C}. They conducted a survey of $^{14}$N/$^{15}$N isotope ratios with HCN, HNC and N$_{2}$H$^{+}$ towards high-mass star-forming regions in different evolutionary stages and found no trend of the $^{14}$N/$^{15}$N ratio with evolutionary stage.  Moreover, all our sources have known kinetic temperatures higher than 10 K (Table 3 and \citetalias{2021ApJS..257...39C}). This implies that the C$^{14}$N/C$^{15}$N ratios are not seriously affected by fractionation effects.  

\begin{figure*}[htbp]
	\centering
	{\includegraphics[width=17cm]{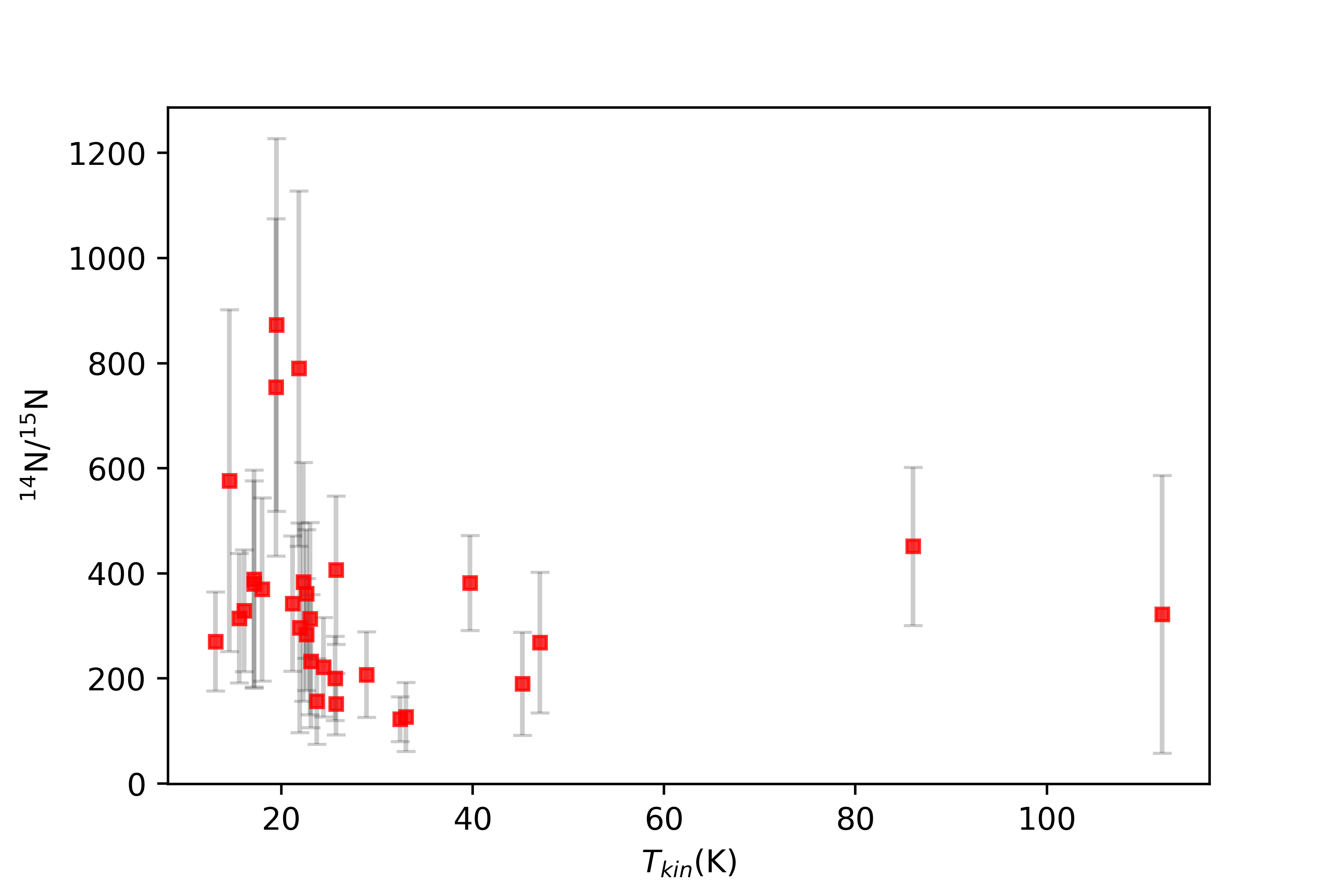}}
	\caption{$^{14}$N/$^{15}$N ratios plotted against gas kinetic temperature. No correlation is found between gas kinetic temperature and nitrogen isotope ratios.}
	\label{fig:fractionation}
\end{figure*}

For our sample, we have plotted our measured C$^{14}$N/C$^{15}$N ratios against the gas kinetic temperature $T_{kin}$ (see Figure \ref{fig:fractionation}), to investigate the N-fractionation effect. Kinetic temperatures were derived from the para-NH$_{3}$ (1, 1) and (2, 2) transitions \citep{2010MNRAS.402.2682H,2011ApJ...741..110D,2016ApJ...822...59S,2021ApJS..257...39C}. No obvious correlation can be found between the nitrogen isotope ratio and $T_{kin}$. 
This indicates that N-fractionation, as a temperature-dependent effect, should be negligible for our measured ratios. 

However, nitrogen fractionation may be scale dependent, possibly indicating a local effect, and observations with different beam sizes may yield different $^{14}$N/$^{15}$N values \citep{2019MNRAS.485.5543C}. Our single-dish telescopes with relatively large beam sizes are likely to include diffuse low density gas that may be influenced by the interstellar radiationn field. Thus higher resolution observations should be useful for probing the N-fractionation effect in more detail, including both the molecular cores and the outskirts.

In summary, 12 of our 35 sources suffer from LTE deviations (G010.62, Section \ref{sec:LTE deviations in CN}) and/or self-absorption (for details, see Section \ref{sec:self-absorptions}), which could question the validity of employing the strongest HF components to obtain the $^{14}$N/$^{15}$N ratio. To minimize these effects on our results, we used the weakest HF component ($J$ = 1/2 -- 1/2 $F$ = 1/2 -- 1/2) of C$^{14}$N rather than the strongest HF component ($J$ = 3/2 -- 1/2 $F$ = 5/2-3/2) to estimate the $^{14}$N/$^{15}$N ratio toward those 12 sources \citep{1984ApJ...283..668C,2002ApJ...578..211S}. It is noteworthy that we continue to utilize the main $F$ = 2 - 1 HF line of C$^{15}$N and the corresponding weighted theoretical $R_{\rm HF}$ value of 34.  The relatively low intensity of the weakest HF component should minimize self-absorption effects. And the weakest HF components of those 12 sources are individually strong enough (with $T_{\rm mb}$ exceeding 0.3 K and signal-to-noise ratios larger than 5) to be used for our analysis. The new results of $^{14}$N/$^{15}$N for these 12 sources are obtained and listed in Table \ref{tab:result} (column 12).

\begin{figure*} [htbp]
	\centering
	{\includegraphics[width=17cm]{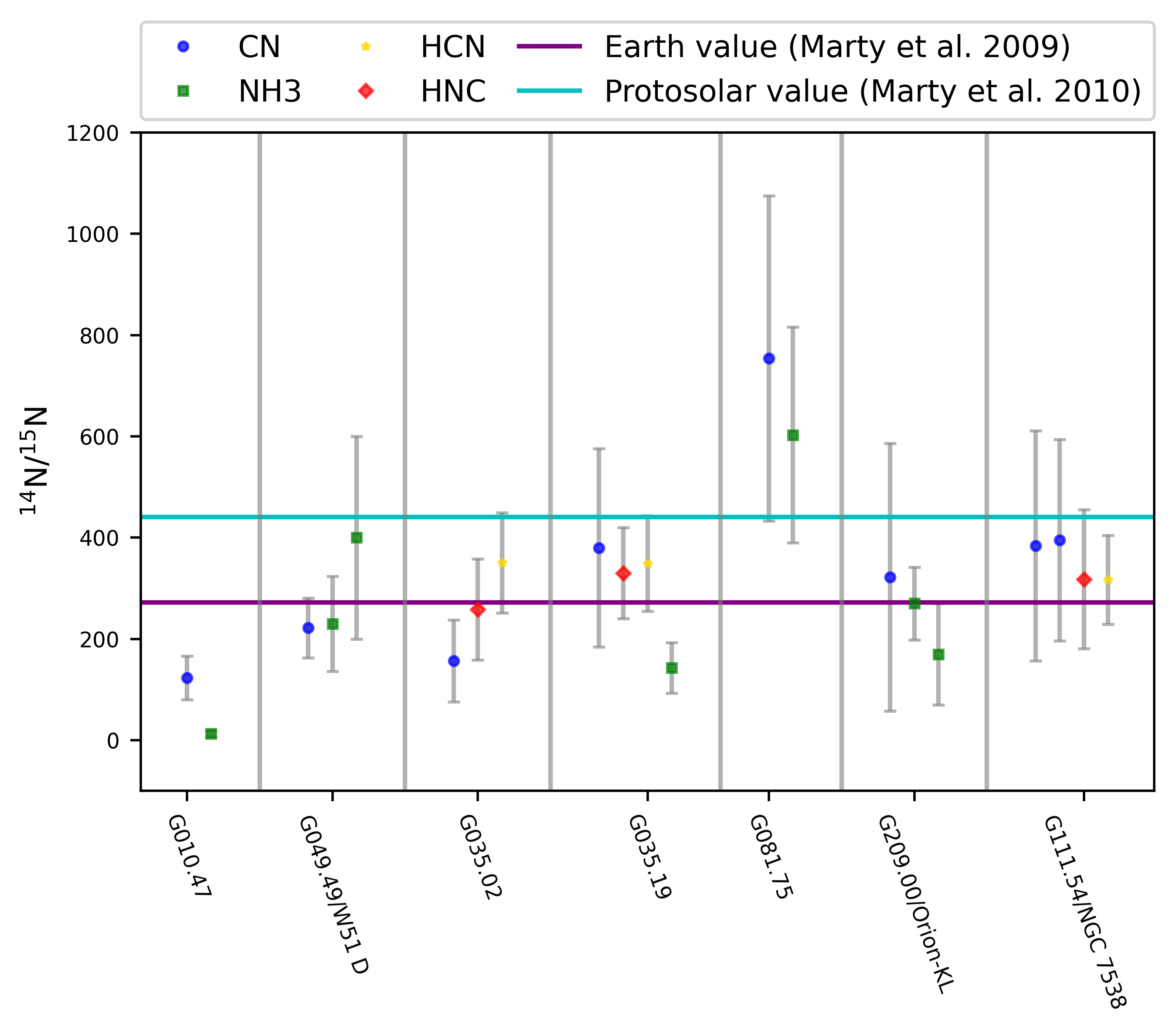}}
	\caption{Visual comparison of the $^{14}$N/$^{15}$N ratios obtained from CN (blue circles), NH$_{3}$ (green squares), HCN (yellow stars) and HNC (red diamonds) in different star forming regions. The purple line shows the nitrogen ratio of the terrestrial atmosphere (TA), derived from N$_{2}$, $\sim$ 272 \citep{2009GeCAS..73R.842M}. The cyan line presents the ratio of 441 $\pm$ 6 measured for the Proto-Solar Nebula (PSN) from the solar wind \citep{2010GeCoA..74..340M}.}
	\label{fig:Comparsion}
\end{figure*}

Previous $^{14}$N/$^{15}$N meaasurements toward sources belonging to our sample were investigated and comparisons were made between them and our $^{14}$N/$^{15}$N results. Out of our 35 sources, 7 sources were measured previously in a variety of molecular species (e.g., CN, NH$_{3}$, HNC, and HCN) by different telescopes (for details, see Table \ref{tab:Compare}). In case of six of these sources, the measured $^{14}$N/$^{15}$N ratios from different tracers, taking into account opacity effects, are consistent with each other within uncertainties (see Figure \ref{fig:Comparsion}). In the case of the seventh source, G010.47 (in the Galactic center region), the measured isotope ratio $^{14}$N/$^{15}$N is $\sim$123 from C$^{14}$N/C$^{15}$N (here, by IRAM 30 m observations) and the $^{14}$NH$_{3}$/$^{15}$NH$_{3}$ value $\sim$13 from Effelsberg 100m. Due to the peculiarity and complexity of the Galactic center region, it was excluded in the current Galactic chemical evolution models \cite[e.g.,][]{2017MNRAS.470..401R,2019MNRAS.490.2838R}. Both high and low $^{14}$N/$^{15}$N ratios measured  in this region may imply strong effects on nitrogen abundance due to both nucleosynthesis and chemical fractionation and more measurements and modeling work are needed  (see details in Section 4.1 from \citetalias{2021ApJS..257...39C}). However, we feel that a ratio of $\sim$13 is unrealistically low in view of all the other measured ratios and should be checked.

\renewcommand\tabcolsep{6.5pt} 
\begin{deluxetable}{llclllc}
	
	\tablecaption{Comparison of our results with $^{14}$N/$^{15}$N ratios from the literature\label{tab:Compare}}
	\tablehead{	\colhead{Object}	&	\colhead{Species}	&	\colhead{$\alpha$(2000)}	&	\colhead{$\delta$(2000)}	&	\colhead{$\frac{^{14}N}{^{15}N}$}	&	\colhead{Telescope/Beam size}	&	\colhead{References}		
	}
	\colnumbers
	\startdata
	G209.00/Orion-KL&CN&05:35:15&-05:23:14&322 $\pm$ 264 &ARO 12 m/64\arcsec& This paper \\
	&NH$_{3}$&05:35:14&-05:22:29&270 $\pm$ 72 &Effelsberg 100 m/40\arcsec& \citetalias{2021ApJS..257...39C} \\
	&NH$_{3}$&05:35:14&-05:22:46&170 $\pm$ 100 &Effelsberg 100 m/40\arcsec& \cite{1985AA...146..134H} \\
	G010.47&CN&18:08:38&-19:51:50&123 $\pm$ 43 &IRAM 30 m/23\arcsec& This paper \\
	&NH$_{3}$&18:08:38&-19:51:49&13 $\pm$ 6 &Effelsberg 100 m/40\arcsec& \citetalias{2021ApJS..257...39C} \\
	G035.02&CN&18:54:00&+02:01:19&157 $\pm$ 81 &IRAM 30 m/23\arcsec& This paper \\
	&HNC&18:54:00&+02:01:19&258 $\pm$ 100 &IRAM 30 m/29\arcsec& \cite{2018MNRAS.478.3693C} \\
	&HCN&18:54:00&+02:01:19&350 $\pm$ 99 &IRAM 30 m/29\arcsec& \cite{2018MNRAS.478.3693C} \\
	G035.19&CN&18:58:13&+01:40:35&380 $\pm$ 196 &IRAM 30 m/23\arcsec& This paper \\
	&HNC&18:58:13&+01:40:36&330 $\pm$ 90 &IRAM 30 m/29\arcsec& \cite{2018MNRAS.478.3693C} \\
	&HCN&18:58:13&+01:40:36&349 $\pm$ 94 &IRAM 30 m/29\arcsec& \cite{2018MNRAS.478.3693C} \\
	&NH$_{3}$&18:58:07&+01:37:11&143 $\pm$ 50 &Effelsberg 100 m/40\arcsec& \citetalias{2021ApJS..257...39C} \\
	G049.49/W51 D&CN&19:23:40&+14:31:05&222 $\pm$ 59 &IRAM 30 m/23\arcsec& This paper \\
	&NH$_{3}$&19:23:39&+14:31:07&230 $\pm$ 94 &Effelsberg 100 m/40\arcsec& \citetalias{2021ApJS..257...39C} \\
	&NH$_{3}$&19:23:39&+14:31:10&400 $\pm$ 200 &Effelsberg 100 m/40\arcsec& \cite{1987AA...173..352M} \\
	G081.75&CN&20:39:01&+42:24:59&754 $\pm$ 321 &IRAM 30 m/23\arcsec& This paper \\
	&NH$_{3}$&20:39:01&+42:24:58&603 $\pm$ 213 &Effelsberg 100 m/40\arcsec& \citetalias{2021ApJS..257...39C}\\
	G111.54/NGC 7538&CN&23:13:45&+61:28:10&384 $\pm$ 227 &IRAM 30 m/23\arcsec& This paper \\
	&CN&23:13:45&+61:28:10&395 $\pm$ 199 &ARO 12 m/64\arcsec& This paper \\
	&HNC&23:13:43&+61:28:10&318 $\pm$ 137&IRAM 30 m/29\arcsec& \cite{2018MNRAS.478.3693C} \\
	&HCN&23:13:43&+61:28:10&317 $\pm$ 88&IRAM 30 m/29\arcsec& \cite{2018MNRAS.478.3693C} \\
	\enddata
	\tablecomments{Column(1): source name; column(2): species; column(3): R.A. (J2000); column(4): decl. (J2000); column(5): resulting nitrogen isotope abundance ratio; column (6): applied telescope and corresponding beam size; column(7): references.}
\end{deluxetable}

\subsection{The Galactic interstellar $^{14}$N/$^{15}$N gradient}\label{sec:gradient}

\begin{figure}[htbp]
	\centering  
		\includegraphics[width=17cm]{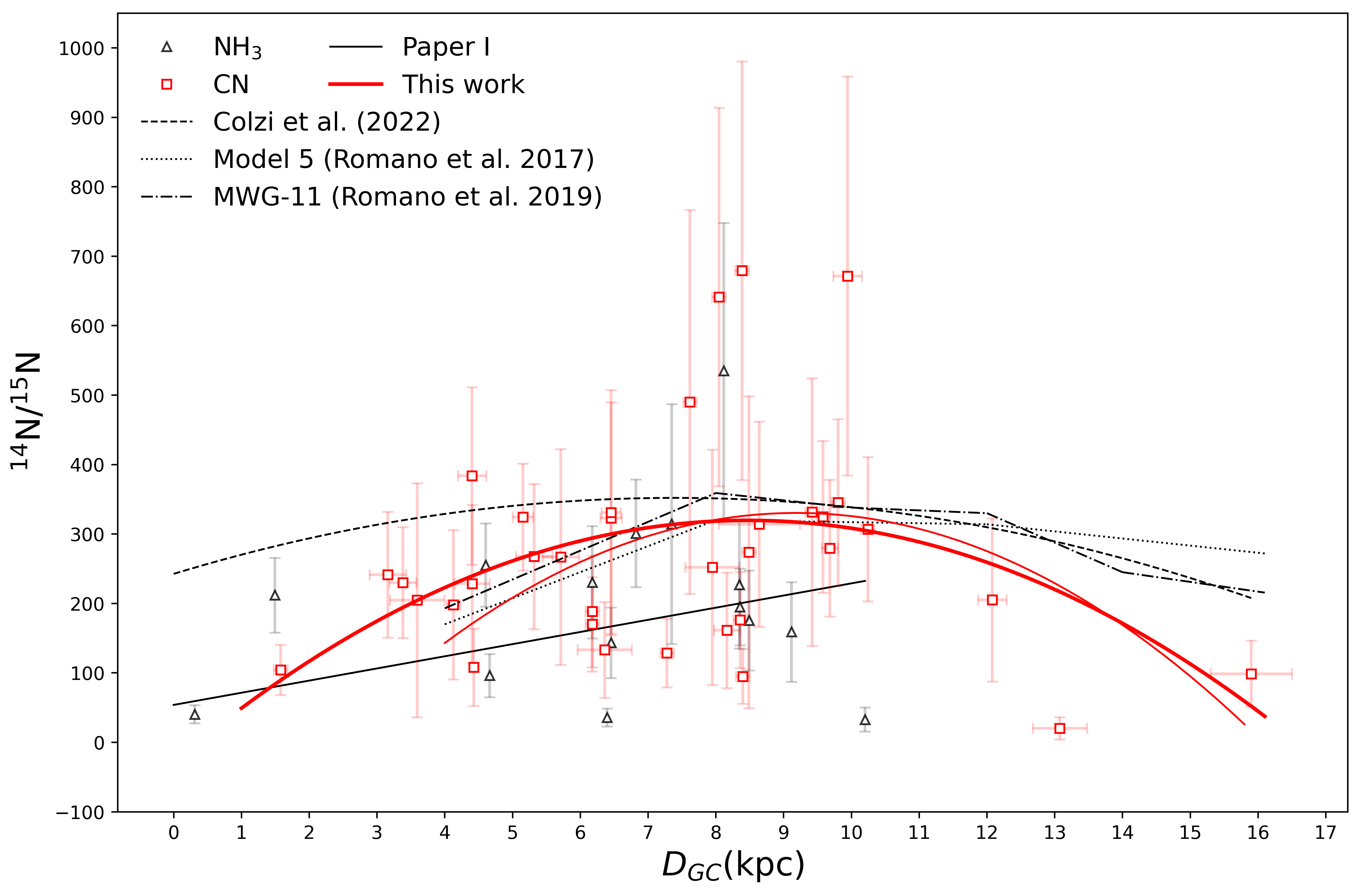}
	\caption{Our C$^{14}$N/C$^{15}$N isotope ratios (hollow red squares) are plotted as a functon of Galactocentric distance. We conducted an unweighted second order polynomial fit for our data (solid red line), yielding: $\frac{{\rm C^{14}N}}{{\rm C^{15}N}} =  (-4.85 \pm 1.89)\;{\rm kpc^{-2}} \times R_{\rm GC}^{2} + (82.11 \pm 31.93) \;{\rm kpc^{-1}} \times R_{\rm GC} - (28.12 \pm 126.62)$. A weaker red solid line indicates a second order polynomal fit to the 23 sources without indications for self-absorption and/or significant deviations from LTE. The hollow black triangles denote the ratios of $^{14}$NH$_{3}$/$^{15}$NH$_{3}$, whcih were fitted out to 10 kpc, showing as a black solid line. The black dashed line shows the modified results from H$^{13}$CN/HC$^{15}$N \citep{2022A&A...667A.151C}. The most recent numerically evauated $^{14}$N/$^{15}$N ratios across the disk of the Milky Way are shown as a black dotted curve \citep[Model 5 from][]{2017MNRAS.470..401R} and as a black dash-dotted curve \citep[MWG-11 from][]{2019MNRAS.490.2838R}.
		}
	\label{fig:CN_Tpeak-G043}
\end{figure}

Our 35 objects in this work are located at various Galactocentric distances ranging from 0 - 16 kpc, allowing us to study the Galactocentric trend of the $^{14}$N/$^{15}$N ratio. Figure \ref{fig:CN_Tpeak-G043} displays our measured C$^{14}$N/C$^{15}$N isotope ratios (hollow red squares) against $R_{\rm GC}$, showing an increasing trend up to $\sim$9 kpc and then a decreasing one toward the outer Galaxy, though there are only a few sources with $R_{\rm GC}$ \textgreater 9 kpc. An unweighted second order polynomial fit was performed for our data, to avoid biasing $^{14}$N/$^{15}$N results towards low values with small error bars. It yields $\frac{{\rm C^{14}N}}{{\rm C^{15}N}} =  (-4.85 \pm 1.89)\;{\rm kpc^{-2}} \times R_{\rm GC}^{2} + (82.11 \pm 31.93) \;{\rm kpc^{-1}} \times R_{\rm GC} - (28.12 \pm 126.62)$. Taking the 12 sources with potential self-absorption and/or LTE deviations out of the sample does not significantly change this trend.
	

For comparison, previous results on the trend of the $^{14}$N/$^{15}$N ratio through different tracers are also investigated, including NH$_{3}$ (\citetalias{2021ApJS..257...39C}) and HCN measurements from \cite{2022A&A...667A.151C}. The results based on NH$_{3}$ observations (hollow black triangles and black solid line) are also plotted in Figure \ref{fig:CN_Tpeak-G043}. Considering that the H$^{13}$CN/HC$^{15}$N results from \cite{2022A&A...667A.151C} may suffer from uncertainties, using relatively early measurements of $^{12}$C/$^{13}$C \citep{2005ApJ...634.1126M} and distance values, we therefore used the most recent $^{12}$C/$^{13}$C ratios \citep{2024MNRAS.527.8151S} and distance values from the Parallax-Based Distance Calculator \citep{2019ApJ...885..131R}, to modify those measurements. The modified results were fitted by a parabolic line, following \cite{2022A&A...667A.151C}, which is also presented in Figure \ref{fig:CN_Tpeak-G043}. We find that measurements of all tracers (NH$_{3}$, HCN and CN)  show the trend of increasing $^{14}$N/$^{15}$N against the galacocentric distance (to $\sim$9 kpc), though measured  $^{14}$N/$^{15}$N values show unsystematic differences between different tracers (Figure \ref{fig:CN_Tpeak-G043}). Toward the outer Galaxy, measured $^{14}$N/$^{15}$N values tend to decrease, though our results from CN show a steeper decreasing trend with respect to results from H$^{13}$CN/HC$^{15}$N \citep{2022A&A...667A.151C}. This causes the low values at large heliocentric distances, shown in Figure \ref{CN_Dsun}, i.e., $^{14}$N/$^{15}$N ratios from distant sources near the Galactic center or in the outer Galaxy are low. The difference on measured $^{14}$N/$^{15}$N results from different tracers might be mainly caused by the often large uncertainties related to measurements implying double isotope ratios (\citetalias{2021ApJS..257...39C}, \citealt{2024MNRAS.527.8151S}).

For comparison, we also plotted predicted $^{14}$N/$^{15}$N ratios from recent GCE models in Figure \ref{fig:CN_Tpeak-G043} (dotted and dash dotted curves, from Model 5 in \cite{2017MNRAS.470..401R} and Model-11 in \cite{2019MNRAS.490.2838R}, respectively). The nucleosynthesis prescriptions in Model 5 adopt the yields for low- and intermediate-mass stars \citep{2010MNRAS.403.1413K}, massive stars \citep{2013ARA&A..51..457N}, super-AGB stars \citep{2014MNRAS.437..195D,2014MNRAS.441..582D}, and novae. MWG-11 adopted new yields for low- and intermediate-mass stars and super-AGB stars \citep{2013MNRAS.431.3642V}, massive stars \citep{2018ApJS..237...13L}, and different initial rotational velocities for massive stars, while maintaining consistent yields for novae. Both GCE models align with our measured C$^{14}$N/C$^{15}$N trend, indicating a rising trend in the C$^{14}$N/C$^{15}$N ratio from the Galactic center region to approximately 9 kpc, followed by a consistent linear decrease with increasing $R_{\rm GC}$ in the outer Galaxy. This trend should be ascribed to the production of $^{15}$N during nova outbursts on long timescales ($\geq$1 Gyr), as reproduced in GCE models of the inner Galaxy \citep[e.g.,][]{2021A&A...653A..72R}, while the decreasing trend in the outer Galaxy may be explained by the strong metal dependence of the $^{14}$N yield \citep{2019MNRAS.490.2838R}, when adopting the stellar yields of  \cite{2013MNRAS.431.3642V,2014MNRAS.437.3274V,2018MNRAS.475.2282V,2020A&A...641A.103V,2021A&A...655A...6V}. Notably, we find that the theoretical gradient over 12 kpc from MWG-11 \citep{2019MNRAS.490.2838R}  is in closer agreement with our measured results, compared to Model 5. This may suggest that different stellar initial rotational velocities influence the nitrogen nucleosynthesis \citep{2018ApJS..237...13L}. To better constrain the $^{14}$N/$^{15}$N gradient, more data are required from the Galactic center region ($\leq$4 kpc) and the outer Galaxy (\textgreater10 kpc). More Galactic disk values with smaller uncertainties would of course also be desirable.

\section{Summary}\label{sec:summary}
We are performing systematic observational studies on different tracers to measure the interstellar $^{14}$N/$^{15}$N ratio across the Galaxy. Here observations of C$^{14}$N and C$^{15}$N  are presented toward our sample of 141 molecular clouds, covering a range from the Galactic center to the outer Galaxy ($\sim$16 kpc). Among 104 IRAM 30m targets, 28 sources were detected in both C$^{14}$N and C$^{15}$N,  while 10 out of 47 ARO targets were detected in both lines. Among those 10 sources observed by both telscopes, three sources (G109.87, G111.54, and G133.94) were detected in both lines. In total, 35 different targets were detected in both lines within our sample of 141 Galactic molecular clouds. Our results including:

(1) Physical parameters derived in those 35 sources with detections of both C$^{14}$N and C$^{15}$N lines include optical depth and excitation temperature. The strongest components of the C$^{14}$N and C$^{15}$N lines containing a variety of hyperfine features, are used to determine $^{14}$N/$^{15}$N abundance ratios toward this sample. No observational bias due to particularly bright sources or effects associated to different linear beam sizes can be found to influence significantly our $^{14}$N/$^{15}$N results. The three sources detected by both the IRAM 30 m and the ARO 12 m telescopes have consistent ratios, within uncertainties.  

(2) Possible contaminating effects influencing $^{14}$N/$^{15}$N were discussed, including nitrogen fractionation, local thermal equilibrium (LTE) deviations and self-absorpton effects. Nitrogen fractionation remains insignificant for our results, as indicated by a non-significant correlation between our abundance ratios and the kinetic temperature. Among our 35 sources, one source (G010.621) shows significant LTE deviations and 12 sources appear to be affected by self-absorption in the C$^{14}$N spectra which affects the reliability of using the strongest C$^{14}$N HF components to determine the $^{14}$N/$^{15}$N ratio. To minimize the latter effect, we used the weakest component of C$^{14}$N when estimating the C$^{14}$N/C$^{15}$N ratio for these sources. 

(3) Our measured $^{14}$N/$^{15}$N isotope ratios from C$^{14}$N and C$^{15}$N measurements show a similar trend as our measurements of NH$_{3}$ and $^{15}$NH$_{3}$ (\citetalias{2021ApJS..257...39C}), i.e., increasing ratios with distance (out to $\sim$9 kpc). Toward the outer Galaxy, measured $^{14}$N/$^{15}$N values tend to decrease. The unweighted second order polynomial fit for our data gives a gradiant of $\frac{^{14}{\rm CN}}{^{15}{\rm CN}} =  (-4.85 \pm 1.89)\;{\rm kpc^{-2}} \times R_{\rm GC}^{2} + (82.11 \pm 31.93) \;{\rm kpc^{-1}} \times R_{\rm GC} - (28.12 \pm 126.62)$. 
This trend is attributed to the production of $^{15}$N during nova outbursts on long timescales (1 Gyr) in the inner Galaxy. The decreasing trend in the outer Galaxy can also be explained by the strong metal dependence of the $^{14}$N yield \citep{2019MNRAS.490.2838R}. Thus plotting the nitrogen isotopes as a function of the distance to the Sun, we find higher values at small distances, while low values are characterizing the large distances of objects mainly belonging to the innermost or outermost parts of the Galaxy. Galactic chemical evolution models match well our measurements, providing a rising trend from the Galactic center region to approximately 9 kpc, followed by a decreasing one toward the outer Galaxy.

%

\begin{acknowledgments}
	We wish to thank the operators and staff at both the IRAM and ARO for their assistance during the observations. This work is supported by the Natural Science Foundation of China (No. 12373021, 12041302) and  the Guangzhou University Graduate Innovation Capability Cultivation Funding Program (2022GDJC-D13). H. Z. Y. would like to thank the China Scholarship Council (CSC) and the Ministry of Science and Higher Education of the Russian Federation (state contract FEUZ-2023-0019) for support.
	Y. T. Y. and Y. X. W. are members of the International Max Planck Research School (IMPRS) for Astronomy and Astrophysics at the Universities of Bonn and Cologne. 
\end{acknowledgments}

\bibliography{sample631}{}

\begin{thebibliography}{}
\expandafter\ifx\csname natexlab\endcsname\relax\def\natexlab#1{#1}\fi
\providecommand{\url}[1]{\href{#1}{#1}}
\providecommand{\dodoi}[1]{doi:~\href{http://doi.org/#1}{\nolinkurl{#1}}}
\providecommand{\doeprint}[1]{\href{http://ascl.net/#1}{\nolinkurl{http://ascl.net/#1}}}
\providecommand{\doarXiv}[1]{\href{https://arxiv.org/abs/#1}{\nolinkurl{https://arxiv.org/abs/#1}}}

\bibitem[{{Adams} \& {Smith}(1981)}]{1981ApJ...247L.123A}
{Adams}, N.~G., \& {Smith}, D. 1981, \apjl, 247, L123, \dodoi{10.1086/183604}

\bibitem[{{Adande} \& {Ziurys}(2012)}]{2012ApJ...744..194A}
{Adande}, G.~R., \& {Ziurys}, L.~M. 2012, \apj, 744, 194,
  \dodoi{10.1088/0004-637X/744/2/194}

\bibitem[{{Calahan} {et~al.}(2018){Calahan}, {Shirley}, {Svoboda}, {Ivanov},
  {Schmid}, {Pulley}, {Lautenbach}, {Zawadzki}, {Bullivant}, {Cook}, {Gray},
  {Henrici}, {Pascale}, {Bosse}, {Chance}, {Choi}, {Dunn}, {Jaime-Frias},
  {Kearsley}, {Kelledy}, {Lewin}, {Mahmood}, {McKinley}, {Mitchell}, \&
  {Robinson}}]{2018ApJ...862...63C}
{Calahan}, J.~K., {Shirley}, Y.~L., {Svoboda}, B.~E., {et~al.} 2018, \apj, 862,
  63, \dodoi{10.3847/1538-4357/aabfea}

\bibitem[{{Charnley} \& {Rodgers}(2002)}]{2002ApJ...569L.133C}
{Charnley}, S.~B., \& {Rodgers}, S.~D. 2002, \apjl, 569, L133,
  \dodoi{10.1086/340484}

\bibitem[{{Chen} {et~al.}(2021){Chen}, {Zhang}, {Henkel}, {Yan}, {Yu}, {Qiu},
  {Tang}, {Wang}, {Liu}, {Wang}, {Zheng}, {Zhao}, \&
  {Zou}}]{2021ApJS..257...39C}
{Chen}, J.~L., {Zhang}, J.~S., {Henkel}, C., {et~al.} 2021, \apjs, 257, 39,
  \dodoi{10.3847/1538-4365/ac205a}

\bibitem[{{Chen} {et~al.}(2013){Chen}, {Gan}, {Ellingsen}, {He}, {Shen}, \&
  {Titmarsh}}]{2013ApJS..206...22C}
{Chen}, X., {Gan}, C.-G., {Ellingsen}, S.~P., {et~al.} 2013, \apjs, 206, 22,
  \dodoi{10.1088/0067-0049/206/2/22}

\bibitem[{{Chen} {et~al.}(2010){Chen}, {Shen}, {Li}, {Xu}, \&
  {He}}]{2010ApJ...710..150C}
{Chen}, X., {Shen}, Z.-Q., {Li}, J.-J., {Xu}, Y., \& {He}, J.-H. 2010, \apj,
  710, 150, \dodoi{10.1088/0004-637X/710/1/150}

\bibitem[{{Chin} {et~al.}(1996){Chin}, {Henkel}, {Whiteoak}, {Langer}, \&
  {Churchwell}}]{1996A&A...305..960C}
{Chin}, Y.~N., {Henkel}, C., {Whiteoak}, J.~B., {Langer}, N., \& {Churchwell},
  E.~B. 1996, \aap, 305, 960.
\newblock \doarXiv{astro-ph/9505067}

\bibitem[{{Cohen} {et~al.}(1980){Cohen}, {Cong}, {Dame}, \&
  {Thaddeus}}]{1980ApJ...239L..53C}
{Cohen}, R.~S., {Cong}, H., {Dame}, T.~M., \& {Thaddeus}, P. 1980, \apjl, 239,
  L53, \dodoi{10.1086/183290}

\bibitem[{{Colzi} {et~al.}(2018{\natexlab{a}}){Colzi}, {Fontani}, {Caselli},
  {Ceccarelli}, {Hily-Blant}, \& {Bizzocchi}}]{2018A&A...609A.129C}
{Colzi}, L., {Fontani}, F., {Caselli}, P., {et~al.} 2018{\natexlab{a}}, \aap,
  609, A129, \dodoi{10.1051/0004-6361/201730576}

\bibitem[{{Colzi} {et~al.}(2019){Colzi}, {Fontani}, {Caselli}, {Leurini},
  {Bizzocchi}, \& {Quaia}}]{2019MNRAS.485.5543C}
---. 2019, \mnras, 485, 5543, \dodoi{10.1093/mnras/stz794}

\bibitem[{{Colzi} {et~al.}(2018{\natexlab{b}}){Colzi}, {Fontani}, {Rivilla},
  {S{\'a}nchez-Monge}, {Testi}, {Beltr{\'a}n}, \&
  {Caselli}}]{2018MNRAS.478.3693C}
{Colzi}, L., {Fontani}, F., {Rivilla}, V.~M., {et~al.} 2018{\natexlab{b}},
  \mnras, 478, 3693, \dodoi{10.1093/mnras/sty1027}

\bibitem[{{Colzi} {et~al.}(2022){Colzi}, {Romano}, {Fontani}, {Rivilla},
  {Bizzocchi}, {Beltran}, {Caselli}, {Elia}, \&
  {Magrini}}]{2022A&A...667A.151C}
{Colzi}, L., {Romano}, D., {Fontani}, F., {et~al.} 2022, \aap, 667, A151,
  \dodoi{10.1051/0004-6361/202244631}

\bibitem[{{Crutcher} {et~al.}(1984){Crutcher}, {Churchwell}, \&
  {Ziurys}}]{1984ApJ...283..668C}
{Crutcher}, R.~M., {Churchwell}, E., \& {Ziurys}, L.~M. 1984, \apj, 283, 668,
  \dodoi{10.1086/162352}

\bibitem[{{Dahmen} {et~al.}(1995){Dahmen}, {Wilson}, \&
  {Matteucci}}]{1995A&A...295..194D}
{Dahmen}, G., {Wilson}, T.~L., \& {Matteucci}, F. 1995, \aap, 295, 194

\bibitem[{{Doherty} {et~al.}(2014{\natexlab{a}}){Doherty}, {Gil-Pons}, {Lau},
  {Lattanzio}, \& {Siess}}]{2014MNRAS.437..195D}
{Doherty}, C.~L., {Gil-Pons}, P., {Lau}, H. H.~B., {Lattanzio}, J.~C., \&
  {Siess}, L. 2014{\natexlab{a}}, \mnras, 437, 195,
  \dodoi{10.1093/mnras/stt1877}

\bibitem[{{Doherty} {et~al.}(2014{\natexlab{b}}){Doherty}, {Gil-Pons}, {Lau},
  {Lattanzio}, {Siess}, \& {Campbell}}]{2014MNRAS.441..582D}
{Doherty}, C.~L., {Gil-Pons}, P., {Lau}, H. H.~B., {et~al.} 2014{\natexlab{b}},
  \mnras, 441, 582, \dodoi{10.1093/mnras/stu571}

\bibitem[{{Dunham} {et~al.}(2011){Dunham}, {Rosolowsky}, {Evans}, {Cyganowski},
  \& {Urquhart}}]{2011ApJ...741..110D}
{Dunham}, M.~K., {Rosolowsky}, E., {Evans}, Neal~J., I., {Cyganowski}, C., \&
  {Urquhart}, J.~S. 2011, \apj, 741, 110, \dodoi{10.1088/0004-637X/741/2/110}

\bibitem[{{Esteban} {et~al.}(2017){Esteban}, {Fang}, {Garc{\'\i}a-Rojas}, \&
  {Toribio San Cipriano}}]{2017MNRAS.471..987E}
{Esteban}, C., {Fang}, X., {Garc{\'\i}a-Rojas}, J., \& {Toribio San Cipriano},
  L. 2017, \mnras, 471, 987, \dodoi{10.1093/mnras/stx1624}

\bibitem[{{Esteban} \& {Garc{\'\i}a-Rojas}(2018)}]{2018MNRAS.478.2315E}
{Esteban}, C., \& {Garc{\'\i}a-Rojas}, J. 2018, \mnras, 478, 2315,
  \dodoi{10.1093/mnras/sty1168}

\bibitem[{{Fontani} {et~al.}(2015){Fontani}, {Caselli}, {Palau}, {Bizzocchi},
  \& {Ceccarelli}}]{2015ApJ...808L..46F}
{Fontani}, F., {Caselli}, P., {Palau}, A., {Bizzocchi}, L., \& {Ceccarelli}, C.
  2015, \apjl, 808, L46, \dodoi{10.1088/2041-8205/808/2/L46}

\bibitem[{{Furuya} \& {Aikawa}(2018)}]{2018ApJ...857..105F}
{Furuya}, K., \& {Aikawa}, Y. 2018, \apj, 857, 105,
  \dodoi{10.3847/1538-4357/aab768}

\bibitem[{{Gong} {et~al.}(2021){Gong}, {Belloche}, {Du}, {Menten}, {Henkel},
  {Li}, {Wyrowski}, \& {Mao}}]{2021A&A...646A.170G}
{Gong}, Y., {Belloche}, A., {Du}, F.~J., {et~al.} 2021, \aap, 646, A170,
  \dodoi{10.1051/0004-6361/202039465}

\bibitem[{{Gravity Collaboration} {et~al.}(2018){Gravity Collaboration},
  {Abuter}, {Amorim}, {Anugu}, {Baub{\"o}ck}, {Benisty}, {Berger}, {Blind},
  {Bonnet}, {Brandner}, {Buron}, {Collin}, {Chapron}, {Cl{\'e}net}, {Coud{\'e}
  Du Foresto}, {de Zeeuw}, {Deen}, {Delplancke-Str{\"o}bele}, {Dembet},
  {Dexter}, {Duvert}, {Eckart}, {Eisenhauer}, {Finger}, {F{\"o}rster
  Schreiber}, {F{\'e}dou}, {Garcia}, {Garcia Lopez}, {Gao}, {Gendron},
  {Genzel}, {Gillessen}, {Gordo}, {Habibi}, {Haubois}, {Haug}, {Hau{\ss}mann},
  {Henning}, {Hippler}, {Horrobin}, {Hubert}, {Hubin}, {Jimenez Rosales},
  {Jochum}, {Jocou}, {Kaufer}, {Kellner}, {Kendrew}, {Kervella}, {Kok},
  {Kulas}, {Lacour}, {Lapeyr{\`e}re}, {Lazareff}, {Le Bouquin}, {L{\'e}na},
  {Lippa}, {Lenzen}, {M{\'e}rand}, {M{\"u}ler}, {Neumann}, {Ott}, {Palanca},
  {Paumard}, {Pasquini}, {Perraut}, {Perrin}, {Pfuhl}, {Plewa}, {Rabien},
  {Ram{\'\i}rez}, {Ramos}, {Rau}, {Rodr{\'\i}guez-Coira}, {Rohloff}, {Rousset},
  {Sanchez-Bermudez}, {Scheithauer}, {Sch{\"o}ller}, {Schuler}, {Spyromilio},
  {Straub}, {Straubmeier}, {Sturm}, {Tacconi}, {Tristram}, {Vincent}, {von
  Fellenberg}, {Wank}, {Waisberg}, {Widmann}, {Wieprecht}, {Wiest},
  {Wiezorrek}, {Woillez}, {Yazici}, {Ziegler}, \& {Zins}}]{2018A&A...615L..15G}
{Gravity Collaboration}, {Abuter}, R., {Amorim}, A., {et~al.} 2018, \aap, 615,
  L15, \dodoi{10.1051/0004-6361/201833718}

\bibitem[{{Guilloteau} \& {Lucas}(2000)}]{2000ASPC..217..299G}
{Guilloteau}, S., \& {Lucas}, R. 2000, in Astronomical Society of the Pacific
  Conference Series, Vol. 217, Imaging at Radio through Submillimeter
  Wavelengths, ed. J.~G. {Mangum} \& S.~J.~E. {Radford}, 299

\bibitem[{{Heays} {et~al.}(2014){Heays}, {Visser}, {Gredel}, {Ubachs}, {Lewis},
  {Gibson}, \& {van Dishoeck}}]{2014A&A...562A..61H}
{Heays}, A.~N., {Visser}, R., {Gredel}, R., {et~al.} 2014, \aap, 562, A61,
  \dodoi{10.1051/0004-6361/201322832}

\bibitem[{{Henkel} {et~al.}(1998){Henkel}, {Chin}, {Mauersberger}, \&
  {Whiteoak}}]{1998A&A...329..443H}
{Henkel}, C., {Chin}, Y.~N., {Mauersberger}, R., \& {Whiteoak}, J.~B. 1998,
  \aap, 329, 443.
\newblock \doarXiv{astro-ph/9710254}

\bibitem[{{Henry} {et~al.}(2010){Henry}, {Kwitter}, {Jaskot}, {Balick},
  {Morrison}, \& {Milingo}}]{2010ApJ...724..748H}
{Henry}, R.~B.~C., {Kwitter}, K.~B., {Jaskot}, A.~E., {et~al.} 2010, \apj, 724,
  748, \dodoi{10.1088/0004-637X/724/1/748}

\bibitem[{{Hermsen} {et~al.}(1985){Hermsen}, {Wilson}, {Walmsley}, \&
  {Batrla}}]{1985AA...146..134H}
{Hermsen}, W., {Wilson}, T.~L., {Walmsley}, C.~M., \& {Batrla}, W. 1985, \aap,
  146, 134

\bibitem[{{Hill} {et~al.}(2010){Hill}, {Longmore}, {Pinte}, {Cunningham},
  {Burton}, \& {Minier}}]{2010MNRAS.402.2682H}
{Hill}, T., {Longmore}, S.~N., {Pinte}, C., {et~al.} 2010, \mnras, 402, 2682,
  \dodoi{10.1111/j.1365-2966.2009.16101.x}

\bibitem[{{Humire} {et~al.}(2020){Humire}, {Thiel}, {Henkel}, {Belloche},
  {Loison}, {Pillai}, {Riquelme}, {Wakelam}, {Langer},
  {Hern{\'a}ndez-G{\'o}mez}, {Mauersberger}, \& {Menten}}]{2020A&A...642A.222H}
{Humire}, P.~K., {Thiel}, V., {Henkel}, C., {et~al.} 2020, \aap, 642, A222,
  \dodoi{10.1051/0004-6361/202038216}

\bibitem[{{Izzard} {et~al.}(2004){Izzard}, {Tout}, {Karakas}, \&
  {Pols}}]{2004MNRAS.350..407I}
{Izzard}, R.~G., {Tout}, C.~A., {Karakas}, A.~I., \& {Pols}, O.~R. 2004,
  \mnras, 350, 407, \dodoi{10.1111/j.1365-2966.2004.07446.x}

\bibitem[{{Karakas}(2010)}]{2010MNRAS.403.1413K}
{Karakas}, A.~I. 2010, \mnras, 403, 1413,
  \dodoi{10.1111/j.1365-2966.2009.16198.x}

\bibitem[{{Kelly} {et~al.}(2015){Kelly}, {Viti}, {Bayet}, {Aladro}, \&
  {Yates}}]{2015A&A...578A..70K}
{Kelly}, G., {Viti}, S., {Bayet}, E., {Aladro}, R., \& {Yates}, J. 2015, \aap,
  578, A70, \dodoi{10.1051/0004-6361/201425502}

\bibitem[{{Larson}(1976)}]{1976MNRAS.176...31L}
{Larson}, R.~B. 1976, \mnras, 176, 31, \dodoi{10.1093/mnras/176.1.31}

\bibitem[{{Li} {et~al.}(2016){Li}, {Zhang}, {Liu}, {Lu}, {Wang}, \&
  {Wang}}]{2016RAA....16...47L}
{Li}, H.-K., {Zhang}, J.-S., {Liu}, Z.-W., {et~al.} 2016, Research in Astronomy
  and Astrophysics, 16, 47, \dodoi{10.1088/1674-4527/16/3/047}

\bibitem[{{Limongi} \& {Chieffi}(2018)}]{2018ApJS..237...13L}
{Limongi}, M., \& {Chieffi}, A. 2018, \apjs, 237, 13,
  \dodoi{10.3847/1538-4365/aacb24}

\bibitem[{{Lis} {et~al.}(2010){Lis}, {Phillips}, {Goldsmith}, {Neufeld},
  {Herbst}, {Comito}, {Schilke}, {M{\"u}ller}, {Bergin}, {Gerin}, {Bell},
  {Emprechtinger}, {Black}, {Blake}, {Boulanger}, {Caux}, {Ceccarelli},
  {Cernicharo}, {Coutens}, {Crockett}, {Daniel}, {Dartois}, {de Luca},
  {Dubernet}, {Encrenaz}, {Falgarone}, {Geballe}, {Godard}, {Giesen},
  {Goicoechea}, {Gry}, {Gupta}, {Hennebelle}, {Hily-Blant}, {Ko{\l}os},
  {Kre{\l}owski}, {Joblin}, {Johnstone}, {Ka{\'z}mierczak}, {Lord}, {Maret},
  {Martin}, {Mart{\'\i}n-Pintado}, {Melnick}, {Menten}, {Monje}, {Mookerjea},
  {Morris}, {Murphy}, {Ossenkopf}, {Pearson}, {P{\'e}rault}, {Persson},
  {Plume}, {Qin}, {Salez}, {Schlemmer}, {Schmidt}, {Sonnentrucker}, {Stutzki},
  {Teyssier}, {Trappe}, {van der Tak}, {Vastel}, {Wang}, {Yorke}, {Yu},
  {Zmuidzinas}, {Boogert}, {Erickson}, {Karpov}, {Kooi}, {Maiwald}, {Schieder},
  \& {Zaal}}]{2010A&A...521L..26L}
{Lis}, D.~C., {Phillips}, T.~G., {Goldsmith}, P.~F., {et~al.} 2010, \aap, 521,
  L26, \dodoi{10.1051/0004-6361/201015072}

\bibitem[{{Loison} {et~al.}(2019){Loison}, {Wakelam}, {Gratier}, \&
  {Hickson}}]{2019MNRAS.484.2747L}
{Loison}, J.-C., {Wakelam}, V., {Gratier}, P., \& {Hickson}, K.~M. 2019,
  \mnras, 484, 2747, \dodoi{10.1093/mnras/sty3293}

\bibitem[{{Loison} {et~al.}(2020){Loison}, {Wakelam}, {Gratier}, \&
  {Hickson}}]{2020MNRAS.498.4663L}
---. 2020, \mnras, 498, 4663, \dodoi{10.1093/mnras/staa2700}

\bibitem[{{Mangum} \& {Shirley}(2015)}]{2015PASP..127..266M}
{Mangum}, J.~G., \& {Shirley}, Y.~L. 2015, \pasp, 127, 266,
  \dodoi{10.1086/680323}

\bibitem[{{Marty} {et~al.}(2009){Marty}, {Zimmermann}, \&
  {Burnard}}]{2009GeCAS..73R.842M}
{Marty}, B., {Zimmermann}, L., \& {Burnard}, P.~G. 2009, Geochimica et
  Cosmochimica Acta Supplement, 73, A842

\bibitem[{{Marty} {et~al.}(2010){Marty}, {Zimmermann}, {Burnard}, {Wieler},
  {Heber}, {Burnett}, {Wiens}, \& {Bochsler}}]{2010GeCoA..74..340M}
{Marty}, B., {Zimmermann}, L., {Burnard}, P.~G., {et~al.} 2010, \gca, 74, 340,
  \dodoi{10.1016/j.gca.2009.09.007}

\bibitem[{{Mauersberger} {et~al.}(1996){Mauersberger}, {Henkel}, {Langer}, \&
  {Chin}}]{1996A&A...313L...1M}
{Mauersberger}, R., {Henkel}, C., {Langer}, N., \& {Chin}, Y.~N. 1996, \aap,
  313, L1, \dodoi{10.48550/arXiv.astro-ph/9607071}

\bibitem[{{Mauersberger} {et~al.}(1987){Mauersberger}, {Henkel}, \&
  {Wilson}}]{1987AA...173..352M}
{Mauersberger}, R., {Henkel}, C., \& {Wilson}, T.~L. 1987, \aap, 173, 352

\bibitem[{{Meynet} \& {Maeder}(2002)}]{2002A&A...390..561M}
{Meynet}, G., \& {Maeder}, A. 2002, \aap, 390, 561,
  \dodoi{10.1051/0004-6361:20020755}

\bibitem[{{Milam} {et~al.}(2005){Milam}, {Savage}, {Brewster}, {Ziurys}, \&
  {Wyckoff}}]{2005ApJ...634.1126M}
{Milam}, S.~N., {Savage}, C., {Brewster}, M.~A., {Ziurys}, L.~M., \& {Wyckoff},
  S. 2005, \apj, 634, 1126, \dodoi{10.1086/497123}

\bibitem[{{Nomoto} {et~al.}(2013){Nomoto}, {Kobayashi}, \&
  {Tominaga}}]{2013ARA&A..51..457N}
{Nomoto}, K., {Kobayashi}, C., \& {Tominaga}, N. 2013, \araa, 51, 457,
  \dodoi{10.1146/annurev-astro-082812-140956}

\bibitem[{{Penzias}(1980)}]{1980Sci...208..663P}
{Penzias}, A.~A. 1980, Science, 208, 663, \dodoi{10.1126/science.208.4445.663}

\bibitem[{{Pickett} {et~al.}(1998){Pickett}, {Poynter}, {Cohen}, {Delitsky},
  {Pearson}, \& {M{\"u}ller}}]{1998JQSRT..60..883P}
{Pickett}, H.~M., {Poynter}, R.~L., {Cohen}, E.~A., {et~al.} 1998, \jqsrt, 60,
  883, \dodoi{10.1016/S0022-4073(98)00091-0}

\bibitem[{{Purcell} {et~al.}(2006){Purcell}, {Balasubramanyam}, {Burton},
  {Walsh}, {Minier}, {Hunt-Cunningham}, {Kedziora-Chudczer}, {Longmore},
  {Hill}, {Bains}, {Barnes}, {Busfield}, {Calisse}, {Crighton}, {Curran},
  {Davis}, {Dempsey}, {Derragopian}, {Fulton}, {Hidas}, {Hoare}, {Lee}, {Ladd},
  {Lumsden}, {Moore}, {Murphy}, {Oudmaijer}, {Pracy}, {Rathborne}, {Robertson},
  {Schultz}, {Shobbrook}, {Sparks}, {Storey}, \&
  {Travouillion}}]{2006MNRAS.367..553P}
{Purcell}, C.~R., {Balasubramanyam}, R., {Burton}, M.~G., {et~al.} 2006,
  \mnras, 367, 553, \dodoi{10.1111/j.1365-2966.2005.09921.x}

\bibitem[{{Reid} {et~al.}(2014){Reid}, {Menten}, {Brunthaler}, {Zheng}, {Dame},
  {Xu}, {Wu}, {Zhang}, {Sanna}, {Sato}, {Hachisuka}, {Choi}, {Immer},
  {Moscadelli}, {Rygl}, \& {Bartkiewicz}}]{2014ApJ...783..130R}
{Reid}, M.~J., {Menten}, K.~M., {Brunthaler}, A., {et~al.} 2014, \apj, 783,
  130, \dodoi{10.1088/0004-637X/783/2/130}

\bibitem[{{Reid} {et~al.}(2019){Reid}, {Menten}, {Brunthaler}, {Zheng}, {Dame},
  {Xu}, {Li}, {Sakai}, {Wu}, {Immer}, {Zhang}, {Sanna}, {Moscadelli}, {Rygl},
  {Bartkiewicz}, {Hu}, {Quiroga-Nu{\~n}ez}, \& {van
  Langevelde}}]{2019ApJ...885..131R}
---. 2019, \apj, 885, 131, \dodoi{10.3847/1538-4357/ab4a11}

\bibitem[{{Rodgers} \& {Charnley}(2008)}]{2008MNRAS.385L..48R}
{Rodgers}, S.~D., \& {Charnley}, S.~B. 2008, \mnras, 385, L48,
  \dodoi{10.1111/j.1745-3933.2008.00431.x}

\bibitem[{{Roman-Duval} {et~al.}(2009){Roman-Duval}, {Jackson}, {Heyer},
  {Johnson}, {Rathborne}, {Shah}, \& {Simon}}]{2009ApJ...699.1153R}
{Roman-Duval}, J., {Jackson}, J.~M., {Heyer}, M., {et~al.} 2009, \apj, 699,
  1153, \dodoi{10.1088/0004-637X/699/2/1153}

\bibitem[{{Romano} {et~al.}(2019){Romano}, {Matteucci}, {Zhang}, {Ivison}, \&
  {Ventura}}]{2019MNRAS.490.2838R}
{Romano}, D., {Matteucci}, F., {Zhang}, Z.-Y., {Ivison}, R.~J., \& {Ventura},
  P. 2019, \mnras, 490, 2838, \dodoi{10.1093/mnras/stz2741}

\bibitem[{{Romano} {et~al.}(2017){Romano}, {Matteucci}, {Zhang},
  {Papadopoulos}, \& {Ivison}}]{2017MNRAS.470..401R}
{Romano}, D., {Matteucci}, F., {Zhang}, Z.~Y., {Papadopoulos}, P.~P., \&
  {Ivison}, R.~J. 2017, \mnras, 470, 401, \dodoi{10.1093/mnras/stx1197}

\bibitem[{{Romano} {et~al.}(2021){Romano}, {Magrini}, {Randich}, {Casali},
  {Bonifacio}, {Jeffries}, {Matteucci}, {Franciosini}, {Spina}, {Guiglion},
  {Chiappini}, {Mucciarelli}, {Ventura}, {Grisoni}, {Bellazzini}, {Bensby},
  {Bragaglia}, {de Laverny}, {Korn}, {Martell}, {Tautvai{\v{s}}ien{\.{e}}},
  {Carraro}, {Gonneau}, {Jofr{\'e}}, {Pancino}, {Smiljanic}, {Vallenari}, {Fu},
  {Guti{\'e}rrez Albarr{\'a}n}, {Jim{\'e}nez-Esteban}, {Montes}, {Damiani},
  {Bergemann}, \& {Worley}}]{2021A&A...653A..72R}
{Romano}, D., {Magrini}, L., {Randich}, S., {et~al.} 2021, \aap, 653, A72,
  \dodoi{10.1051/0004-6361/202141340}

\bibitem[{{Roueff} {et~al.}(2015){Roueff}, {Loison}, \&
  {Hickson}}]{2015A&A...576A..99R}
{Roueff}, E., {Loison}, J.~C., \& {Hickson}, K.~M. 2015, \aap, 576, A99,
  \dodoi{10.1051/0004-6361/201425113}

\bibitem[{{Savage} {et~al.}(2002){Savage}, {Apponi}, {Ziurys}, \&
  {Wyckoff}}]{2002ApJ...578..211S}
{Savage}, C., {Apponi}, A.~J., {Ziurys}, L.~M., \& {Wyckoff}, S. 2002, \apj,
  578, 211, \dodoi{10.1086/342468}

\bibitem[{{Spezzano} {et~al.}(2022){Spezzano}, {Caselli}, {Sipil{\"a}}, \&
  {Bizzocchi}}]{2022A&A...664L...2S}
{Spezzano}, S., {Caselli}, P., {Sipil{\"a}}, O., \& {Bizzocchi}, L. 2022, \aap,
  664, L2, \dodoi{10.1051/0004-6361/202244301}

\bibitem[{{Sun} {et~al.}(2024){Sun}, {Zhang}, {Wang}, {Lin}, {Papadopoulos},
  {Romano}, {Feng}, {Sun}, {Zhang}, \& {Matteucci}}]{2024MNRAS.527.8151S}
{Sun}, Y., {Zhang}, Z.-Y., {Wang}, J., {et~al.} 2024, \mnras, 527, 8151,
  \dodoi{10.1093/mnras/stad3643}

\bibitem[{{Svoboda} {et~al.}(2016){Svoboda}, {Shirley}, {Battersby},
  {Rosolowsky}, {Ginsburg}, {Ellsworth-Bowers}, {Pestalozzi}, {Dunham},
  {Evans}, {Bally}, \& {Glenn}}]{2016ApJ...822...59S}
{Svoboda}, B.~E., {Shirley}, Y.~L., {Battersby}, C., {et~al.} 2016, \apj, 822,
  59, \dodoi{10.3847/0004-637X/822/2/59}

\bibitem[{{Terzieva} \& {Herbst}(2000)}]{2000MNRAS.317..563T}
{Terzieva}, R., \& {Herbst}, E. 2000, \mnras, 317, 563,
  \dodoi{10.1046/j.1365-8711.2000.03618.x}

\bibitem[{{Ventura} {et~al.}(2020){Ventura}, {Dell'Agli}, {Lugaro}, {Romano},
  {Tailo}, \& {Yag{\"u}e}}]{2020A&A...641A.103V}
{Ventura}, P., {Dell'Agli}, F., {Lugaro}, M., {et~al.} 2020, \aap, 641, A103,
  \dodoi{10.1051/0004-6361/202038289}

\bibitem[{{Ventura} {et~al.}(2013){Ventura}, {Di Criscienzo}, {Carini}, \&
  {D'Antona}}]{2013MNRAS.431.3642V}
{Ventura}, P., {Di Criscienzo}, M., {Carini}, R., \& {D'Antona}, F. 2013,
  \mnras, 431, 3642, \dodoi{10.1093/mnras/stt444}

\bibitem[{{Ventura} {et~al.}(2014){Ventura}, {di Criscienzo}, {D'Antona},
  {Vesperini}, {Tailo}, {Dell'Agli}, \& {D'Ercole}}]{2014MNRAS.437.3274V}
{Ventura}, P., {di Criscienzo}, M., {D'Antona}, F., {et~al.} 2014, \mnras, 437,
  3274, \dodoi{10.1093/mnras/stt2126}

\bibitem[{{Ventura} {et~al.}(2018){Ventura}, {Karakas}, {Dell'Agli},
  {Garc{\'\i}a-Hern{\'a}ndez}, \& {Guzman-Ramirez}}]{2018MNRAS.475.2282V}
{Ventura}, P., {Karakas}, A., {Dell'Agli}, F., {Garc{\'\i}a-Hern{\'a}ndez},
  D.~A., \& {Guzman-Ramirez}, L. 2018, \mnras, 475, 2282,
  \dodoi{10.1093/mnras/stx3338}

\bibitem[{{Ventura} {et~al.}(2021){Ventura}, {Dell'Agli}, {Romano}, {Tosi},
  {Limongi}, {Chieffi}, {Castellani}, {Tailo}, {Lugaro}, {Marini}, \&
  {Yag{\"u}e Lopez}}]{2021A&A...655A...6V}
{Ventura}, P., {Dell'Agli}, F., {Romano}, D., {et~al.} 2021, \aap, 655, A6,
  \dodoi{10.1051/0004-6361/202141017}

\bibitem[{{Visser} {et~al.}(2018){Visser}, {Bruderer}, {Cazzoletti},
  {Facchini}, {Heays}, \& {van Dishoeck}}]{2018A&A...615A..75V}
{Visser}, R., {Bruderer}, S., {Cazzoletti}, P., {et~al.} 2018, \aap, 615, A75,
  \dodoi{10.1051/0004-6361/201731898}

\bibitem[{{Viti} {et~al.}(2019){Viti}, {Fontani}, {Jim{\'e}nez-Serra}, \&
  {Holdship}}]{2019MNRAS.486.4805V}
{Viti}, S., {Fontani}, F., {Jim{\'e}nez-Serra}, I., \& {Holdship}, J. 2019,
  \mnras, 486, 4805, \dodoi{10.1093/mnras/stz1172}

\bibitem[{{Weaver}(1970)}]{1970IAUS...38..126W}
{Weaver}, H. 1970, in The Spiral Structure of our Galaxy, ed. W.~{Becker} \&
  G.~I. {Kontopoulos}, Vol.~38, 126

\bibitem[{{Wilson} \& {Rood}(1994)}]{1994ARA&A..32..191W}
{Wilson}, T.~L., \& {Rood}, R. 1994, \araa, 32, 191,
  \dodoi{10.1146/annurev.aa.32.090194.001203}

\bibitem[{{Wirstr{\"o}m} \& {Charnley}(2018)}]{2018MNRAS.474.3720W}
{Wirstr{\"o}m}, E.~S., \& {Charnley}, S.~B. 2018, \mnras, 474, 3720,
  \dodoi{10.1093/mnras/stx3030}

\bibitem[{{Wouterloot} {et~al.}(1993){Wouterloot}, {Brand}, \&
  {Fiegle}}]{1993A&AS...98..589W}
{Wouterloot}, J.~G.~A., {Brand}, J., \& {Fiegle}, K. 1993, \aaps, 98, 589

\bibitem[{{Wouterloot} {et~al.}(2008){Wouterloot}, {Henkel}, {Brand}, \&
  {Davis}}]{2008A&A...487..237W}
{Wouterloot}, J.~G.~A., {Henkel}, C., {Brand}, J., \& {Davis}, G.~R. 2008,
  \aap, 487, 237, \dodoi{10.1051/0004-6361:20078156}

\bibitem[{{Xiang} {et~al.}(2017){Xiang}, {Liu}, {Shi}, {Yuan}, {Huang}, {Luo},
  {Zhang}, {Zhao}, {Zhang}, {Ren}, {Chen}, {Wang}, {Li}, {Huo}, {Zhang},
  {Wang}, {Zhang}, {Hou}, \& {Wang}}]{2017MNRAS.464.3657X}
{Xiang}, M.~S., {Liu}, X.~W., {Shi}, J.~R., {et~al.} 2017, \mnras, 464, 3657,
  \dodoi{10.1093/mnras/stw2523}

\bibitem[{{Yan} {et~al.}(2019){Yan}, {Zhang}, {Henkel}, {Mufakharov}, {Jia},
  {Tang}, {Wu}, {Li}, {Zeng}, {Wang}, {Li}, {Huang}, \&
  {Jian}}]{2019ApJ...877..154Y}
{Yan}, Y.~T., {Zhang}, J.~S., {Henkel}, C., {et~al.} 2019, \apj, 877, 154,
  \dodoi{10.3847/1538-4357/ab17d6}

\bibitem[{{Yan} {et~al.}(2023){Yan}, {Henkel}, {Kobayashi}, {Menten}, {Gong},
  {Zhang}, {Yu}, {Yang}, {Xie}, \& {Wang}}]{2023A&A...670A..98Y}
{Yan}, Y.~T., {Henkel}, C., {Kobayashi}, C., {et~al.} 2023, \aap, 670, A98,
  \dodoi{10.1051/0004-6361/202244584}

\bibitem[{{Yu} {et~al.}(2020){Yu}, {Zhang}, {Henkel}, {Yan}, {Liu}, {Tang},
  {Langer}, {Luan}, {Chen}, {Wang}, {Deng}, \& {Zou}}]{2020ApJ...899..145Y}
{Yu}, H.~Z., {Zhang}, J.~S., {Henkel}, C., {et~al.} 2020, \apj, 899, 145,
  \dodoi{10.3847/1538-4357/aba8f1}

\bibitem[{{Zhang} {et~al.}(2020{\natexlab{a}}){Zhang}, {Yan}, {Liu}, {Yu},
  {Chen}, \& {Henkel}}]{2020IAUGA..30..278Z}
{Zhang}, J.~S., {Yan}, Y.~T., {Liu}, W., {et~al.} 2020{\natexlab{a}}, in IAU
  General Assembly, 278--279, \dodoi{10.1017/S1743921319004381}

\bibitem[{{Zhang} {et~al.}(2015){Zhang}, {Sun}, {Riquelme}, {Henkel}, {Lu},
  {Zhang}, {Wang}, {Wang}, \& {Li}}]{2015ApJS..219...28Z}
{Zhang}, J.~S., {Sun}, L.~L., {Riquelme}, D., {et~al.} 2015, \apjs, 219, 28,
  \dodoi{10.1088/0067-0049/219/2/28}

\bibitem[{{Zhang} {et~al.}(2020{\natexlab{b}}){Zhang}, {Liu}, {Yan}, {Yu},
  {Liu}, {Zheng}, {Romano}, {Zhang}, {Wang}, {Chen}, {Wang}, {Zhang}, {Lu},
  {Chen}, {Zou}, {Yang}, {Wen}, \& {Lu}}]{2020ApJS..249....6Z}
{Zhang}, J.~S., {Liu}, W., {Yan}, Y.~T., {et~al.} 2020{\natexlab{b}}, \apjs,
  249, 6, \dodoi{10.3847/1538-4365/ab9112}

\bibitem[{{Zou} {et~al.}(2023){Zou}, {Zhang}, {Henkel}, {Romano}, {Liu},
  {Zheng}, {Yan}, {Chen}, {Wang}, \& {Zhao}}]{2023ApJS..268...56Z}
{Zou}, Y.~P., {Zhang}, J.~S., {Henkel}, C., {et~al.} 2023, \apjs, 268, 56,
  \dodoi{10.3847/1538-4365/acee6b}

\end{thebibliography}
\bibliographystyle{aasjournal}

\appendix


The appendix (Table \ref{tab:samplelist}) presents observation parameters of the whole sample.


\startlongtable
\renewcommand\tabcolsep{14.0pt} 
		\begin{deluxetable}{lccccc}
			
			\tablecaption{Our source list for C$^{14}$N and C$^{15}$N observations \label{tab:samplelist}}
			
			\tablehead{
				\colhead{Object}      & \colhead{Telescope} & \colhead{$\alpha$(2000) $\delta$(2000)}                                     & \colhead{time}  & \colhead{Molecule}                                    & \colhead{r.m.s.} \\
				\colhead{}
				&      \colhead{}     & \colhead{($^h \; ^m \; ^s$)   ($^{\circ} \; ^{\prime} \; ^{\prime\prime}$)} & \colhead{(min)} &                    \colhead{}                         & \colhead{(mK)}   \\ 
			}
			\decimalcolnumbers
			\startdata
			G121.29     & IRAM & 00:36:47.35 63:29:02.1  & 20  & \textbf{C$^{14}$N, $N$=1 -- 0} & 22.02  \\
			&      &                         & 35  & \textbf{C$^{15}$N, $N$=1 -- 0} & 14.52  \\
			G123.06     & IRAM & 00:52:24.70 56:33:50.5  & 16  & \textbf{C$^{14}$N, $N$=1 -- 0} & 26.94  \\
			&      &                         & 198 & \textbf{C$^{15}$N, $N$=1 -- 0} & 5.35   \\
			WB380       & IRAM & 01:07:50.70 65:21:21.4  & 38  & \textbf{C$^{14}$N, $N$=1 -- 0} & 23.40  \\
			&      &                         & 19  & \textbf{C$^{15}$N, $N$=1 -- 0} & 20.82  \\
			WB391       & IRAM & 01:19:26.49 65:45:44.8  & 76  & \textbf{C$^{14}$N, $N$=1 -- 0} & 76.41  \\
			&      &                         & 76  & C$^{15}$N, $N$=1 -- 0                           & 13.01  \\
			G133.94     & IRAM & 02:27:03.81 61:52:25.2  & 20  & \textbf{C$^{14}$N, $N$=1 -- 0} & 10.41  \\
			&      &                         & 20  & \textbf{C$^{15}$N, $N$=1 -- 0} & 13.35  \\
			& ARO  & 02:27:03.81 61:52:25.2  & 40  & \textbf{C$^{14}$N, $N$=1 -- 0} & 14.96  \\
			&      &                         & 45  & \textbf{C$^{15}$N, $N$=1 -- 0} & 8.37   \\
			W3OH        & IRAM & 02:27:04.18 61:52:25.4  & 18  & \textbf{C$^{14}$N, $N$=1 -- 0} & 323.00 \\
			&      &                         & 18  & C$^{15}$N, $N$=1 -- 0                           & 42.15  \\
			WB434       & IRAM & 02:41:29.28 60:43:26.9  & 76  & C$^{14}$N, $N$=1 -- 0                           & 13.17  \\
			&      &                         & 76  & C$^{15}$N, $N$=1 -- 0                           & 10.42  \\
			G135.27     & IRAM & 02:43:28.56 62:57:08.3  & 58  & \textbf{C$^{14}$N, $N$=1 -- 0} & 3.83   \\
			&      &                         & 58  & \textbf{C$^{15}$N, $N$=1 -- 0} & 7.39   \\
			& ARO  & 02:43:28.56 62:57:08.3  & 14  & C$^{14}$N, $N$=1 -- 0                           & 52.62  \\
			&      &                         &     & C$^{15}$N, $N$=1 -- 0                           &        \\
			WB437       & IRAM & 02:43:58.63 62:56:08.6  & 38  & \textbf{C$^{14}$N, $N$=1 -- 0} & 71.69  \\
			&      &                         & 38  & C$^{15}$N, $N$=1 -- 0                           & 17.54  \\
			HALO2       & IRAM & 02:44:58.56 60:57:08.3  & 46  & \textbf{C$^{14}$N, $N$=1 -- 0} & 30.73  \\
			&      &                         & 46  & C$^{15}$N, $N$=1 -- 0                           & 88.00  \\
			WB440       & IRAM & 02:46:07.58 62:46:31.4  & 38  & \textbf{C$^{14}$N, $N$=1 -- 0} & 29.84  \\
			&      &                         & 38  & C$^{15}$N, $N$=1 -- 0                           & 19.09  \\
			WB477       & IRAM & 03:17:28.75 60:32:25.9  & 114 & \textbf{C$^{14}$N, $N$=1 -- 0} & 26.47  \\
			&      &                         & 114 & C$^{15}$N, $N$=1 -- 0                           & 8.16   \\
			WB501       & IRAM & 03:52:27.53 57:48:33.1  & 38  & \textbf{C$^{14}$N, $N$=1 -- 0} & 64.56  \\
			&      &                         & 38  & C$^{15}$N, $N$=1 -- 0                           & 16.21  \\
			WB515       & IRAM & 04:01:54.70 54:25:44.0  & 38  & C$^{14}$N, $N$=1 -- 0                           & 81.81  \\
			&      &                         & 38  & C$^{15}$N, $N$=1 -- 0                           & 17.24  \\
			WB529       & IRAM & 04:06:25.49 53:21:49.2  & 76  & \textbf{C$^{14}$N, $N$=1 -- 0} & 50.11  \\
			&      &                         & 76  & C$^{15}$N, $N$=1 -- 0                           & 9.27   \\
			G160.14     & IRAM & 05:01:40.24 47:07:19.0  & 174 & \textbf{C$^{14}$N, $N$=1 -- 0} & 10.66  \\
			&      &                         & 87  & C$^{15}$N, $N$=1 -- 0                           & 8.46   \\
			& ARO  & 05:01:40.24 47:07:19.0  & 40  & \textbf{C$^{14}$N, $N$=1 -- 0} & 13.48  \\
			&      &                         & ... & C$^{15}$N, $N$=1 -- 0                           &        \\
			G168.06     & IRAM & 05:17:13.74 39:22:19.9  & 174 & \textbf{C$^{14}$N, $N$=1 -- 0} & 11.07  \\
			&      &                         & 87  & C$^{15}$N, $N$=1 -- 0                           & 7.41   \\
			& ARO  & 05:17:13.74 39:22:19.9  & 24  & \textbf{C$^{14}$N, $N$=1 -- 0} & 16.62  \\
			&      &                         & ... & C$^{15}$N, $N$=1 -- 0                           &        \\
			G174.20     & IRAM & 05:30:48.01 33:47:54.5  & 60  & \textbf{C$^{14}$N, $N$=1 -- 0} & 14.36  \\
			&      &                         & 90  & \textbf{C$^{15}$N, $N$=1 -- 0} & 8.27   \\
			G209.00     & IRAM & 05:35:15.80 -05:23:14.1 & 4   & \textbf{C$^{14}$N, $N$=1 -- 0} & 125.00 \\
			&      &                         & 210 & \textbf{C$^{15}$N, $N$=1 -- 0} & 17.30  \\
			G209.19     & IRAM & 05:35:28.80 -05:23:43.1 & 114 & \textbf{C$^{14}$N, $N$=1 -- 0} & 12.25  \\
			&      &                         & 58  & C$^{15}$N, $N$=1 -- 0                           & 9.30   \\
			G173.48     & IRAM & 05:39:13.06 35:45:51.2  & 36  & \textbf{C$^{14}$N, $N$=1 -- 0} & 15.13  \\
			&      &                         & 120 & \textbf{C$^{15}$N, $N$=1 -- 0} & 6.47   \\
			G182.67     & IRAM & 05:39:28.42 24:56:31.9  & 48  & \textbf{C$^{14}$N, $N$=1 -- 0} & 8.46   \\
			&      &                         & 24  & C$^{15}$N, $N$=1 -- 0                           & 11.07  \\
			& ARO  & 05:39:28.42 24:56:31.9  & 12  & \textbf{C$^{14}$N, $N$=1 -- 0} & 23.77  \\
			&      &                         & ... & C$^{15}$N, $N$=1 -- 0                           &        \\
			G192.60     & IRAM & 06:12:54.01 17:59:23.2  & 12  & \textbf{C$^{14}$N, $N$=1 -- 0} & 43.03  \\
			&      &                         & 50  & \textbf{C$^{15}$N, $N$=1 -- 0} & 8.75   \\
			G196.45     & IRAM & 06:14:37.64 13:49:36.6  & 174 & \textbf{C$^{14}$N, $N$=1 -- 0} & 11.18  \\
			&      &                         & 87  & C$^{15}$N, $N$=1 -- 0                           & 7.83   \\
			& ARO  & 06:14:37.64 13:49:36.6  & 40  & \textbf{C$^{14}$N, $N$=1 -- 0} & 17.08  \\
			&      &                         & ... & C$^{15}$N, $N$=1 -- 0                           &        \\
			G211.59     & IRAM & 06:52:45.32 01:40:23.0  & 40  & \textbf{C$^{14}$N, $N$=1 -- 0} & 19.28  \\
			&      &                         & 45  & \textbf{C$^{15}$N, $N$=1 -- 0} & 15.69  \\
			SGRC        & IRAM & 17:44:46.80 -29:28:24.5 & 58  & \textbf{C$^{14}$N, $N$=1 -- 0} & 77.86  \\
			&      &                         & 29  & C$^{15}$N, $N$=1 -- 0                           & 14.54  \\
			SGRA        & IRAM & 17:45:40.54 -29:00:16.2 & 48  & \textbf{C$^{14}$N, $N$=1 -- 0} & 268.00 \\
			&      &                         & 24  & C$^{15}$N, $N$=1 -- 0                           & 19.77  \\
			G000.37     & IRAM & 17:46:21.40 -28:35:39.8 & 114 & \textbf{C$^{14}$N, $N$=1 -- 0} & 21.42  \\
			&      &                         & 58  & C$^{15}$N, $N$=1 -- 0                           & 12.71  \\
			G000.67     & IRAM & 17:47:20.00 -28:22:40.0 & 60  & \textbf{C$^{14}$N, $N$=1 -- 0} & 177.00 \\
			&      &                         & 30  & C$^{15}$N, $N$=1 -- 0                           & 28.54  \\
			1.3COMPLEX  & IRAM & 17:48:00.64 -27:56:08.1 & 38  & \textbf{C$^{14}$N, $N$=1 -- 0} & 65.90  \\
			&      &                         & 38  & C$^{15}$N, $N$=1 -- 0                           & 31.53  \\
			G001.28     & IRAM & 17:48:21.90 -27:48:19.0 & 70  & \textbf{C$^{14}$N, $N$=1 -- 0} & 17.25  \\
			&      &                         & 70  & C$^{15}$N, $N$=1 -- 0                           & 16.77  \\
			SGRD        & IRAM & 17:48:42.24 -28:01:27.7 & 114 & \textbf{C$^{14}$N, $N$=1 -- 0} & 76.45  \\
			&      &                         & 58  & C$^{15}$N, $N$=1 -- 0                           & 12.28  \\
			G001.14     & IRAM & 17:48:48.54 -28:01:11.3 & 18  & \textbf{C$^{14}$N, $N$=1 -- 0} & 61.39  \\
			&      &                         & 18  & \textbf{C$^{15}$N, $N$=1 -- 0} & 67.85  \\
			G002.70     & IRAM & 17:51:45.97 -26:35:57.0 & 70  & \textbf{C$^{14}$N, $N$=1 -- 0} & 14.43  \\
			&      &                         & 70  & C$^{15}$N, $N$=1 -- 0                           & 11.22  \\
			G011.49     & IRAM & 17:51:45.97 -26:35:57.0 & 70  & C$^{14}$N, $N$=1 -- 0                           & 39.99  \\
			&      &                         & 70  & C$^{15}$N, $N$=1 -- 0                           & 26.30  \\
			M53-03      & IRAM & 17:59:17.80 -24:24:38.0 & 23  & \textbf{C$^{14}$N, $N$=1 -- 0} & 95.95  \\
			&      &                         & 23  & C$^{15}$N, $N$=1 -- 0                           & 18.06  \\
			M5.3-0.3    & IRAM & 17:59:28.80 -24:24:38.0 & 74  & \textbf{C$^{14}$N, $N$=1 -- 0} & 97.01  \\
			&      &                         & 74  & C$^{15}$N, $N$=1 -- 0                           & 15.75  \\
			G005.88     & IRAM & 18:00:30.28 -24:04:04.5 & 89  & \textbf{C$^{14}$N, $N$=1 -- 0} & 37.84  \\
			&      &                         & 89  & \textbf{C$^{15}$N, $N$=1 -- 0} & 13.34  \\
			G007.47     & IRAM & 18:02:13.18 -22:27:58.9 & 87  & \textbf{C$^{14}$N, $N$=1 -- 0} & 32.33  \\
			&      &                         & 87  & C$^{15}$N, $N$=1 -- 0                           & 23.99  \\
			G009.61     & IRAM & 18:06:14.13 -20:31:44.3 & 18  & \textbf{C$^{14}$N, $N$=1 -- 0} & 39.69  \\
			&      &                         & 18  & C$^{15}$N, $N$=1 -- 0                           & 54.41  \\
			G009.62     & IRAM & 18:06:14.66 -20:31:31.7 & 58  & \textbf{C$^{14}$N, $N$=1 -- 0} & 35.46  \\
			&      &                         & 30  & \textbf{C$^{15}$N, $N$=1 -- 0} & 15.12  \\
			G010.47     & IRAM & 18:08:38.22 -19:51:50.2 & 58  & \textbf{C$^{14}$N, $N$=1 -- 0} & 45.60  \\
			&      &                         & 58  & \textbf{C$^{15}$N, $N$=1 -- 0} & 26.29  \\
			G010.62     & IRAM & 18:10:17.98 -19:54:04.6 & 58  & \textbf{C$^{14}$N, $N$=1 -- 0} & 29.54  \\
			&      &                         & 29  & C$^{15}$N, $N$=1 -- 0                           & 19.16  \\
			G010.621    & IRAM & 18:10:28.56 -19:55:48.7 & 89  & \textbf{C$^{14}$N, $N$=1 -- 0} & 23.43  \\
			&      &                         & 89  & \textbf{C$^{15}$N, $N$=1 -- 0} & 23.26  \\
			G012.88     & IRAM & 18:11:51.44 -17:31:29.4 & 70  & \textbf{C$^{14}$N, $N$=1 -- 0} & 18.37  \\
			&      &                         & 70  & \textbf{C$^{15}$N, $N$=1 -- 0} & 18.89  \\
			G012.02     & IRAM & 18:12:01.84 -18:31:55.8 & 60  & \textbf{C$^{14}$N, $N$=1 -- 0} & 22.17  \\
			&      &                         & 30  & C$^{15}$N, $N$=1 -- 0                           & 15.10  \\
			G012.81     & IRAM & 18:14:14.06 -17:55:11.3 & 60  & \textbf{C$^{14}$N, $N$=1 -- 0} & 41.54  \\
			&      &                         & 60  & \textbf{C$^{15}$N, $N$=1 -- 0} & 25.05  \\
			W33         & IRAM & 18:14:14.39 -17:55:49.9 & 18  & \textbf{C$^{14}$N, $N$=1 -- 0} & 651.00 \\
			&      &                         & 18  & C$^{15}$N, $N$=1 -- 0                           & 28.17  \\
			G013.87     & IRAM & 18:14:35.83 -16:45:35.8 & 120 & \textbf{C$^{14}$N, $N$=1 -- 0} & 17.13  \\
			&      &                         & 120 & \textbf{C$^{15}$N, $N$=1 -- 0} & 7.60   \\
			G014.63     & IRAM & 18:19:15.54 -16:29:45.7 & 70  & \textbf{C$^{14}$N, $N$=1 -- 0} & 20.21  \\
			&      &                         & 35  & C$^{15}$N, $N$=1 -- 0                           & 26.83  \\
			G016.58     & IRAM & 18:21:09.08 -14:31:48.5 & 70  & \textbf{C$^{14}$N, $N$=1 -- 0} & 16.28  \\
			&      &                         & 35  & C$^{15}$N, $N$=1 -- 0                           & 16.10  \\
			G22.89+0.39 & IRAM & 18:31:34.33 -08:44:50.7 & 18  & C$^{14}$N, $N$=1 -- 0                           & 34.30  \\
			&      &                         & 18  & C$^{15}$N, $N$=1 -- 0                           & 19.00  \\
			G23.54      & IRAM & 18:33:19.47 -08:14:24.7 & 18  & \textbf{C$^{14}$N, $N$=1 -- 0} & 51.47  \\
			&      &                         & 18  & C$^{15}$N, $N$=1 -- 0                           & 19.82  \\
			G023.25     & IRAM & 18:34:31.24 -08:42:47.3 & 138 & \textbf{C$^{14}$N, $N$=1 -- 0} & 21.95  \\
			&      &                         & 138 & C$^{15}$N, $N$=1 -- 0                           & 13.26  \\
			G023.43     & IRAM & 18:34:39.18 -08:31:25.4 & 35  & \textbf{C$^{14}$N, $N$=1 -- 0} & 24.94  \\
			&      &                         & 35  & \textbf{C$^{15}$N, $N$=1 -- 0} & 14.10  \\
			G24.39+0.04 & IRAM & 18:35:37.39 -07:34:40.4 & 18  & \textbf{C$^{14}$N, $N$=1 -- 0} & 44.07  \\
			&      &                         & 18  & C$^{15}$N, $N$=1 -- 0                           & 19.62  \\
			G027.36     & IRAM & 18:41:51.05 -05:01:43.4 & 58  & \textbf{C$^{14}$N, $N$=1 -- 0} & 38.59  \\
			&      &                         & 29  & C$^{15}$N, $N$=1 -- 0                           & 23.68  \\
			G029.86     & IRAM & 18:45:59.57 -02:45:06.5 & 60  & \textbf{C$^{14}$N, $N$=1 -- 0} & 14.94  \\
			&      &                         & 60  & \textbf{C$^{15}$N, $N$=1 -- 0} & 16.24  \\
			G029.95     & IRAM & 18:46:03.74 -02:39:22.3 & 120 & \textbf{C$^{14}$N, $N$=1 -- 0} & 12.85  \\
			&      &                         & 120 & \textbf{C$^{15}$N, $N$=1 -- 0} & 14.52  \\
			G031        & IRAM & 18:48:12.39 -01:26:30.7 & 38  & \textbf{C$^{14}$N, $N$=1 -- 0} & 144.00 \\
			&      &                         & 38  & C$^{15}$N, $N$=1 -- 0                           & 18.40  \\
			G032.04     & IRAM & 18:49:36.57 -00:45:45.5 & 70  & \textbf{C$^{14}$N, $N$=1 -- 0} & 19.27  \\
			&      &                         & 70  & C$^{15}$N, $N$=1 -- 0                           & 21.24  \\
			G34.3+0.2   & IRAM & 18:53:18.40 01:14:56.0  & 18  & \textbf{C$^{14}$N, $N$=1 -- 0} & 793.00 \\
			&      &                         & 18  & C$^{15}$N, $N$=1 -- 0                           & 29.17  \\
			G033        & IRAM & 18:53:32.56 00:31:39.1  & 9   & \textbf{C$^{14}$N, $N$=1 -- 0} & 45.40  \\
			&      &                         & 9   & C$^{15}$N, $N$=1 -- 0                           & 27.43  \\
			G035.02     & IRAM & 18:54:00.65 02:01:19.2  & 70  & \textbf{C$^{14}$N, $N$=1 -- 0} & 33.41  \\
			&      &                         & 70  & \textbf{C$^{15}$N, $N$=1 -- 0} & 38.17  \\
			G037.42     & IRAM & 18:54:14.34 04:41:39.6  & 70  & \textbf{C$^{14}$N, $N$=1 -- 0} & 11.07  \\
			&      &                         & 35  & C$^{15}$N, $N$=1 -- 0                           & 15.58  \\
			G034.04-0.3 & IRAM & 18:54:33.70 00:50:41.2  & 18  & \textbf{C$^{14}$N, $N$=1 -- 0} & 28.61  \\
			&      &                         & 18  & C$^{15}$N, $N$=1 -- 0                           & 19.26  \\
			G035.14     & IRAM & 18:58:12.62 01:40:50.5  & 18  & \textbf{C$^{14}$N, $N$=1 -- 0} & 32.49  \\
			&      &                         & 18  & \textbf{C$^{15}$N, $N$=1 -- 0} & 22.59  \\
			G035.19     & IRAM & 18:58:13.05 01:40:35.6  & 70  & \textbf{C$^{14}$N, $N$=1 -- 0} & 17.61  \\
			&      &                         & 35  & \textbf{C$^{15}$N, $N$=1 -- 0} & 14.16  \\
			G040.62     & IRAM & 19:06:01.62 06:46:36.1  & 70  & \textbf{C$^{14}$N, $N$=1 -- 0} & 17.25  \\
			&      &                         & 70  & C$^{15}$N, $N$=1 -- 0                           & 19.90  \\
			G043.16     & IRAM & 19:10:13.41 09:06:12.8  & 75  & \textbf{C$^{14}$N, $N$=1 -- 0} & 72.31  \\
			&      &                         & 38  & C$^{15}$N, $N$=1 -- 0                           & 17.07  \\
			G043        & IRAM & 19:14:26.39 09:22:36.5  & 18  & \textbf{C$^{14}$N, $N$=1 -- 0} & 27.29  \\
			&      &                         & 18  & C$^{15}$N, $N$=1 -- 0                           & 14.29  \\
			G048.60     & IRAM & 19:20:31.17 13:55:25.2  & 70  & \textbf{C$^{14}$N, $N$=1 -- 0} & 14.03  \\
			&      &                         & 35  & C$^{15}$N, $N$=1 -- 0                           & 15.65  \\
			G052.10     & IRAM & 19:23:37.32 17:29:10.4  & 70  & \textbf{C$^{14}$N, $N$=1 -- 0} & 10.16  \\
			&      &                         & 70  & C$^{15}$N, $N$=1 -- 0                           & 10.72  \\
			G049.48     & IRAM & 19:23:39.82 14:31:04.9  & 56  & \textbf{C$^{14}$N, $N$=1 -- 0} & 24.79  \\
			&      &                         & 56  & \textbf{C$^{15}$N, $N$=1 -- 0} & 28.20  \\
			G049.49     & IRAM & 19:23:40.50 14:31:05.5  & 18  & \textbf{C$^{14}$N, $N$=1 -- 0} & 30.52  \\
			&      &                         & 18  & \textbf{C$^{15}$N, $N$=1 -- 0} & 21.01  \\
			G059.78     & IRAM & 19:43:11.24 23:44:03.2  & 25  & \textbf{C$^{14}$N, $N$=1 -- 0} & 35.81  \\
			&      &                         & 25  & \textbf{C$^{15}$N, $N$=1 -- 0} & 38.24  \\
			G069.54     & IRAM & 20:10:09.07 31:31:35.9  & 20  & \textbf{C$^{14}$N, $N$=1 -- 0} & 22.61  \\
			&      &                         & 20  & \textbf{C$^{15}$N, $N$=1 -- 0} & 26.71  \\
			G078.12     & IRAM & 20:14:26.07 41:13:32.6  & 30  & \textbf{C$^{14}$N, $N$=1 -- 0} & 30.76  \\
			&      &                         & 30  & \textbf{C$^{15}$N, $N$=1 -- 0} & 32.15  \\
			G075.29     & IRAM & 20:16:16.01 37:35:45.8  & 70  & \textbf{C$^{14}$N, $N$=1 -- 0} & 76.74  \\
			&      &                         & 70  & C$^{15}$N, $N$=1 -- 0                           & 12.59  \\
			& ARO  & 20:16:16.01 37:35:45.8  & 40  & \textbf{C$^{14}$N, $N$=1 -- 0} & 15.71  \\
			&      &                         & ... & C$^{15}$N, $N$=1 -- 0                           &        \\
			G073.65     & IRAM & 20:16:21.93 35:36:06.0  & 58  & \textbf{C$^{14}$N, $N$=1 -- 0} & 29.02  \\
			&      &                         & 29  & C$^{15}$N, $N$=1 -- 0                           & 19.13  \\
			& ARO  & 20:16:21.93 35:36:06.0  & 80  & \textbf{C$^{14}$N, $N$=1 -- 0} & 14.30  \\
			&      &                         & ... & C$^{15}$N, $N$=1 -- 0                           &        \\
			G078.88     & IRAM & 20:29:24.82 40:11:19.5  & 70  & \textbf{C$^{14}$N, $N$=1 -- 0} & 12.56  \\
			&      &                         & 30  & \textbf{C$^{15}$N, $N$=1 -- 0} & 14.24  \\
			G081.75     & IRAM & 20:39:01.99 42:24:59.2  & 30  & \textbf{C$^{14}$N, $N$=1 -- 0} & 25.48  \\
			&      &                         & 30  & \textbf{C$^{15}$N, $N$=1 -- 0} & 23.51  \\
			WB018       & IRAM & 20:58:23.00 48:32:48.0  & 38  & \textbf{C$^{14}$N, $N$=1 -- 0} & 23.48  \\
			&      &                         & 38  & C$^{15}$N, $N$=1 -- 0                           & 19.05  \\
			WB021       & IRAM & 21:01:34.93 48:55:01.0  & 38  & \textbf{C$^{14}$N, $N$=1 -- 0} & 21.60  \\
			&      &                         & 38  & C$^{15}$N, $N$=1 -- 0                           & 16.25  \\
			WB022       & IRAM & 21:02:01.90 48:02:08.0  & 76  & \textbf{C$^{14}$N, $N$=1 -- 0} & 16.23  \\
			&      &                         & 76  & C$^{15}$N, $N$=1 -- 0                           & 13.03  \\
			WB042       & IRAM & 21:09:11.10 53:34:27.0  & 38  & \textbf{C$^{14}$N, $N$=1 -- 0} & 28.12  \\
			&      &                         & 38  & C$^{15}$N, $N$=1 -- 0                           & 13.10  \\
			G092.67     & IRAM & 21:09:21.73 52:22:37.0  & 60  & \textbf{C$^{14}$N, $N$=1 -- 0} & 20.25  \\
			&      &                         & 60  & \textbf{C$^{15}$N, $N$=1 -- 0} & 37.96  \\
			WB044       & IRAM & 21:09:41.30 47:58:07.0  & 38  & \textbf{C$^{14}$N, $N$=1 -- 0} & 20.44  \\
			&      &                         & 38  & C$^{15}$N, $N$=1 -- 0                           & 16.75  \\
			WB045       & IRAM & 21:09:46.53 48:10:59.4  & 38  & \textbf{C$^{14}$N, $N$=1 -- 0} & 21.99  \\
			&      &                         & 38  & C$^{15}$N, $N$=1 -- 0                           & 16.22  \\
			IRAS21156   & IRAM & 21:17:14.01 51:54:16.8  & 11  & \textbf{C$^{14}$N, $N$=1 -- 0} & 60.34  \\
			&      &                         & 11  & C$^{15}$N, $N$=1 -- 0                           & 22.23  \\
			WB066       & IRAM & 21:18:52.80 55:03:22.6  & 38  & \textbf{C$^{14}$N, $N$=1 -- 0} & 22.99  \\
			&      &                         & 38  & C$^{15}$N, $N$=1 -- 0                           & 13.87  \\
			WB081       & IRAM & 21:27:33.00 56:05:09.0  & 38  & \textbf{C$^{14}$N, $N$=1 -- 0} & 26.22  \\
			&      &                         & 38  & C$^{15}$N, $N$=1 -- 0                           & 14.65  \\
			IRAS21410   & IRAM & 21:42:52.91 50:23:11.4  & 11  & \textbf{C$^{14}$N, $N$=1 -- 0} & 48.23  \\
			&      &                         & 11  & C$^{15}$N, $N$=1 -- 0                           & 26.17  \\
			G105.41     & IRAM & 21:43:06.48 66:06:55.3  & 124 & \textbf{C$^{14}$N, $N$=1 -- 0} & 25.50  \\
			&      &                         & 124 & C$^{15}$N, $N$=1 -- 0                           & 29.28  \\
			IRAS21418   & IRAM & 21:43:30.72 54:16:45.0  & 11  & \textbf{C$^{14}$N, $N$=1 -- 0} & 44.76  \\
			&      &                         & 11  & C$^{15}$N, $N$=1 -- 0                           & 24.68  \\
			WB123       & IRAM & 21:46:07.12 57:26:31.8  & 19  & \textbf{C$^{14}$N, $N$=1 -- 0} & 19.49  \\
			&      &                         & 19  & C$^{15}$N, $N$=1 -- 0                           & 13.63  \\
			WB124       & IRAM & 21:46:36.80 57:12:25.0  & 38  & \textbf{C$^{14}$N, $N$=1 -- 0} & 14.82  \\
			&      &                         & 38  & C$^{15}$N, $N$=1 -- 0                           & 10.97  \\
			WB132       & IRAM & 21:50:12.60 56:59:24.0  & 38  & \textbf{C$^{14}$N, $N$=1 -- 0} & 17.37  \\
			&      &                         & 38  & C$^{15}$N, $N$=1 -- 0                           & 13.43  \\
			WB136       & IRAM & 21:53:38.80 56:27:53.0  & 38  & \textbf{C$^{14}$N, $N$=1 -- 0} & 19.41  \\
			&      &                         & 38  & C$^{15}$N, $N$=1 -- 0                           & 15.16  \\
			WB144       & IRAM & 21:57:44.59 58:21:06.0  & 38  & \textbf{C$^{14}$N, $N$=1 -- 0} & 31.01  \\
			&      &                         & 38  & C$^{15}$N, $N$=1 -- 0                           & 13.57  \\
			WB163       & IRAM & 22:16:28.60 60:03:49.0  & 9   & \textbf{C$^{14}$N, $N$=1 -- 0} & 30.45  \\
			&      &                         & 9   & C$^{15}$N, $N$=1 -- 0                           & 22.35  \\
			WB171       & IRAM & 22:19:17.97 63:18:52.9  & 38  & \textbf{C$^{14}$N, $N$=1 -- 0} & 23.27  \\
			&      &                         & 38  & \textbf{C$^{15}$N, $N$=1 -- 0} & 14.12  \\
			WB173       & IRAM & 22:19:27.70 63:32:56.0  & 38  & \textbf{C$^{14}$N, $N$=1 -- 0} & 104.00 \\
			&      &                         & 38  & C$^{15}$N, $N$=1 -- 0                           & 16.59  \\
			WB176       & IRAM & 22:21:22.50 63:51:13.0  & 38  & \textbf{C$^{14}$N, $N$=1 -- 0} & 43.08  \\
			&      &                         & 38  & C$^{15}$N, $N$=1 -- 0                           & 15.99  \\
			WB182       & IRAM & 22:28:29.30 62:59:44.0  & 38  & \textbf{C$^{14}$N, $N$=1 -- 0} & 93.32  \\
			&      &                         & 38  & C$^{15}$N, $N$=1 -- 0                           & 14.05  \\
			WB184       & IRAM & 22:28:52.20 64:13:43.0  & 38  & \textbf{C$^{14}$N, $N$=1 -- 0} & 21.47  \\
			&      &                         & 38  & C$^{15}$N, $N$=1 -- 0                           & 20.67  \\
			WB191       & IRAM & 22:32:46.00 58:28:22.0  & 38  & \textbf{C$^{14}$N, $N$=1 -- 0} & 71.44  \\
			&      &                         & 38  & C$^{15}$N, $N$=1 -- 0                           & 15.34  \\
			WB196       & IRAM & 22:35:08.20 69:10:42.0  & 38  & \textbf{C$^{14}$N, $N$=1 -- 0} & 55.97  \\
			&      &                         & 38  & C$^{15}$N, $N$=1 -- 0                           & 15.50  \\
			WB195       & IRAM & 22:35:09.00 65:41:29.0  & 38  & \textbf{C$^{14}$N, $N$=1 -- 0} & 58.93  \\
			&      &                         & 38  & C$^{15}$N, $N$=1 -- 0                           & 15.65  \\
			G109.87     & IRAM & 22:56:18.05 62:01:49.5  & 60  & \textbf{C$^{14}$N, $N$=1 -- 0} & 23.62  \\
			&      &                         & 60  & \textbf{C$^{15}$N, $N$=1 -- 0} & 19.02  \\
			& ARO  & 22:56:18.05 62:01:49.5  & 20  & \textbf{C$^{14}$N, $N$=1 -- 0} & 22.31  \\
			&      &                         & 75  & \textbf{C$^{15}$N, $N$=1 -- 0} & 8.92   \\
			G111.54     & IRAM & 23:13:45.36 61:28:10.5  & 20  & \textbf{C$^{14}$N, $N$=1 -- 0} & 14.69  \\
			&      &                         & 20  & \textbf{C$^{15}$N, $N$=1 -- 0} & 14.92  \\
			& ARO  & 23:13:45.36 61:28:10.5  & 20  & \textbf{C$^{14}$N, $N$=1 -- 0} & 18.20  \\
			&      &                         & 25  & \textbf{C$^{15}$N, $N$=1 -- 0} & 12.48  \\
			G122.01     & ARO  & 00:44:58.39 55:46:47.6  & 20  & \textbf{C$^{14}$N, $N$=1 -- 0} & 24.23  \\
			&      &                         & ... & C$^{15}$N, $N$=1 -- 0                           &        \\
			G123.062    & ARO  & 00:52:24.19 56:33:43.1  & 12  & \textbf{C$^{14}$N, $N$=1 -- 0} & 30.55  \\
			&      &                         & ... & C$^{15}$N, $N$=1 -- 0                           &        \\
			G134.62     & ARO  & 02:22:51.71 58:35:11.4  & 28  & C$^{14}$N, $N$=1 -- 0                           & 20.51  \\
			&      &                         & ... & C$^{15}$N, $N$=1 -- 0                           &        \\
			G136.84     & ARO  & 02:49:33.60 60:48:27.6  & 40  & \textbf{C$^{14}$N, $N$=1 -- 0} & 17.09  \\
			&      &                         & ... & C$^{15}$N, $N$=1 -- 0                           &        \\
			G170.65     & ARO  & 05:20:22.07 36:37:56.6  & 56  & \textbf{C$^{14}$N, $N$=1 -- 0} & 11.80  \\
			&      &                         & ... & C$^{15}$N, $N$=1 -- 0                           &        \\
			G176.51     & ARO  & 05:37:52.13 32:00:03.9  & 28  & \textbf{C$^{14}$N, $N$=1 -- 0} & 23.57  \\
			&      &                         & ... & C$^{15}$N, $N$=1 -- 0                           &        \\
			G183.72     & ARO  & 05:40:24.22 23:50:54.7  & 12  & \textbf{C$^{14}$N, $N$=1 -- 0} & 19.94  \\
			&      &                         & ... & C$^{15}$N, $N$=1 -- 0                           &        \\
			G192.16     & ARO  & 05:58:13.53 16:31:58.9  & 12  & \textbf{C$^{14}$N, $N$=1 -- 0} & 19.29  \\
			&      &                         & ... & C$^{15}$N, $N$=1 -- 0                           &        \\
			G213.70     & ARO  & 06:07:47.85 -06:22:56.5 & 8   & \textbf{C$^{14}$N, $N$=1 -- 0} & 51.53  \\
			&      &                         & ... & C$^{15}$N, $N$=1 -- 0                           &        \\
			G188.94     & ARO  & 06:08:53.34 21:38:29.1  & 8   & \textbf{C$^{14}$N, $N$=1 -- 0} & 36.22  \\
			&      &                         & ... & C$^{15}$N, $N$=1 -- 0                           &        \\
			G188.79     & ARO  & 06:09:06.97 21:50:41.4  & 8   & \textbf{C$^{14}$N, $N$=1 -- 0} & 36.82  \\
			&      &                         & ... & C$^{15}$N, $N$=1 -- 0                           &        \\
			G217.79     & ARO  & 07:04:04.82 -03:50:50.6 & 44  & C$^{14}$N, $N$=1 -- 0                           & 12.41  \\
			&      &                         & ... & C$^{15}$N, $N$=1 -- 0                           &        \\
			G239.35     & ARO  & 07:22:58.32 -25:46:03.0 & 44  & \textbf{C$^{14}$N, $N$=1 -- 0} & 14.76  \\
			&      &                         & ... & C$^{15}$N, $N$=1 -- 0                           &        \\
			G229.57     & ARO  & 07:23:01.77 -14:41:34.3 & 40  & \textbf{C$^{14}$N, $N$=1 -- 0} & 27.98  \\
			&      &                         & ... & C$^{15}$N, $N$=1 -- 0                           &        \\
			G232.62     & ARO  & 07:32:09.78 -16:58:12.8 & 40  & \textbf{C$^{14}$N, $N$=1 -- 0} & 14.54  \\
			&      &                         & ... & C$^{15}$N, $N$=1 -- 0                           &        \\
			G236.81     & ARO  & 07:44:28.23 -20:08:30.6 & 40  & \textbf{C$^{14}$N, $N$=1 -- 0} & 28.59  \\
			&      &                         & ... & C$^{15}$N, $N$=1 -- 0                           &        \\
			G240.31     & ARO  & 07:44:51.96 -24:07:41.3 & 40  & \textbf{C$^{14}$N, $N$=1 -- 0} & 51.91  \\
			&      &                         & ... & C$^{15}$N, $N$=1 -- 0                           &        \\
			G090.92     & ARO  & 21:09:12.96 50:01:03.6  & 40  & \textbf{C$^{14}$N, $N$=1 -- 0} & 22.60  \\
			&      &                         & ... & C$^{15}$N, $N$=1 -- 0                           &        \\
			G097.53     & ARO  & 21:32:12.43 55:53:49.6  & 40  & \textbf{C$^{14}$N, $N$=1 -- 0} & 21.27  \\
			&      &                         & ... & C$^{15}$N, $N$=1 -- 0                           &        \\
			G095.29     & ARO  & 21:39:40.50 51:20:32.8  & 40  & \textbf{C$^{14}$N, $N$=1 -- 0} & 15.63  \\
			&      &                         & ... & C$^{15}$N, $N$=1 -- 0                           &        \\
			G094.60     & ARO  & 21:39:58.27 50:14:20.9  & 40  & \textbf{C$^{14}$N, $N$=1 -- 0} & 14.98  \\
			&      &                         & ... & C$^{15}$N, $N$=1 -- 0                           &        \\
			G100.37     & ARO  & 22:16:10.36 52:21:34.1  & 40  & \textbf{C$^{14}$N, $N$=1 -- 0} & 16.58  \\
			&      &                         & ... & C$^{15}$N, $N$=1 -- 0                           &        \\
			G108.20     & ARO  & 22:49:31.47 59:55:42.0  & 40  & \textbf{C$^{14}$N, $N$=1 -- 0} & 16.01  \\
			&      &                         & ... & C$^{15}$N, $N$=1 -- 0                           &        \\
			G108.42     & ARO  & 22:49:58.87 60:17:56.6  & 40  & C$^{14}$N, $N$=1 -- 0                           & 17.90  \\
			&      &                         & ... & C$^{15}$N, $N$=1 -- 0                           &        \\
			G108.59     & ARO  & 22:52:38.31 60:00:51.8  & 20  & \textbf{C$^{14}$N, $N$=1 -- 0} & 21.79  \\
			&      &                         & ... & C$^{15}$N, $N$=1 -- 0                           &        \\
			G110.19     & ARO  & 22:57:29.80 62:29:46.8  & 20  & \textbf{C$^{14}$N, $N$=1 -- 0} & 27.38  \\
			&      &                         & ... & C$^{15}$N, $N$=1 -- 0                           &        \\
			G108.47     & ARO  & 23:02:32.08 56:57:51.4  & 32  & \textbf{C$^{14}$N, $N$=1 -- 0} & 21.04  \\
			&      &                         & ... & C$^{15}$N, $N$=1 -- 0                           &        \\
			G111.25     & ARO  & 23:16:10.34 59:55:28.6  & 20  & \textbf{C$^{14}$N, $N$=1 -- 0} & 19.83  \\
			&      &                         & ... & C$^{15}$N, $N$=1 -- 0                           &        \\
			G111.23     & ARO  & 23:17:20.78 59:28:46.9  & 20  & \textbf{C$^{14}$N, $N$=1 -- 0} & 20.24  \\
			&      &                         & ... & C$^{15}$N, $N$=1 -- 0                           &        \\
			G115.05     & ARO  & 23:44:03.28 61:47:22.1  & 20  & \textbf{C$^{14}$N, $N$=1 -- 0} & 19.06  \\
			&      &                         & ... & C$^{15}$N, $N$=1 -- 0                           &       
			\enddata
		\end{deluxetable}
		\tablecomments{Column (1): source name; Column(2): used Telescope; Column(3): Right ascension (J2000) and Declination (J2000); Column(4): total integration time; Column(5): molecular line detections in boldface; Column(6): the rms noise level with a spectral resolution of $\sim$1.1 km$\,$s for IRAM 30 m and 0.8 km$\,$s$^{-1}$ for ARO 12 m observations.}	





\end{CJK*}
\end{document}